\documentclass{statsoc}

\usepackage[a4paper]{geometry}
\usepackage{amssymb, amsmath, multirow}
\usepackage{etoolbox}

\makeatletter
\patchcmd{\@makecaption}
  {\parbox}
  {\advance\@tempdima-\fontdimen2} 
  {}{}
\makeatother  

\usepackage{graphicx, afterpage}
\usepackage[caption = false]{subfig} 
\usepackage{enumerate}
\usepackage{comment}
\usepackage{natbib, float}

\usepackage{epsfig, latexsym, url, xcolor, bbm}

\makeatletter
\patchcmd{\@makecaption}
  {\parbox}
  {\advance\@tempdima-\fontdimen2} 
  {}{}
\makeatother

\usepackage{float} 
\floatstyle{plain}
\newfloat{Algorithm}{thp}{lop}
\floatname{Algorithm}{Algorithm}

\RequirePackage[colorlinks,citecolor=blue,urlcolor=blue]{hyperref}

\newtheorem{theorem}{Theorem}
\newtheorem{corollary}{Corollary}

\newtheorem{lemma}{Lemma}

\newtheorem{thma}{Bonnet's Theorem (Stein's Lemma)}

\newtheorem{thmb}{Price's Theorem}

\newcommand\ev{{\text{ev}} }   
\newcommand\df{{\text{d}}}   
\newcommand\e{{\rm e}}                             
\newcommand\E{{\text{E}}}           
\newcommand\dg{{\text{dg}}}         
\newcommand\Var{{\text{Var}}}   
\newcommand\blockdiag{{\text{blockdiag}}}                           
\newcommand\mL{{\mathcal{L}}}

\newcommand\Cov{{\text{Cov}}}   
   
\newcommand\diag{{\text{diag}}}   
\newcommand\N{{\text{N}}} 
 
\newcommand\G{{\mathcal{G}}} 
\newcommand\F{{\mathcal{F}}} 
\newcommand\tr{\text{tr}}   
\newcommand\logit{\text{logit}} 
\newcommand\KL{\text{KL}} 
\newcommand\MCMC{\text{MCMC}} 
\newcommand\MMD{\text{MMD}} 
\newcommand\tll{\text{tll}}   
\newcommand{\vc}{\text{vec}}  
\newcommand{\vech}{\text{vech}}   
\newcommand{\Prob}{\text{P}}   
\newcommand{\dH}{\bar{\bar{H}}}  

\usepackage{amsmath}

\newcommand{\ra}[1]{\renewcommand{\arraystretch}{#1}}
\graphicspath{{img/}}

\title[Natural gradient updates in Gaussian variational approximation]{Analytic natural gradient updates for Cholesky factor in Gaussian variational approximation}
\author[Linda Tan]{Linda S. L. Tan”}
\address{Department of Statistics and Data Science, National University of Singapore,
117546 Singapore.}
\email{statsll@nus.edu.sg}

\begin{document}

\begin{abstract}
Natural gradients can improve convergence in stochastic variational inference significantly but inverting the Fisher information matrix is daunting in high dimensions. Moreover, in Gaussian variational approximation, natural gradient updates of the precision matrix do not ensure positive definiteness. To tackle this issue, we derive analytic natural gradient updates of the Cholesky factor of the covariance or precision matrix, and consider sparsity constraints representing different posterior correlation structures. Stochastic normalized natural gradient ascent with momentum is proposed for implementation in generalized linear mixed models and deep neural networks.
\end{abstract}


\section{Introduction}
Variational inference is fast and provides an attractive alternative to Markov chain Monte Carlo (MCMC) methods for approximating intractable posterior distributions in Bayesian analysis. Stochastic gradient methods \citep{Robbins1951} further enabled variational inference for high-dimensional models and large data sets \citep{Sato2001, Hoffman2013, Salimans2013}. Euclidean gradients are often used in optimizing the variational objective function, called evidence lower bound. However, the steepest ascent direction in the parameter space, where distance between densities is measured using Kullback-Leibler (KL) divergence, is actually given by the natural gradient \citep{Amari1998}, which is obtained by premultiplying the Euclidean gradient with the inverse Fisher information. Stochastic optimization based on natural gradients can be more robust, with the ability to escape plateaus, yielding faster convergence \citep{Rattray1998}. \cite{Martens2020} showed that natural gradient descent can be viewed as a second order optimization method, with the Fisher information in place of the Hessian.

Various alternatives to natural gradient descent similarly account for geometry of the evidence lower bound. The KL proximal-point algorithm \citep{Khan2015} uses KL divergence as a proximal term, and is equivalent to natural gradient descent for conditionally-conjugate exponential-family models. \cite{Martens2015} used a Kronecker-factored approximation to the Fisher information matrix in natural gradient descent for neural networks, while \cite{Zhang2018} approximated the negative log-likelihood Hessian in the noisy natural gradient with the Fisher information matrix in point estimation. The variational predictive natural gradient \citep{Tang2019} improves on the natural gradient by accounting for curvature of the expected log-likelihood in the evidence lower bound. \cite{Kim2023} provide convergence guarantees for proximal stochastic gradient descent in black-box variational inference. 

Computing natural gradients is complicated, but natural gradient updates for conjugate exponential family models can be simpler than Euclidean ones \citep{Hoffman2013}. For variational densities in the minimal exponential family \citep{Wainwright2008}, natural gradient of the evidence lower bound with respect to the natural parameter is just the Euclidean gradient with respect to the expected sufficient statistics \citep{Khan2017}. In Gaussian variational approximation \citep{Opper2009}, stochastic natural gradient updates of the mean and precision, which depend respectively on the first and second derivatives of the log posterior density \citep{Khan2018} can be derived using the theorems of \cite{Bonnet1964} and \cite{Price1958}, but the precision matrix's update does not ensure positive definiteness. 

To handle the positive definite constraint, \cite{Khan2017} use back-tracking line search. \cite{Khan2018} use the generalized Gauss-Newton approximation for the log likelihood Hessian in variational online Gauss-Newton (VOGN), which is later applied to deep networks by \cite{Osawa2019}. \cite{Ong2018a} compute natural gradients for Cholesky factor of the precision matrix by solving a linear system numerically. For one-one transformations of the natural parameter, the inverse Fisher information can be computed as a Jacobian-vector product via automatic differentiation \citep{Salimbeni2018}. Using a factor structure for the covariance, \cite{Tran2020} compute natural gradients via a conjugate gradient linear solver based on a block diagonal approximation of the Fisher information. \cite{Lin2020} use Riemannian gradient descent with a retraction map to obtain a precision matrix update that has an additional term to ensure positive definiteness. \cite{Tran2021} derive an update of the covariance matrix based on an approximation of the natural gradient, and a popular retraction for the manifold of symmetric positive definite matrices. \cite{Lin2021} use local parameter coordinates to derive natural gradient updates for low-rank, square root and sparse matrices.

We consider Cholesky decomposition of the covariance or precision matrix, and derive the inverse Fisher information in closed form to obtain analytic natural gradient updates. Our derivation is based on properties of elimination and commutation matrices for handling lower triangular matrices \citep{Magnus1980} and is independent of \cite{Lin2021}'s local parameter coordinates approach. Extending \cite{Lin2019a}, we obtain unbiased natural gradient estimates with respect to the Cholesky factor in terms of the first or second order derivative of the log posterior via Stein's Lemma \citep{Stein1981}. First order gradient estimates are more efficient computationally and storage wise, but second order estimates have lower variance close to the mode and are especially useful for high-dimensional dense covariance matrices. Compared with Euclidean updates of the mean and Cholesky factor \citep{Titsias2014}, natural gradient updates require more computation, but can improve convergence significantly. Updating the Cholesky factor instead of the precision matrix also has advantages in terms of storage, computation (finding the determinant and inverse) and simulations. 

Gaussian variational approximation has been applied in many contexts, such as likelihood-free inference \citep{Ong2018a}, deep Bayesian neural networks \citep{Khan2018}, exponential random graph models \citep{Tan2020a} and factor copula models \citep{Nguyen2020}. A Gaussian variational approximation can be specified for variables which have undergone independent parametric transformations, resulting in a Gaussian copula variational approximation \citep{Han2016, Smith2020}. To accommodate constrained, skewed or heavy-tailed variables, \cite{Lin2019b} developed natural gradient variational inference for mixture of exponential family distributions. The multivariate Gaussian can also be used as a prior for parameters in mean-field variational approximation for other types of latent variables, to expand the variational model hierarchically and induce dependencies among latent variables \citep{Ranganath2016}. Our natural gradient updates are also useful in these contexts. 

For high-dimensional models, we derive efficient natural gradient updates under sparsity constraints, where (i) the covariance matrix is a block diagonal representing the variational Bayes assumption \citep{Attias1999}, and (ii) the precision matrix has a sparse structure mirroring posterior conditional independence in a hierarchical model where local variables are independent given global ones \citep{Tan2018}. Sparsity constraints can be easily imposed on Euclidean gradients by setting relevant entries to zero, but the same does not apply to natural gradients due to premultiplication by the inverse Fisher information. In comparison, the automatic differentiation variational inference algorithm in Stan \citep{Kucukelbir2017} only allows Gaussian variational approximations with a diagonal or full covariance matrix, and uses Euclidean gradients to update the Cholesky factor in stochastic gradient ascent. 

Finally, we demonstrate that the learning rate, Adam \citep{Kingma2015}, which can be interpreted as a sign-based approach with per dimension variance adaptation \citep{Balles2018}, is incompatible with natural gradients as it neglects their scale information. We propose stochastic normalized natural gradient ascent with momentum (Snngm) as an alternative, which is shown to converge if the objective function is $L$-Lipschitz smooth with bounded gradients. Our approach differs from \cite{Cutkosky2020} in normalization of the natural gradient instead of momentum of the Euclidean gradients. Stochastic normalized gradient descent is suited to non-convex optimization as it can overcome plateaus/cliffs in objective functions \citep{Hazan2015}. 

Section \ref{sec_SVI} introduces variational inference, where intractable lower bounds are optimized using stochastic gradient ascent, and Section \ref{sec_natgrad} motivates the use of natural gradients. Section \ref{sec_GVA} derives the natural gradient update of the natural parameter in Gaussian variational approximation, and compares it with Euclidean gradients and other parametrizations. In Section \ref{sec_Gauss}, we present natural gradient updates of the mean and Cholesky factor of the covariance/precision matrix, and unbiased estimates based on the first or second derivative of the log posterior for intractable lower bounds. Incompatibility of Adam with natural gradients is demonstrated in Section \ref{sec_Normalized SGD} and Snngm is proposed for implementation. Section \ref{sec imposing sparsity} presents natural gradient updates for high-dimensional models where sparsity constraints are imposed. Performance of natural gradient updates is investigated using generalized linear mixed models (GLMMs) and deep neural networks in Section \ref{sec_applications}. We conclude with a discussion of future work in Section \ref{sec_Conclusion}.

\section{Stochastic gradient variational inference} \label{sec_SVI}
Let $p(y|\theta)$ denote the likelihood of variables $\theta \in \mathbb{R}^d$ given observed data $y$, and $p(\theta)$ be a prior density for $\theta$. In variational inference, an intractable posterior distribution, $p(\theta|y) = p(y|\theta)p(\theta)/p(y)$, is approximated by a more tractable density $q_\lambda(\theta)$ with parameter $\lambda$, that minimizes the KL divergence between $q_\lambda(\theta)$ and $p(\theta|y)$. As
\[
\log p(y) = \underbrace{\int q_\lambda(\theta) \log \frac{q_\lambda(\theta)}{p(\theta|y)} \df \theta}_{\text{KL divergence}} 
+ \underbrace{\int q_\lambda(\theta) \log \frac{p(y, \theta)}{q_\lambda(\theta)} \df \theta}_{\text{Evidence lower bound, $\mL(\lambda)$}},
\]
minimizing the KL divergence is equivalent to maximizing the lower bound on the log marginal likelihood, $\log p(y)$, with respect to $\lambda$. Let $\mL(\lambda) = \E_q [h(\theta)]$ denote the lower bound, where $h(\theta) =\log p(y, \theta) - \log q_\lambda (\theta)$. When $\mL$ is intractable, stochastic gradient ascent can be used for optimization. Starting with an initial estimate of $\lambda$, an update 
\[
\lambda \leftarrow \lambda + \rho_t \, \widehat{\nabla}_{\lambda} \mL,
\]
is performed at iteration $t$, where $\widehat{\nabla}_{\lambda} \mL$ is an unbiased estimate of the Euclidean gradient $\nabla_{\lambda} \mL$. Applying chain rule, $\nabla_\lambda \mL = \int \{\nabla_\lambda q_\lambda (\theta)\} h(\theta) d \theta$, since $\E_q[\nabla_\lambda \log q_\lambda (\theta)] = 0$. Under regularity conditions, the algorithm will converge to a local maximum of $\mL$ if the stepsize $\rho_t$ satisfies $\sum_{t=1}^\infty \rho_t = \infty$ and $\sum_{t=1}^\infty \rho_t^2 < \infty$ \citep{Spall2003}.

\section{Natural gradient} \label{sec_natgrad}
Search for the optimal $\lambda$ is performed in the parameter space of $q_\lambda(\theta)$, which has its own curvature, and the Euclidean metric may not be appropriate for measuring the distance between densities indexed by different $\lambda$s. For instance, N(0.1, 1000) and N(0.2, 1000) are similar, while N(0.1, 0.1) and N(0.2, 0.1) are vastly different, but both pairs have the same Euclidean distance. \cite{Amari2016} defines the distance between $\lambda$ and $\lambda +d\lambda$ as $2\KL(q_\lambda \| q_{\lambda + d\lambda})$ for small $d\lambda$. Using a second order Taylor expansion, this is approximately equal to 
\[
2\E_q \left[ \log q_\lambda(\theta) - \{ \log q_\lambda(\theta) + d\lambda^\top   \nabla_\lambda \log q_\lambda(\theta) + \tfrac{1}{2} d\lambda^\top   \nabla_\lambda^2 \log q_\lambda (\theta) d\lambda \} \right] 
= d\lambda^\top   F_\lambda d\lambda,
\]
where $F_\lambda= -\E_q [\nabla_\lambda^2 \log q_\lambda (\theta)]$ is the Fisher information of $q_\lambda(\theta)$. Thus, the distance between $\lambda$ and $\lambda + d\lambda$ is not $d\lambda^\top   d\lambda$ as in a Euclidean space, but $d\lambda^\top   F_\lambda d\lambda$. The set of all distributions $q_\lambda(\theta)$ is a manifold and the KL divergence provides the manifold with a Riemannian structure, with Fisher norm $\| d\lambda \|_{F_\lambda} = \sqrt{d\lambda^\top F_\lambda d\lambda}$ if $F_\lambda$ is positive definite.

Steepest ascent direction of $\mL$ at $\lambda$ is defined as the vector $a$ that maximizes $\mL(\lambda + a)$, where $\|a\|_{F_\lambda}$ is equal to a small constant $\epsilon>0$ \citep{Amari1998}. Using Lagrange multipliers, 
\[
\begin{aligned}
\mathfrak{L} &= \mL(\lambda + a) - \alpha (\|a\|^2_{F_\lambda} - \epsilon^2) 
\approx \mL(\lambda) +  a^\top   \nabla_\lambda \mL - \alpha (a^\top   F_\lambda a -\epsilon^2).
\end{aligned}
\]
Setting $\nabla_{a} \mathfrak{L}  \approx \nabla_\lambda \mL - 2\alpha F_\lambda a$ to zero yields $a = \epsilon (\widetilde{\nabla}_\lambda \mL)/ \|\widetilde{\nabla}_\lambda \mL \|_{F_\lambda}$. Thus, steepest ascent direction in the parameter space is given by the natural gradient, 
\[
\widetilde{\nabla}_\lambda \mL = F_\lambda ^{-1} \nabla_\lambda \mL.
\]
Replacing the unbiased Euclidean gradient estimate with that of the natural gradient yields the stochastic natural gradient update, $\lambda \leftarrow \lambda + \rho_t \, F_{\lambda}^{-1} \,\widehat{\nabla}_{\lambda} \mL$.

Another motivation for using the natural gradient is that, provided $q_\lambda (\theta)$ is a good approximation to $p(\theta|y)$, then close to the mode,
\[
\nabla^2_\lambda \mL = \int \nabla_\lambda^2 q_\lambda(\theta) \left\{\log p(y, \theta) - \log q_\lambda(\theta) \right\}  \df \theta - F_\lambda \approx -F_\lambda,
\]
as the first term is approximately zero. Thus the natural gradient update resembles Newton-Raphson, a {\em second-order} optimization method, where $\lambda \leftarrow \lambda - (\nabla_\lambda^2 \mL)^{-1} \nabla_\lambda \mL$. If $\xi \equiv \xi(\lambda)$ is a smooth invertible reparametrization of $q_\lambda(\theta)$ and $J = \nabla_\xi \lambda$, then 
\begin{equation} \label{reparametrized natural gradient}
\widetilde{\nabla}_\xi \mL = F_\xi^{-1} \nabla_\xi \mL 
= ( J F_\lambda J^\top  )^{-1} J \nabla_\lambda \mL
= (\nabla_\lambda \xi)^\top   \widetilde{\nabla}_\lambda \mL.
\end{equation}

\section{Gaussian variational approximation} \label{sec_GVA}
A popular option for $q_\lambda(\theta)$ is the multivariate Gaussian, $\N(\mu, \Sigma)$, which is a member of the exponential family. It can be written as
\begin{equation} \label{exp family}
q_\lambda(\theta) = \exp \left\{ s(\theta)^\top   \lambda -  A(\lambda) \right\}, 
\quad s(\theta) = \begin{bmatrix} \theta \\ \vech(\theta\theta^\top ) \end{bmatrix},
\quad \lambda = \begin{bmatrix} \Sigma^{-1} \mu \\  -\frac{1}{2} D^\top \vc(\Sigma^{-1}) \end{bmatrix},
\end{equation}
where $s(\theta)$ is the sufficient statistic, $\lambda$ is the natural parameter and $A(\lambda) = \tfrac{1}{2} \mu^\top   \Sigma^{-1} \mu + \tfrac{1}{2} \log |\Sigma| + \frac{d}{2} \log(2\pi)$ is the log-partition function. For any square matrix $X$, $\vc(X)$ is the vector obtained by stacking columns of $X$ from left to right and $\vech(X)$ is obtained from $\vc(X)$ by omitting supradiagonal elements. If $X$ is symmetric, $D$ is the duplication matrix such that $D \vech(X) = \vc(X)$. For \eqref{exp family}, $m = \E [s(\theta)] = \nabla_\lambda A(\lambda)$ and $\Var[s(\theta)] = \nabla_\lambda^2 A(\lambda)  = \nabla_\lambda m = F_\lambda$. Applying chain rule, $\nabla_{\lambda} \mL  = \nabla_\lambda m  \nabla_m \mL = F_\lambda \nabla_m \mL$ \citep{Khan2017}. Thus, the natural gradient,
\begin{equation*} 
\widetilde{\nabla}_\lambda \mL = \nabla_m \mL 
= \begin{bmatrix} \nabla_\mu \mL - 2(\nabla_{\Sigma}\mL) \mu \\ D^\top   \vc(\nabla_{\Sigma} \mL)
 \end{bmatrix},
\end{equation*}
where $\vc(\nabla_{\Sigma} \mL) = \nabla_{\vc(\Sigma)} \mL$, can be obtained without finding $F_\lambda^{-1}$ explicitly. The derivation is in the supplement S1 and the natural gradient update for $\lambda$ is in Table \ref{Tab a}. 

\begin{table}
\caption{\label{Tab a} Natural gradient updates. $\nabla_\mu \mL$ and $\nabla_\Sigma \mL$ are evaluated at $(\mu^{(t)}, \Sigma^{(t)})$.} 
\centering \ra{1.2}
\fbox{\resizebox{\columnwidth}{!}{
\begin{tabular}{ccc}
$\kappa$ & $\xi$ & $\lambda$ (natural parameter)   \\ \hline
$\Sigma^{(t+1)} = \Sigma^{(t)} + 2 \rho_t  \Sigma^{(t)} \nabla_\Sigma \mL \Sigma^{(t)}$ & 
$\Sigma^{-1(t+1)} =\Sigma^{-1(t)} - 2 \rho_t  \nabla_\Sigma \mL$ & 
$\Sigma^{-1(t+1)} = \Sigma^{-1(t)} - 2 \rho_t  \nabla_\Sigma \mL$
\\ 
$\mu^{(t+1)} = \mu^{(t)} + \rho_t \Sigma^{(t)} \nabla_\mu \mL$ & 
$\mu^{(t+1)} = \mu^{(t)} + \rho_t \Sigma^{(t)} \nabla_\mu \mL$ & 
$\mu^{(t+1)} = \mu^{(t)}  + \rho_t \Sigma^{(t+1)} \nabla_\mu \mL$ \\ 
\end{tabular}}}
\end{table}

Consider other parametrizations, $\kappa = (\mu^\top, \vech(\Sigma)^\top )^\top$ and $\xi = (\mu^\top, \vech(\Sigma^{-1})^\top  )^\top$, which are one-one transformations of $\lambda$. The natural gradients $\widetilde{\nabla}_\kappa \mL$ and $\widetilde{\nabla}_\xi \mL$ are derived using \eqref{reparametrized natural gradient} in the supplement S1, and corresponding updates are in Table \ref{Tab a}. The update for $\xi$ is almost identical to $\lambda$, except that the update of $\mu$ for $\lambda$ relies on the updated $\Sigma$, unlike that for $\xi$. The Fisher information of $\kappa$ and $\xi$ are block diagonal matrices, which imply that $\kappa$ and $\xi$ are orthogonal parametrizations. However, it is only through the non-orthogonal parametrization $\lambda$, that we discover that the updated $\Sigma$ can be used to improve the update of $\mu$, due to its curvature.

\subsection{An illustration using Poisson  loglinear model} \label{sec_Poisson}
To gain insights on how natural gradient updates in Table \ref{Tab a} compare with the Euclidean gradient update, 
\[
\mu^{(t+1)} = \mu^{(t)} + \rho_t \nabla_\mu \mL, \quad \Sigma^{(t+1)} = \Sigma^{(t)} + \rho_t  \nabla_\Sigma \mL,
\]
consider the loglinear model. Let $y_i \sim \text{Poisson}(\delta_i)$ and $\log \delta_i = x_i^\top \theta$ for $i=1, \dots, n$, where $x_i$ and $\theta$ denote covariates and regression coefficients respectively. Consider a prior, $\theta \sim \N(0, \sigma_0^2 I)$ where $\sigma_0^2 = 100$, and a Gaussian approximation $\N(\mu, \Sigma)$ of the true posterior of $\theta$. The lower bound $\mL$ is tractable and hence its curvature can be studied easily. Expressions of $\mL$, $\nabla_\mu \mL$ and $\nabla_\Sigma \mL$ are given in the supplement S2. 

\begin{figure}[htb!]
\centering
\includegraphics[height=107pt, angle=0]{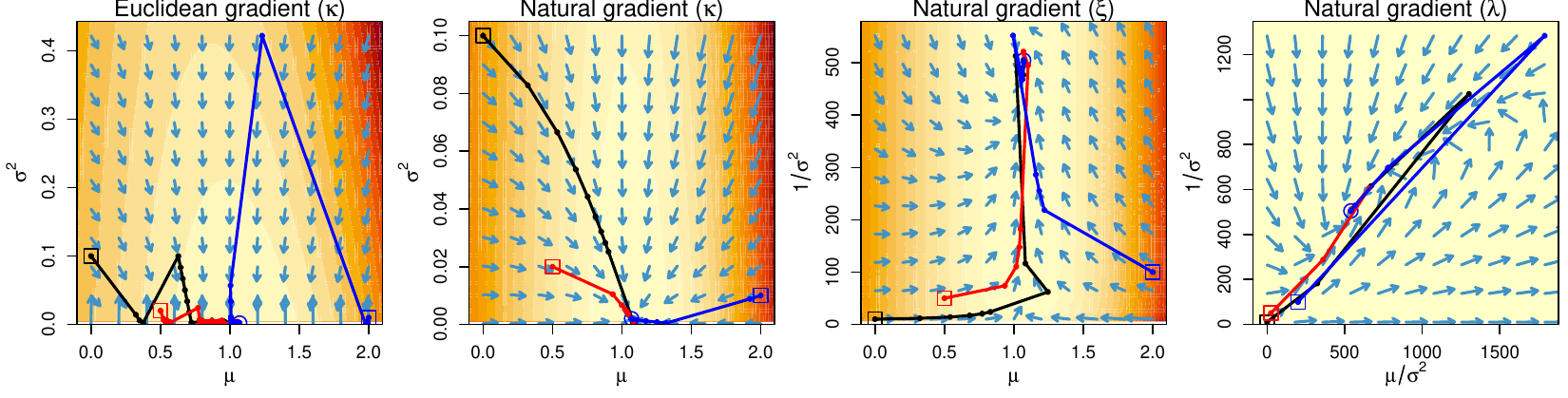}
\caption{Gradient vector field and trajectories for gradient ascent from 3 starting points.}
\label{Fig a}
\end{figure}

To visualize the gradient vector field, we consider intercept-only models and write $\Sigma$ as $\sigma^2$. Variational parameters $(\mu, \sigma^2)$ are estimated using gradient ascent and the largest possible stepsize $\rho_t \in \{1, 0.1, 0.01, ...\}$ is used in each iteration, provided that the update of $\sigma^2$ is positive and $\mL$ is increasing. We use as observations the number of satellites of 173 female horseshoe crabs \citep[Table 3.2,][]{Agresti2018}. Figure \ref{Fig a} shows the gradient vector field and gradient ascent trajectories from 3 starting points marked by squares. $\mL$ is maximized at $(\mu, \sigma^2) = (1.07, 0.002)$, which is marked by a circle. Number of iterations to converge and smallest $\rho_t$ used are reported in Table \ref{Tab b}.

\begin{table}
\caption{\label{Tab b} Loglinear model: number of iterations and smallest $\rho_t$ used.} 
\centering
\fbox{\begin{tabular}{ccccc}
Starting point & Euclidean ($\kappa$)  & Natural ($\kappa$) & Natural ($\xi$) & Natural ($\lambda$) \\ \hline
(0, 0.1) & 141 (1e-05) & 15 (0.01) & 11 (0.01) & 6 (1) \\ 
(0.5, 0.02) & 107 (1e-06) & 12 (0.1) & 8 (0.1) & 5 (1) \\ 
(2, 0.01) & 115 (1e-06) & 9 (0.01) & 8 (0.1) & 5 (1) \\ 
\end{tabular}}
\end{table}

The first two plots show a sharp contrast between the Euclidean and natural gradient vector fields for $\kappa = (\mu, \sigma^2)$, especially when $\sigma^2$ is close to zero. While natural gradients are collectively directed at the mode, Euclidean gradients have strong vertical components, causing longer zigzag trajectories. Number of iterations required for Euclidean gradients is an order of magnitude larger than natural gradients, and a much smaller stepsize has to be used at some point to avoid a negative $\sigma^2$ update or $\mL$ decreasing. Natural gradient ascent for $\lambda$ is most efficient, likely because natural gradient updates of stepsize 1 correspond to fixed point iteration for the natural parameter \citep{Tan2013}. In the one-dimensional case, we can reparametrize $\sigma^2 = \log\{1 + \exp(w)\}$ for instance, and update $w$ without using smaller stepsize to ensure positiveness, but such techniques are hard to apply in higher dimensions. We chose $d=1$ for visualization but issues of ensuring positive definiteness of $\Sigma$ occur similarly in higher dimensions.

\section{Natural gradient updates for mean and Cholesky factor} \label{sec_Gauss}
In many applications, the mean and Cholesky factor of the covariance/precision matrix in Gaussian variational approximation are updated using stochastic gradient ascent, which avoids positive definite constraints and allows flexibility in choice of stepsize, reduction in computation/storage costs and ease in simulations. However, existing analytic updates for Cholesky factors are based on Euclidean gradients \citep{Titsias2014, Tan2018}. We derive the natural gradient counterparts by considering $\lambda = (\mu^\top, \vech(C)^\top)^\top$ where $\Sigma = CC^\top$, or $\lambda = (\mu^\top, \vech(T)^\top)^\top $ where $\Sigma^{-1} = TT^\top $, such that $C$ and $T$ are lower triangular matrices. In these cases, $\lambda$ is not the natural parameter and we need to find the inverse Fisher information explicitly to get the natural gradient. The Fisher information for both cases turn out to be block diagonal matrices of the same form. Hence the inverse can be found using a common result in (ii) of Lemma \ref{lem inv}, while (iii) is useful in simplifying $F_\lambda^{-1} \nabla_\lambda \mL$. The Fisher information and natural gradient for these two parametrizations are presented in Theorem \ref{thm inv}. 

Let $X \in \mathbb{R}^{d \times d}$. In Lemma \ref{lem inv}, $K$ is the commutation matrix such that $K \vc(X) = \vc(X^\top )$, and $L$ is the elimination matrix such that $L \vc(X) = \vech(X)$. If $X$ is lower triangular, then $L^\top \vech(X) = \vc(X)$. \cite{Magnus1980} highlight that one should differentiate with respect to $\vech(X)$ instead of $\vc(X)$ if $X$ is symmetric or lower triangular. The transformation matrices, $K$, $L$ and $N = (K+ I_{d^2})/2$, are useful in handling the redundancy of supradiagonal elements. Let $\bar{X}$ be the lower triangular matrix derived from $X$ by replacing supradiagonal elements by zero, while $\bar{\bar{X}}$ is obtained from $\bar{X}$ by halving the diagonal.

\begin{lemma} \label{lem inv}
Let $C$ be any $d \times d$ lower triangular matrix and 
\[
\mathfrak{I}(C) = L\{ (C^{-1} \otimes  C^{-\top})K + (I_d \otimes C^{-\top}C^{-1})  \} L^\top .
\]
\begin{enumerate}[(i)]
\item $\mathfrak{I}(C) = 2 L ( I_d \otimes C^{-\top})N ( I_d \otimes C^{-1})L^\top $.
\item $\mathfrak{I}(C)^{-1} = \tfrac{1}{2}L ( I_d \otimes C) L^\top  (L N L^\top  )^{-1}L (I_d \otimes C^\top  ) L^\top $.
\item $\mathfrak{I}(C)^{-1} \vech(G) = \vech(C \dH)$ for any $d \times d$ matrix $G$, where $H = C^\top  \bar{G}$.
\end{enumerate}
\end{lemma}

\begin{theorem} \label{thm inv}
(i) If $\lambda = (\mu^\top, \vech(C)^\top)^\top $, let $H = C^\top \bar{G}$ and $\nabla_{\vech(C)} \mL = \vech(G)$. The natural gradient update at iteration $t$ is 
\[
\begin{aligned}
\mu^{(t+1)} &= \mu^{(t)} + \rho_t \Sigma^{(t)} \nabla_\mu \mL^{(t)}, \\
C^{(t+1)} &= C^{(t)} + \rho_t C^{(t)} \dH^{(t)},
\end{aligned}
\;\; {\text since} \;\;
F_\lambda= \begin{bmatrix} \Sigma^{-1} & 0 \\ 0 & \mathfrak{I}(C) \end{bmatrix} \;\; \text{and} \;\;
\widetilde{\nabla}_\lambda \mL = \begin{bmatrix} \Sigma \nabla_\mu \mL \\ \vech(C \dH) \end{bmatrix}.
\]
(ii) If $\lambda = (\mu^\top, \vech(T)^\top)^\top $, let $H = T^\top \bar{G}$ and $ \nabla_{\vech(T)} \mL = \vech(G)$. The natural gradient update at iteration $t$ is 
\[
\begin{aligned}
\mu^{(t+1)} &= \mu^{(t)} + \rho_t \Sigma^{(t)} \nabla_\mu \mL^{(t)}, \\
T^{(t+1)} &= T^{(t)} + \rho_t T^{(t)} \dH^{(t)},
\end{aligned}
\;\; {\text since} \;\;
F_\lambda= \begin{bmatrix} \Sigma^{-1} & 0 \\ 0 & \mathfrak{I}(T) \end{bmatrix} \;\; \text{and} \;\;
\widetilde{\nabla}_\lambda \mL = \begin{bmatrix} \Sigma \nabla_\mu \mL \\ \vech(T \dH) \end{bmatrix}.
\]
\end{theorem}

From Theorem \ref{thm inv}, if $C$ is a diagonal matrix, then $G$ is also diagonal and $H = CG$ while $\dH = \frac{1}{2} CG$. Thus the update for $C$ simplifies to $C \leftarrow C + \frac{\rho_t}{2} C^2 G$. Inspired by the superior performance of the natural parameter in Section \ref{sec_Poisson}, we consider a one-one transformation of $\lambda = (\mu^\top, \vech(T)^\top)^\top $ in Corollary \ref{cor inv}, which reveals that the updated $T$ can be used to update $\mu$. Unfortunately, it is not possible to obtain similar updates for $C$. Proofs of Lemma \ref{lem inv}, Theorem \ref{thm inv} and Corollary \ref{cor inv} are given in the supplement S3. 

\begin{corollary} \label{cor inv}
Let $\xi = ((T^\top  \mu)^\top , \vech(T)^\top )^\top $, $H = T^\top \bar{G}$ and $\nabla_{\vech(T)} \mL = \vech(G)$. The natural gradient update of $\xi$ at iteration $t$ is
\[
\begin{aligned}
T^{(t+1)} &= T^{(t)} + \rho_t T^{(t)}\dH^{(t)},  \\
\mu^{(t+1)} &= \mu^{(t)} + \rho_t {T^{(t+1)}}^{-\top} {T^{(t)}}^{-1} \nabla_\mu \mL^{(t)}.
\end{aligned}
\]
\end{corollary}

\begin{figure}[htb!]
\centering
\includegraphics[width=0.6\textwidth]{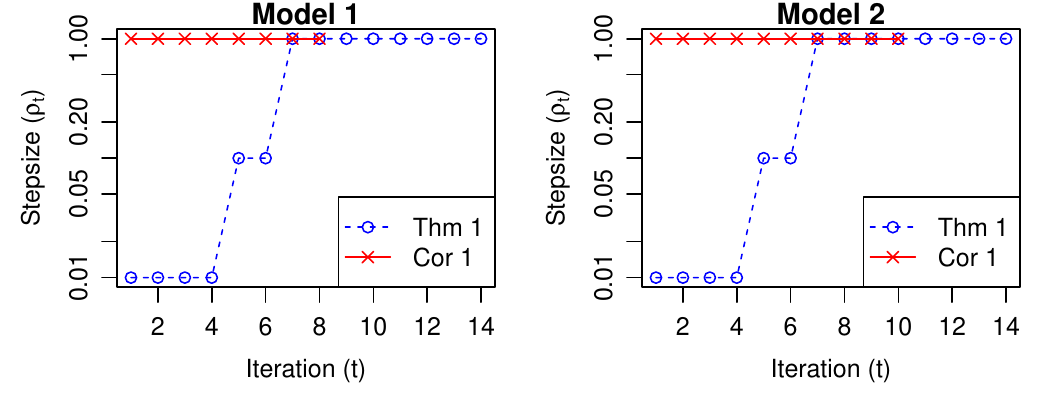}
\caption{Stepsize $\rho_t$ used at each iteration in natural gradient update of $(\mu, T)$.}
\label{Fig b}
\end{figure}

To investigate the differences between updates of $(\mu, T)$ in Theorem \ref{thm inv} and Corollary \ref{cor inv}, consider the loglinear model in Section \ref{sec_Poisson} again, this time fitting Model 1: {\tt Sa $\sim$ Width}, and Model 2: {\tt Sa $\sim$  Color + Width}. The largest stepsize $\rho_t \in \{1, 0.1, 0.01, ...\}$ is used provided $\mL$ is increasing. Figure \ref{Fig b} shows that updates in Corollary \ref{cor inv} converge faster and are more resilient to larger stepsizes.

\subsection{Stochastic natural gradient updates}
If $\nabla_\lambda \mL$ is not analytic, stochastic natural gradient ascent can be performed using an unbiased estimate. The Fisher information in the natural gradient $\widetilde{\nabla}_\lambda \mL = F_\lambda^{-1} \nabla_\lambda \mL$ can be computed exactly given $\lambda$, and stochasticity arises only from substitution of $\nabla_\lambda \mL$ with $\widehat{\nabla}_\lambda \mL$. Hence, an unbiased estimate of the natural gradient is obtained if $\E_q(\widehat{\nabla}_\lambda \mL) = \nabla_\lambda \mL$. 

For updates of $(\mu, \Sigma)$, \cite{Khan2018} obtained unbiased estimates of $\nabla_\mu \mL$ and $\nabla_{\Sigma}\mL$ using the Theorems of \cite{Bonnet1964} and \cite{Price1958}, which are stated below. The proofs can be found in \cite{Lin2019a}. ACL is an abbreviation for absolute continuity on almost every straight line \citep{Leoni2017}. The second equality  in Bonnet's Theorem is also known as Stein's Lemma \citep{Stein1981}. As $\mL = \E_q [h(\theta)]$, if $\theta$ is a sample generated from $q_\lambda(\theta)$ at iteration $t$, then the stochastic natural gradient update of natural parameter $\lambda$ is   
\begin{equation*} \label{VOGN}
{\Sigma^{(t+1)}}^{-1} = {\Sigma^{(t)}}^{-1} - \rho_t \nabla^2_\theta h(\theta), 
\quad 
\mu^{(t+1)} = \mu^{(t)} + \rho_t \Sigma^{(t+1)} \nabla_{\theta} h(\theta).
\end{equation*}
The above result follows from Table \ref{Tab a} by replacing $\nabla_\mu \mL$ and $\nabla_\Sigma \mL$ by their unbiased estimates, $\nabla_\theta h(\theta)$ and $\tfrac{1}{2} \nabla_\theta^2 h(\theta)$, from Bonnet's and Price's Theorems.

\begin{thma} If $\theta \sim N(\mu, \Sigma)$ and $h: \mathbbm{R}^d \rightarrow \mathbbm{R}$ is locally ACL and continuous, then $\nabla_\mu  \E_q [h(\theta)]  = \E_q \left[\Sigma^{-1} (\theta-\mu) h(\theta) \right] =  \E_q[\nabla_\theta h(\theta)]$.
\end{thma}

\begin{thmb}
If $\theta \sim N(\mu, \Sigma)$, $h: \mathbbm{R}^d \rightarrow \mathbbm{R}$ is continuously differentiable, $\nabla_\theta h(\theta)$ is locally ACL and $E_q[h(\theta)]$ is well-defined, then
\[
\nabla_\Sigma  \E_q [h(\theta)] = \tfrac{1}{2} \E_q \left[\Sigma^{-1}  (\theta-\mu) \nabla_\theta h(\theta)^\top  \right] = \tfrac{1}{2} \E_q \left [\nabla_\theta^2 h(\theta) \right].
\]
\end{thmb}

For Cholesky factors, Price's Theorem cannot be applied directly but there are several alternatives. The score function method uses $ \nabla_\lambda q_\lambda (\theta) = q_\lambda (\theta) \nabla_\lambda \log q_\lambda (\theta)$ to write $\nabla_\lambda \mL =  \E_q[\nabla_\lambda \log q_\lambda (\theta) h(\theta)]$, but tends to have high variance leading to slow convergence, and variance reduction techniques are required \citep{Paisley2012, Ranganath2014, Ruiz2016b}. The reparametrization trick \citep{Kingma2014} introduces a differentiable transformation $\theta = \mathcal{T}_\lambda(z)$ so that the density $\phi(z)$ of $z$ is independent of $\lambda$. Making a variable substitution and applying chain rule,
$\nabla_\lambda \mL = \E_{\phi} [\nabla_\lambda \theta \,\nabla_\theta h(\theta)]$. Reparametrization trick's gradients typically have lower variance than the score function method \citep{Xu2019}, but yields unbiased estimates of $\nabla_{\vech(C)} \mL$ and $\nabla_{\vech(T)} \mL$ in terms of the {\em first derivative} of $h(\theta)$. We propose alternative unbiased estimates in terms of the {\em second derivative} of $h(\theta)$ in Theorem \ref{thm stein}. Our results extend  Bonnet's and Price's Theorems to gradients with respect to the Cholesky factor of the covariance/precision matrix. Lemma \ref{lem stein} is instrumental in proving Theorem \ref{thm stein} and all proofs are given in the supplement S3. 

\begin{lemma} \label{lem stein}
If $\theta \sim N(\mu, \Sigma)$ and $h: \mathbbm{R}^d \rightarrow \mathbbm{R}$ is locally ACL and continuous, then $\E_q \left[ \{\Sigma^{-1} (\theta-\mu)  (\theta - \mu) ^\top   - I_d \}  h(\theta)\right] =  \E_q \left[ \nabla_\theta h(\theta)(\theta - \mu) ^\top  \right] $.
\end{lemma}

\begin{theorem} \label{thm stein}
Suppose $h: \mathbbm{R}^d \rightarrow \mathbbm{R}$ is continuously differentiable, and $h$ and $\nabla_\theta h(\theta)$ are locally ACL. Let $\theta \sim N(\mu, \Sigma)$, $\Sigma = CC^\top $ and $\Sigma^{-1} = TT^\top $ where $C$ and $T$ are lower triangular matrices, $z = C^{-1}(\theta - \mu)$ and $v = T^{-1} \nabla_\theta h(\theta)$. 
\[
\begin{gathered}
\nabla_{\vech(C)} \mL = E_q \vech(\G_1) = E_q \vech(\F_1), \;\; {\text where} \;\; \G_1 =  \nabla_\theta h(\theta) z^\top, \;\; \F_1 = \nabla_\theta^2 h(\theta) C. \\
\nabla_{\vech(T)} \mL = E_q \vech(\G_2) = E_q \vech(\F_2),\;\; {\text where} \;\; \G_2 = - (\theta - \mu) v^\top, \;\; \F_2 = -\Sigma \nabla_\theta^2 h(\theta) T^{-\top}. 
\end{gathered}
\]
\end{theorem}

Theorem \ref{thm stein} is obtained by first finding $\nabla _{\vech(C)} \mL$ and $\nabla _{\vech(T)} \mL$ using the score function method, which yields unbiased estimates in terms of $h(\theta)$. Applying Bonnet's Theorem (Stein's Lemma), we get estimates in terms of $\nabla_\theta h(\theta)$, which are {\em identical} to those obtained from the reparametrization trick. Finally, estimates in terms of $\nabla_\theta^2 h(\theta)$ are obtained using Price's Theorem. The reparametrization trick is thus connected to the score function method via Stein's Lemma. Since Price's Theorem can be derived from Bonnet's Theorem by applying Stein's Lemma, we are applying Stein's Lemma repeatedly to obtain unbiased estimates in terms of even higher derivatives of $h(\theta)$.

Second order estimates are expensive computationally, but can be advantageous if $\nabla^2_\theta h(\theta)$ is not too complex as they are more stable close to the optimum where $\mL$ will be approximately quadratic. Suppose $\ell(\theta) = \log p(y, \theta)$ is well approximated by a second order Taylor expansion about its mode $\hat{\theta}$. Then 
\[
\begin{gathered}
h(\theta) \approx \ell(\hat{\theta}) + \tfrac{1}{2} (\theta - \hat{\theta})^\top  \nabla_{\theta}^2 \ell(\hat{\theta}) (\theta - \hat{\theta}) + \tfrac{d}{2}\log(2\pi) + \tfrac{1}{2}\log|\Sigma| + \tfrac{1}{2}(\theta - \mu) \Sigma^{-1} (\theta - \mu), \\
\nabla_\theta h(\theta) \approx \nabla_{\theta}^2 \ell(\hat{\theta}) (\theta - \hat{\theta}) + \Sigma^{-1} (\theta - \mu),  \qquad 
\nabla_{\theta}^2 h(\theta) \approx \nabla_{\theta}^2 \ell(\hat{\theta}) + \Sigma^{-1}.
\end{gathered}
\]
While $\F_1$ and $\F_2$ are independent of $\theta$ and hence have almost zero variation, $\G_1$ and $\G_2$ are subjected to stochasticity due to simulation of $\theta$ from $q_\lambda(\theta)$. Later experiments indicate that this is important for inferring high-dimensional dense covariance matrices in the Gaussian variational approximation. Instead of computing $\nabla^2_\theta h(\theta)$, we can also reduce cost by using approximations. \cite{Khan2015} designed updates involving $\nabla^2_\theta h(\theta)$ for GLMs that scale linearly with the number of observations if $d$ is large, while \cite{Khan2018} use the generalized Gauss-Newton \citep{Graves2011} or gradient magnitude \citep{Bottou2016} approximation for $\nabla_\theta^2 h(\theta)$. Besides second order gradient estimates, control variates can also be used for variance reduction \citep{Ranganath2014}.

Stochastic variational algorithms obtained using Theorem \ref{thm stein} are outlined in Tables \ref{Alg1} and \ref{Alg2}, and we have applied Corollary \ref{cor inv} in the update of $\mu$ in Algorithm 2N. Algorithms based on Euclidean and natural gradients are placed side-by-side for ease of comparison. Those based on natural gradients have an additional step for computing $\dH$, and the updates involve some form of scaling, which can help to improve convergence.

\begin{table} 
\caption{Stochastic variational algorithms for updating $(\mu, C)$. \label{Alg1}} 
\centering
\fbox{\begin{tabular}{l|l}
Algorithm 1E (Euclidean gradient) &  Algorithm 1N (Natural gradient) \\ \hline
\multicolumn{2}{c}{
\begin{minipage}{0.76\textwidth}
\vspace{1mm}
Initialize $\mu$ and $C$. For $t=1,2,\dots$,
\begin{enumerate} [1.]
\item Generate $z \sim \N(0, I_d)$ and compute $\theta = C z + \mu$.
\item Find $\bar{G}$ where $G = \nabla_\theta h(\theta) z^\top $ or 
$G = \nabla_\theta^2 h(\theta) C$. 
\end{enumerate}
\end{minipage}
} \\ \hline
\begin{minipage}[t]{0.3\textwidth}
\vspace{-2mm}
\begin{enumerate}[1.]
\setcounter{enumi}{2}
\item Update $\mu \leftarrow \mu + \rho_t \nabla_\theta h(\theta)$.
\item Update $C \leftarrow C + \rho_t \bar{G}$.
\end{enumerate}
 \end{minipage} &
\begin{minipage}[t]{0.36\textwidth}
\vspace{-3mm}
\begin{enumerate} [1.]
\setcounter{enumi}{2}
\item Update $\mu \leftarrow \mu + \rho_t  C C^\top  \nabla_\theta h(\theta)$.
\item Find $\dH$ where $H =  C^\top  \bar{G}$.
\item Update $C \leftarrow C+ \rho_t C \dH$.
\end{enumerate}
 \end{minipage}
\vspace{1mm}
\end{tabular}}
\end{table}

\begin{table} 
\caption{\label{Alg2} Stochastic variational algorithms for updating $(\mu, T)$.}
\centering
\fbox{\begin{tabular}{l|l}
Algorithm 2E (Euclidean gradient) & Algorithm 2N (Natural gradient) \\ \hline
\multicolumn{2}{c}{
\begin{minipage}{0.85\textwidth}
\vspace{1mm}
Initialize $\mu$ and $T$. For $t=1,2,\dots$,
\begin{enumerate}[1.]
\item Generate $z \sim \N(0, I_d)$ and compute $\theta = T^{-\top}z + \mu$.
\item Find $\bar{G}$ where $G = -T^{-\top} z v^\top $, $v = T^{-1} \nabla_\theta h(\theta)$ or $G = -T^{-\top} T^{-1} \nabla_\theta^2 h(\theta) T^{-\top}$. 
\end{enumerate}
\end{minipage}
} \\ \hline
\begin{minipage}[t]{0.34\textwidth}
\vspace{-2mm}
\begin{enumerate}[1.]
\setcounter{enumi}{2}
\item Update $\mu \leftarrow \mu + \rho_t \nabla_\theta h(\theta)$.
\item Update $T \leftarrow T + \rho_t \bar{G}$.
\end{enumerate}
 \end{minipage} &
\begin{minipage}[t]{0.33\textwidth}
\vspace{-3mm}
\begin{enumerate}[1.]
\setcounter{enumi}{2}
\item Find $\dH$ where $H =  T^\top  \bar{G}$. 
\item Update $T \leftarrow T + \rho_t T \dH$.
\item Update $\mu \leftarrow \mu + \rho_t  T^{-\top} v$.
\end{enumerate}
\end{minipage} 
\vspace{0.5mm}
\end{tabular}}
\end{table}

\section{Choice of stepsize in stochastic natural gradient ascent} \label{sec_Normalized SGD}
For high-dimensional models, it is often important to use an adaptive stepsize $\rho_t$ that is robust to noisy gradients. Some popular approaches include Adagrad \citep{Duchi2011}, Adadelta \citep{Zeiler2012} and Adam \citep{Kingma2015}, which compute {\em elementwise} adaptive learning rates using past gradients. Adam (Table \ref{Adam and Snngm}) introduces momentum by computing the exponential moving average of the gradient ($m_t$) and squared gradient ($v_t$), and corrects for the bias due to initializing $m_t$ and $v_t$ at 0 using $\widehat{m}_t$ and $\widehat{v}_t$. The effective step is $\alpha \widehat{m}_t/(\sqrt{\widehat{v}_t} + \epsilon)$, where $\epsilon$ is a small constant added to avoid division by zero. Despite its wide applicability, we observe that use of natural gradients with Adam fails to yield significant improvement in convergence compared to Euclidean gradients.

\begin{table}
\caption{\label{Adam and Snngm} Adam, Snngm and Nagm (scalar functions are performed elementwise on vectors).}
\centering
\fbox{\begin{tabular}{l|l|l}
Adam \citep{Kingma2015} & Snngm & Nagm \\ \hline
\begin{minipage}[t]{0.35\textwidth}
\vspace{-2mm}
Initialize $m_0 = 0$, $v_0=0$ and $\lambda^{(1)}$. \\
$\alpha =0.001$, $\beta_1=0.9$, $\beta_2=0.999$, $\epsilon =10^{-8}$.
For $t=1,2,\dots$,
\begin{enumerate}[1.]
\item Compute gradient $g_t$.
\item $m_t = \beta_1 m_{t-1} + (1- \beta_1) g_t$.
\item $v_t = \beta_2 v_{t-1} + (1- \beta_2) g_t^2$.
\item $\widehat{m}_t = m_t/(1-\beta_1^t)$ \\
 $\widehat{v}_t = v_t/(1-\beta_2^t)$.
\item $\lambda^{(t+1)} = \lambda^{(t)} + \alpha \widehat{m}_t/(\sqrt{\widehat{v}_t} + \epsilon)$.
\end{enumerate}
\end{minipage} &
\begin{minipage}[t]{0.31\textwidth}
\vspace{-2mm}
Initialize $m_0 = 0$ and $\lambda^{(1)}$. \\
$\alpha = \alpha_0 \sqrt{\ell_\lambda}$ where $\ell_\lambda$ is length of $\lambda$, $\beta = 0.9$. For $t=1,2,\dots$,
\begin{enumerate}[1.]
\item Compute natural gradient \\ estimate $\widetilde{g}_t$.
\item $m_t = \beta m_{t-1} + (1-\beta)\frac{ \widetilde{g}_t}{\|\widetilde{g}_t\|}$.
\item $\widehat{m}_t = m_t/(1-\beta^t)$.
\item $\lambda^{(t+1)} = \lambda^{(t)} + \alpha \widehat{m}_t$.
\end{enumerate}
 \end{minipage} & 
\begin{minipage}[t]{0.28\textwidth}
\vspace{-2mm}
Initialize $m_0 = 0$ and $\lambda^{(1)}$. \\
$\beta = 0.9$, $\mathcal{T}=5(10^5)$. \\
For $t=1,2,\dots$,
\begin{enumerate}[1.]
\item Compute Euclidean \\ gradient estimate $\hat{g}_t$.
\item $\hat{g}_t \leftarrow \min\left(1, \frac{\mathcal{T}}{\|\hat{g}_t\|} \right) \hat{g}_t$
\item $m_t = \beta m_{t-1} + (1-\beta) \hat{g}_t$.
\item $\lambda^{(t+1)} = \lambda^{(t)} + \alpha F_t^{-1} m_t$.
\end{enumerate}
 \end{minipage}
\end{tabular}}
\end{table}

Indeed, Adam can be interpreted as a {\em sign-based} variance adapted approach \citep{Balles2018}, since (ignoring $\epsilon$) the update step can be expressed as 
\[
\alpha \frac{\text{sign}(\widehat{m}_t)}{\sqrt{\widehat{v}_t/\widehat{m}_t^2}} \approx \alpha \frac{\text{sign}(\widehat{m}_t)}{\sqrt{\E(g_t^2)/\E(g_t)^2}} 
= \alpha \frac{\text{sign}(\widehat{m}_t)}{\sqrt{1 + \Var(g_t)/\E(g_t)^2}}.
\]
The update direction is dominated by the sign of $\widehat{m}_t$, while the per dimension magnitude is bounded by $\alpha$, and reduced when there is high uncertainty (measured by the relative variance). If we replace the Euclidean gradient estimate $\widehat{g}_t$ by natural gradient estimate $\widetilde{g}_t$, Adam will update by focusing on the sign information in $\widetilde{g}_t$ while the scaling is largely neglected. Loss of scale information is compounded by per dimension variance adaption. 

We explore two alternatives in Table \ref{Adam and Snngm} that retain scale information: a basic approach Nagm that estimates the natural gradient based on the exponential moving average (momentum) of the Euclidean gradient, and {\em stochastic normalized natural gradient ascent with momentum} (Snngm). For Nagm, $F_t = F_\lambda(\lambda^{(t)})$ and gradient clipping is used to avoid exploding gradients. In practice, Nagm is applied to $\mu$ and $C$ separately as a smaller learning rate must be used for $C$ to avoid divergence. We use an $\alpha$ that is 10 times and 100 times smaller than that for $\mu$ for diagonal and dense covariance matrices respectively.

\subsection{Snngm}
Excluding momentum by setting $\beta = 0$, the update step in Snngm is $\alpha \widetilde{g}_t /\|\widetilde{g}_t \|$, where $\| \cdot \|$ denotes the Euclidean norm. The norm of this step is fixed at $\alpha$, and the effective stepsize is $\rho_t = \alpha /\|\widetilde{g}_t \|$, which is common for all parameters to preserve the scaling information. In the initial optimization stage when $\lambda$ is far from the mode, $\rho_t$ will be small as the gradient tends to be large. This is important for stability especially if the initialization is far from the mode. As $\lambda$ approaches the optimum, $\rho_t$ increases as the gradient tends to zero. Normalized natural gradient ascent can thus avoid slow convergence close to the mode and evade saddle points \citep{Hazan2015}. As the true natural gradient is unknown, we inject momentum using the exponential moving average for robustness against noisy gradients. As $\|\widetilde{g}_t\|$ tends to increase with the length of $\lambda$, scaling up $\alpha$ proportionally can  prevent the stepsize from becoming too small in high-dimensions. 

\begin{figure}[htb!]
\centering
\includegraphics[width=0.78\textwidth]{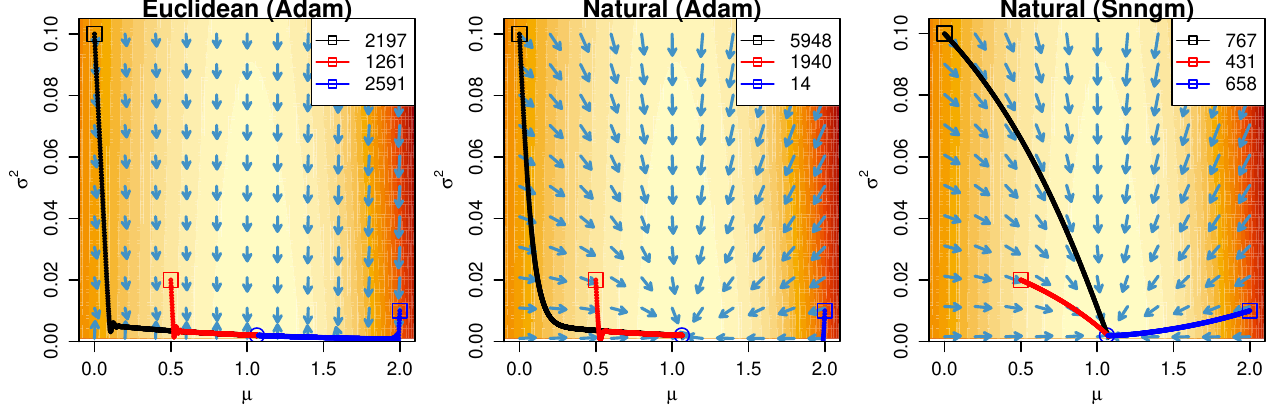}
\caption{Gradient vector field and trajectories of Adam and Snngm ($\alpha = 0.001 \sqrt{2}$) from three starting points. Legend shows the total number of iterations.}
\label{Fig c}
\end{figure}

To illustrate the difference between Adam and Snngm, consider the intercept-only loglinear model in Section \ref{sec_Poisson} again. Figure \ref{Fig c} shows that natural gradients did not yield any improvements in Adam. Instead, more iterations were required, and the run starting from (2, 0.01) was terminated due to a negative $\sigma^2$ update. The second plot also shows that Adam does not follow the flow of natural gradients closely unlike Snngm in the third plot. This is likely caused by the loss of scale information in Adam. On the other hand, the number of iterations was reduced by about three times using Snngm.

\begin{figure}[htb!]
\centering 
\includegraphics[height=100pt]{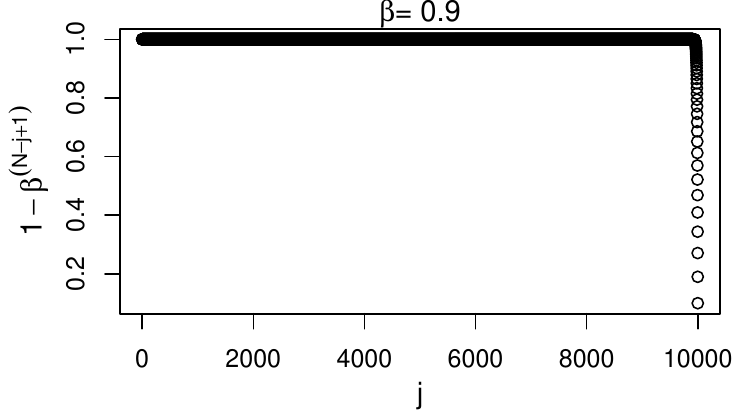}
\caption{Value of $1 - \beta^{N-j+1}$ for $j=1, \dots, N$ where $N=10000$.}
\label{Fig2}
\end{figure}

We analyze the convergence of Snngm under assumptions (A1)--(A4), of which (A1)--(A3) are similar to that made by \cite{Defossez2020} in proving convergence of Adam. 
\begin{description}
\item[(A1)] $\mL(\lambda) \leq \mL^*$ $\forall\; \lambda \in \mathbb{R}^d$.
\item[(A2)] $\| \widehat{\nabla}_\lambda \mL (\lambda) \| \leq R$ $\forall\; \lambda \in \mathbb{R}^d$.
\item[(A3)] $\mL(\lambda)$ is $L$-Lipschitz smooth: $\exists$ a constant $L > 0$ such that $\| \nabla_\lambda \mL(\lambda') - \nabla_\lambda \mL(\lambda) \| \leq L \| \lambda' - \lambda \|$ $\forall \; \lambda, \lambda' \in \mathbb{R}^d$.
\item[(A4)] $0 < R_1 \leq \ev(F_\lambda) \leq R_2$ $\forall \lambda \in \mathbb{R}^d$, where $\ev(F_\lambda)$ denotes the eigenvalues of $F_\lambda$.
\end{description}
Following \cite{Defossez2020}, let $\tau$ be a random index such that $\Prob(\tau=j) \propto 1 - \beta^{N-j+1}$ for $j  \in \{1, \dots, N\}$. The proportionality constant of the distribution of $\tau$ is,
\[
C = \sum_{j=1}^N (1-\beta^{N-j+1}) = N - \frac{\beta(1-\beta^N  )}{1-\beta} \geq  N - \frac{\beta}{1-\beta} = \widetilde{N}.
\]
For the distribution of $\tau$, almost all values of $j$ are sampled uniformly, except the last few which are sampled less often. Figure \ref{Fig2} shows the value of $1 - \beta^{N-j+1}$ for $N=10000$ and $\beta = 0.9$. All values are greater than 0.99 except for the last 43 of them. 

Theorem \ref{thm NSNGAM} provides bounds for the expected squared norm of the gradient at iteration $\tau$ in Snngm. The proof is given in the supplement S4. If we assume $N \gg \beta/(1-\beta)$, then $\widetilde{N} \approx N$. Setting $\alpha = 1/\sqrt{N}$ then yields an $O(1/\sqrt{N})$ convergence rate. 

\begin{theorem} \label{thm NSNGAM}
In Snngm, under assumptions (A1)--(A4) and for any $N > \beta/(1-\beta)$, 
\begin{equation*} 
\E\|g_\tau \|^2 \leq \frac{R R_2}{ R_1} \left\{ \frac{\mL^* - \mL(\lambda^{(1)})}{\widetilde{N} \alpha} +  \frac{N L\alpha}{\widetilde{N}}  \left( \frac{\beta}{1-\beta} + \frac{1}{2} \right) \right\}.
\end{equation*}
\end{theorem}

In later experiments, we compare Snngm and Nagm with Adam based on Euclidean or natural gradients (nAdam), variational online Gauss-Newton \citep[VOGN,][]{Khan2018} and ProxGenAdam \citep[PGA,][]{Kim2023}. VOGN uses the natural gradient update for the natural parameter of the Gaussian density. It ensures positive-definiteness of the precision matrix by using the generalized Gauss-Newton (GGN) approximation for the Hessian of the log-likelihood, and considers bias-corrected momentum for $\mu$ to derive Adam-like updates. PGA uses Euclidean gradients, and combines a variant of proximal stochastic gradient descent \citep{Domke2020} called ProxGen \citep{Yun2021} with Adam, where updates for diagonal elements of $C$ are modified to keep them away from zero. We derive similar modified updates of $T$ for PGA in the supplement S5, where VOGN and PGA are also described in more detail. Hyperparameters for Adam and PGA are fixed at default values in Table \ref{Adam and Snngm}. For Snngm, Nagm and VOGN, $\alpha$ requires tuning while other hyperparameters are generally set at default values (refer to code for values used).

\section{Imposing sparsity} \label{sec imposing sparsity}
For high-dimensional models, it is useful to impose sparsity on the Cholesky factor of the covariance or precision matrix to increase efficiency. For Algorithms 1E and 2E, updates of sparse Cholesky factors can be obtained by extracting entries in the Euclidean gradients that correspond to nonzero entries in the Cholesky factor, but the same may not apply to natural gradients due to premultiplication by the Fisher information. Suppose $\lambda = (\lambda_1^\top  , \lambda_2^\top  )^\top  $ and the Fisher information, Euclidean gradient and natural gradient are
\begin{align*}
F_\lambda = \begin{bmatrix} F_{11} & F_{12} \\ F_{21} & F_{22} \end{bmatrix}, \quad 
g = \begin{bmatrix} g_1 \\ g_2 \end{bmatrix}, \quad 
\widetilde{g} = F_\lambda^{-1} g =  \begin{bmatrix} \widetilde{g}_1 \\ \widetilde{g}_2 \end{bmatrix},
\end{align*}
respectively. By block matrix inversion, $\widetilde{g}_1 = F_{11}^{-1}g_1 - F_{11}^{-1} F_{12} \widetilde{g}_2$. If we fix $\lambda_2 = 0$ ($\lambda_2$ is no longer an unknown parameter), then the natural gradient for updating $\lambda_1$ is $F_{11}^{-1} g_1$, which is equal to $\widetilde{g}_1 + F_{11}^{-1} F_{12} \widetilde{g}_2$, but not $\widetilde{g}_1$, unless $F$ is a block diagonal matrix. 

In this section, we derive efficient natural gradient updates of the Cholesky factors in two cases, (i) the covariance matrix has a block diagonal structure corresponding to the variational Bayes assumption, and (ii) the precision matrix reflects the posterior conditional independence structure in a hierarchical model where local variables are independent conditional on the global variables.

\subsection{Block diagonal covariance}
Let $q_\lambda (\theta) = \prod_{i=1}^K q_{\lambda_i} (\theta_i)$ for some partitioning $\theta = (\theta_1^\top  , \dots, \theta_K^\top  )^\top $, where $\theta_i \sim \N(\mu_i, \Sigma_i)$. Then $\Sigma = \blockdiag(\Sigma_1, \dots, \Sigma_K)$ and $\mu = (\mu_1^\top, \dots, \mu_K^\top  )^\top  $. Let $C_i C_i^\top  $ be the Cholesky decomposition of $\Sigma_i$, where $C_i$ is a lower triangular matrix for $i=1, \dots, K$, and $C =  \blockdiag(C_1, \dots, C_K)$. For the parametrization $\lambda = (\mu^\top  , \vech(C_1)^\top  , \dots,  \vech(C_K)^\top  )^\top $, the Fisher information, $F_\lambda = \blockdiag(\Sigma^{-1}, \mathfrak{I}(C_1), \dots, \mathfrak{I}(C_K))$, where $\mathfrak{I}(\cdot)$ is defined in Lemma \ref{lem inv}. Let $\nabla_{\vech (C_i)} \mL = \vech(G_i)$ and $H_i = C_i^\top \bar{G}_i$ for $i=1, \dots, K$. Then it follows from Lemma \ref{lem inv} that the natural gradient,
\[
\widetilde{\nabla}_{\lambda} \mL = F_\lambda^{-1} \nabla_\lambda \mL =  \begin{bmatrix}
\Sigma & 0 & \dots & 0 \\
0 & \mathfrak{I}(C_1)^{-1} & \dots & 0 \\
\vdots & \vdots & \ddots & \vdots \\
0 & 0 & \dots & \mathfrak{I}(C_K)^{-1} \\
\end{bmatrix}
\begin{bmatrix} \nabla_\mu \mL \\ \vech(\bar{G}_1)  \\ \vdots \\ \vech(\bar{G}_K) \end{bmatrix}
= \begin{bmatrix} \Sigma \nabla_\mu \mL \\ \vech (C_1 \dH_1) \\  \vdots \\  \vech (C_K \dH_K) \end{bmatrix}.
\]

\begin{figure}[htb!]
\centering
\includegraphics[width=0.4\textwidth]{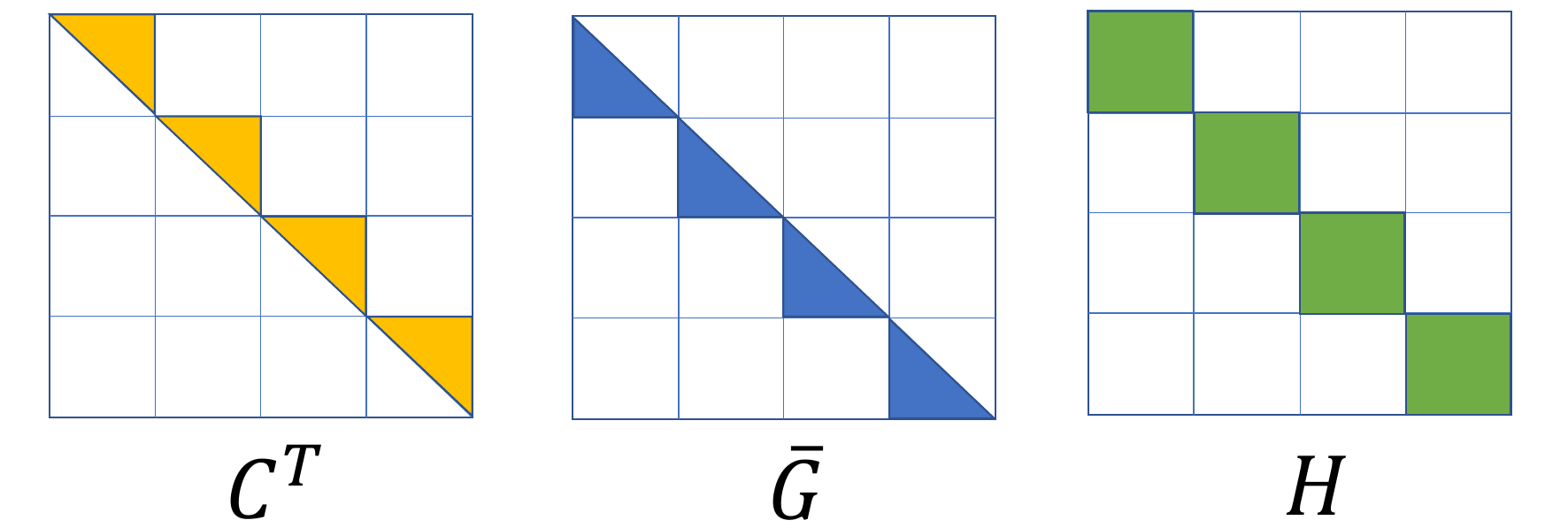}
\caption{Shaded regions represent nonzero entries in $C^\top  $, $\bar{G}$ and $H = C^\top  \bar{G}$ ($K=4$).}
\label{Fig1}
\end{figure}

This expression reveals the sparse structures of matrices that underlie computation of the natural gradient. Let $G = \blockdiag(G_1, \dots, G_K)$ and $H = \blockdiag(H_1, \dots, H_K)$. Then $H = C^\top  \bar{G}$ and $C\dH = \blockdiag(C_1 \dH_1, \dots, C_K \dH_K)$. Thus $C$, $\bar{G}$, $\dH$ and $C\dH$ have the same sparse block lower triangular structure (see Figure \ref{Fig1}), which is useful in improving storage and computational efficiency.

If the Euclidean gradient $\nabla_\lambda \mL$ is intractable, then unbiased estimates of $\nabla_{\vech(C_i)} \mL$ can be obtained using Theorem \ref{thm stein}. As $C$ is a block diagonal matrix, we only extract entries in $\G_1$ and $\F_1$ that correspond to $C_1, \dots, C_K$ on the block diagonal. For $i=1, \dots, K$,
\[
\nabla_{\vech(C_i)} \mL = \E_q \vech(\nabla_{\theta_i} h(\theta) z_i^\top) = \E_q \vech(\nabla_{\theta_i}^2 h(\theta) C_i),
\;\; \text{where} \;\; z_i = C_i^{-1}(\theta_i - \mu_i).
\]
The resulting stochastic variational algorithm 1S is outlined in Table \ref{Algc}.

\begin{table}
\caption{\label{Algc}  Stochastic natural gradient algorithms incorporating sparsity.}
\centering
\fbox{\begin{tabular}{l|l}
Algorithm 1S (Update $\mu$ and $C$) & Algorithm 2S (Update $\mu$ and $T$) \\ \hline
\begin{minipage}[t]{0.47\textwidth}
\vspace{-1mm}
Initialize $\mu$ and $C= \blockdiag(C_1, \dots, C_K)$. \\
For $t=1,2,\dots$,
\begin{enumerate}[1.]
\item Generate $z = (z_1, \dots, z_K)^\top \sim \N(0, I_d)$ \\
and compute $\theta = Cz + \mu$.
\item Find $\bar{G} = \blockdiag(\bar{G}_1, \dots, \bar{G}_K)$ where\\
$G_i = \nabla_{\theta_i} h(\theta) z_i^\top$ or $\nabla_{\theta_i}^2 h(\theta) C_i$.
\item Compute $\dH$ where $H = C^\top   \bar{G}$.
\item Update
$\mu \leftarrow \mu + \rho_t  C C^\top   \nabla_\theta h(\theta)$.
\item Update $C \leftarrow C + \rho_t C \dH$.
\end{enumerate}
\end{minipage} &
\begin{minipage}[t]{0.47\textwidth}
\vspace{-1mm}
Initialize $\mu$ and $T$ in \eqref{sparse T}. 
For $t=1,2,\dots$,
\begin{enumerate}[1.]
\item Generate $z \sim \N(0, I_d)$ and \\
compute $\theta = T^{-\top}z + \mu$.
\item Find $\bar{G}$, which is given by $-u v^\top $ or $-T_d^{-\top} T^{-1} \nabla_\theta^2 h(\theta)T^{-\top}$, where elements corresponding to zeros in $T$ are set to 0.
\item Compute $\dH$ where $H = T_d^{T}\bar{G}$.
\item Update $T \leftarrow T + \rho_t T \dH$.
\item Update $\mu \leftarrow \mu + \rho_t  T^{-\top} v$.
\end{enumerate}
 \end{minipage}
\end{tabular}}
\end{table}

\subsection{Sparse precision matrix} \label{sec_sparse precision matrix}
Consider a hierarchical model where the local variables specific to individual observations, $\theta_1, \dots, \theta_n$, are independent of each other given the global variables shared across all observations, $\theta_g$. Then the joint density is of the form,
\begin{equation} \label{eq_joint density}
p(y,\theta) = p(\theta_g) \prod_{i=1}^n p(y_i|\theta_i, \theta_g) p(\theta_i|\theta_g),
\end{equation}
where $y=(y_1, \dots, y_n)^\top $, $\theta = (\theta_1^\top  , \dots, \theta_n^\top  , \theta_g^\top  )^\top  $ and $p(\theta_g)$ is a prior density for the global variables. To reflect the fact $\theta_1, \dots, \theta_n$ are conditionally independent given $\theta_g$ a posteriori, let the Cholesky factor of $\Sigma^{-1}$ be of the form
\begin{equation} \label{sparse T}
T = \begin{bmatrix}
T_{1}  & \dots & 0 & 0\\
\vdots & \ddots & \vdots & \vdots \\
0  & \dots & T_{n} &0 \\
T_{g1} & \dots & T_{gn} & T_{g}
\end{bmatrix},
\end{equation}
where $T_1, \dots, T_n, T_g$ are lower triangular matrices. Let $\mu = (\mu_1^\top  , \dots, \mu_n^\top  , \mu_g^\top  )^\top  $ be the corresponding partitioning. Consider
\[
\lambda = (\mu^\top  , \vech(T_1)^\top  , \vc(T_{g1})^\top  , \dots,  \vech(T_n)^\top  ,\vc(T_{gn})^\top  , \vech(T_g)^\top  )^\top  . 
\]
Then the Fisher information is a block diagonal matrix which can be inverted analytically. The natural gradient update for $T$ in \eqref{sparse T} is presented in Theorem \ref{thm4} and the proof, which relies on Lemma \ref{lem inv}, is in the supplement S6. Note that $\bar{G}$, $\dH$ and $T\dH$ have the same sparse structure as $T$, which allows efficient storage.

\begin{theorem} \label{thm4}
For $i=1, \dots, n$, let $\nabla_{\vech(T_i)} \mL = \vech(A_i)$, $\nabla_{\vc(T_{gi})} \mL = \vc(G_{gi})$ and $G_i = A_i + T_i^{-\top} T_{gi}^\top  G_{gi}$. In addition, let $\nabla_{\vech(T_g)} \mL = \vech(G_g)$ and $T_d = \blockdiag(T_1, \dots, T_n, T_g)$. The natural gradient update for $T$ in \eqref{sparse T} at iteration $t$ is 
\[
T^{(t+1)} = T^{(t)} + \rho_t T^{(t)} \dH, \;\; \text{where} \;\;
H = T_d^\top  \bar{G} \;\; \text{and} \;\; 
G = \begin{bmatrix}
G_1  &  \dots & 0 & 0 \\
\vdots & \ddots & \vdots & \vdots \\
0 & \dots  & G_n  & 0 \\
G_{g1} & \dots & G_{gn}  & G_g 
\end{bmatrix}.
\]
\end{theorem}

\subsection{Stochastic natural gradient for sparse precision matrix}
If $\nabla_\lambda \mL$ is not tractable, an unbiased estimate of $\bar{G}$ can be obtained from Theorem \ref{thm stein} by extracting entries in $\G_2$ and $\F_2$ that correspond to nonzero entries in $T$. As the $\theta_i$s are conditionally independent a posteriori, $\nabla_{\theta_i, \theta_j}^2 h(\theta) = 0$ if $i \neq j$ and $\nabla_\theta^2 h(\theta)$ is also sparse. Let $u = (u_1^\top , \dots, u_n^\top , u_g^\top )^\top  = T_d^{-\top} T^\top  (\theta - \mu)$ and $v = (v_1^\top  , \dots, v_n^\top  , v_g^\top  )^\top  = T^{-1} \nabla_\theta h(\theta)$. In the supplement S6, we show that an unbiased estimate of $\bar{G}$ is given by $- uv^\top$ or $-T_d^{-\top} T^{-1} \nabla_\theta^2 h(\theta)T^{-\top}$, where elements that correspond to zeros in $T$ are set to zero.
The overall procedure is outlined in Algorithm 2S (Table \ref{Algc}). Compared with 2N, the computation of $\bar{G}$ and $\dH$ differ in sparsity and usage of $T_d$ instead of $T$ in some places.

\section{Applications} \label{sec_applications}
We apply proposed methods to GLMs, GLMMs and deep neural networks. Variational approximation accuracy is assessed using maximum mean discrepancy \cite[MMD, ][]{Gretton2012} when it is feasible to sample from the posterior via MCMC, or the predictive (test) log-likelihood when test sets are available. MCMC is performed in RStan, where two chains are run in parallel with the first half of the iterations discarded as burn-in. Given samples  $\{\theta_\MCMC^{(s)}\}_{s=1}^S$ from MCMC and $\{\theta^{(s)}\}_{s=1}^S$ from a variational density, we compute $M = -\log\{  \max(\MMD, 0) + 10^{-5} \}$ \citep{Zhou2023}, where 
\begin{equation*}
	\MMD = \frac{1}{S(S-1)} \sum_{i\neq j}^S \{ k(\theta^{(i)}, \theta^{(j)}) + 
	k(\theta_\MCMC^{(i)}, \theta_\MCMC^{(j)}) - k(\theta^{(i)}, \theta_\MCMC^{(j)}) - k(\theta^{(j)}, \theta_\MCMC^{(i)}) \},
\end{equation*}
and $k(\cdot, \cdot)$ is the radial basis kernel function. We set $S=1000$ and report the average $\overline{M}$ over 50 evaluations of $M$, where a higher value indicates a more accurate posterior approximation. Given a test set $\{y_k^*\}_{k=1}^K$, the test log-likelihood is computed as 
\begin{equation*}
\begin{aligned}
\tll = \frac{1}{K} \sum_{k=1}^K \log \left( \frac{1}{S} \sum_{s=1}^S p(y_k^*|\theta^{(s)}) \right).
\end{aligned}
\end{equation*}
For classification problems, we compute the accuracy (acc) for the test set as well, which is the fraction of correctly classified instances (expressed as a percentage). The most likely class is assigned based on posterior predictive probabilities.

We also include some comparisons with real-valued non-volume preserving \citep[real NVP,][]{Dinh2017} flows, which are highly flexible and can potentially yield better posterior approximations than variational Gaussians, with some tradeoff in computation time. The scale and translation functions in real NVP are taken as multilayer perceptrons with ReLU activations, each having a single hidden layer containing $2d$ units. The flow length is set as 8 and real NVP flows are trained using the PyTorch package {\tt normflows} \citep{Stimper2023} using 20,000 iterations with $\N(0, I_d)$ as base density. We initialize $\mu=0$ in all cases except for the weights in deep neural networks where Xavier initialization was applied, and $C = I_d/\sqrt{n}$ where $n$ is the number of observations. The prior $\theta \sim \N(0, \sigma_0^2 I_d)$ where $\sigma_0 = 10$ is used in all models except deep net GLMs, which use adaptive shrinkage priors for regularization. Our code is written in Julia \citep{Bezanson2017} and is available as supplement. All experiments are run on an Intel Core i9-9900K CPU @ 3.60GHz and runtimes are in seconds.

\subsection{Simulations from loglinear model}
\begin{table}
\caption{\label{Tab 7} Loglinear model: test log-likelihood and runtime in ().} 
\centering 
\fbox{
\begin{small}
\begin{tabular}{l|cc|cc}
& \multicolumn{2}{c|}{Diagonal Covariance} & \multicolumn{2}{c}{Full Covariance}\\
& $d=50$ & $d=100$ & $d=50$ & $d=100$ \\ \hline
Adam & -1.107 (3.1) & -1.307 (5.6) & -1.107 (4.8) & -1.306 (9.3) \\
PGA & -1.095 (3.1) & -1.162 (5.6) & -1.095 (4.8) & -1.159 (9.4) \\
nAdam & -1.108 (3.2) & -1.482 (5.6) & -1.331 (4.9) & -3.848 (9.7) \\
Snngm & -1.094 (3.1) & -1.153 (5.6) & -1.093 (4.9) & -1.151 (9.6) \\
Nagm & -1.094 (3.1) & -1.155 (5.6) & -1.093 (4.9) & -1.151 (9.7) \\
VOGN & -1.094 (3.1) & -1.154 (5.6) & -1.093 (4.8) & -1.151 (9.4) 
\end{tabular}
\end{small}}
\end{table}

Consider the loglinear model (containing an intercept) in Section \ref{sec_Poisson} with tractable lower bounds (see supplement S2), to study the impact of natural gradients apart from the use of higher derivatives for variance control. We generate a training and test set each containing $10^5$ observations from the model with $d \in \{50, 100\}$, by simulating each covariate from $U(0,1)$ and letting $\{\theta_i\}$ be evenly distributed in $[-1,1]$. For VOGN, $\nabla_\Sigma \E_q[\log p(y_i|\theta)]$ is used in place of the GGN approximation as it is positive definite. The data is processed in minibatches of 5000. 

\begin{figure}
\begin{center}
\includegraphics[width=\textwidth]{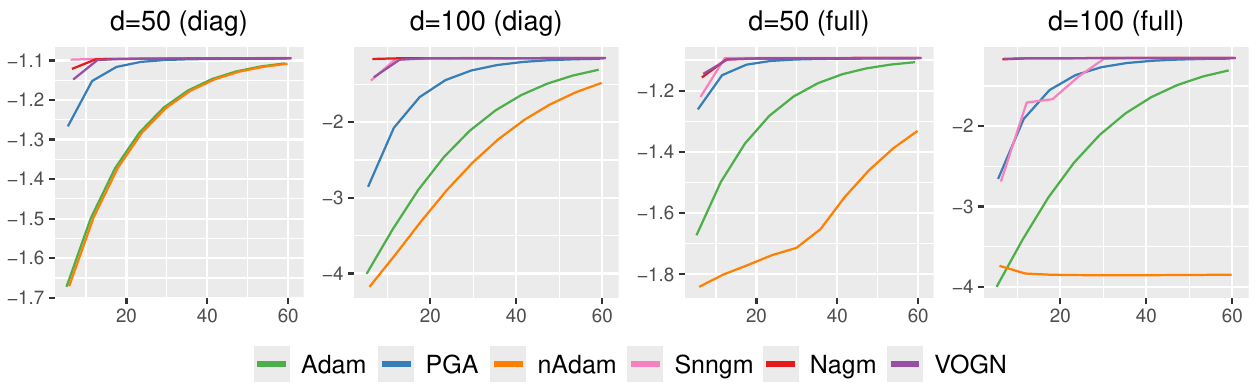}
\end{center}
\caption{\label{Fig 6} Loglinear model: Test log-likelihood against number of epochs.}
\end{figure}

Table \ref{Tab 7} shows that the top three test log-likelihoods in each case are consistently achieved by Snngm, Nagm and VOGN, with PGA closely behind. The results of Adam and nAdam are lower in comparison, with nAdam being worse than Adam. For the loglinear model with {\em tractable gradients}, there is no difference in runtime across different algorithms. Figure \ref{Fig 6} shows that Nagm, VOGN and Snngm are among the fastest to converge followed by PGA. Adam and nAdam are much slower than the rest, and nAdam converges slowly or may even fail to converge (for the full covariance case), showing that replacing Euclidean by natural gradients may lead to worse performance in Adam. 

\subsection{Logistic regression} \label{sec_log_reg}
Next, consider logistic regression where the lower bound is intractable but the Hessian $\nabla_\theta^2 h(\theta)$ can be computed analytically. The expression of $h(\theta)$ and its derivatives are given in the supplement S7. Given $\{(x_i, y_i)|i=1, \dots, n\}$ where $x_i \in \mathbb{R}^d$ and $y_i \in \{0,1\}$, it is assumed that $y_i \sim \text{Bernoulli}(p_i)$, where $\logit(p_i) = x_i^\top \theta$ and $x_i$ contains an intercept. 

\setlength{\tabcolsep}{3.6pt}
\begin{table}
\caption{\label{Tab 8} Logistic regression: test log-likelihood, test accuracy (\%), maximum mean discrepancy ($\overline{M}$) and runtime ($t$).} 
\centering \ra{1.05}
\fbox{\resizebox{\columnwidth}{!}{
\begin{tabular}{l|ccc|ccc|cccc|cccc|cc|cc}
& \multicolumn{6}{c|}{Simulation} & \multicolumn{8}{c|}{a4a} & \multicolumn{4}{c}{German} \\ \cline{2-19}
& \multicolumn{3}{c|}{Diagonal}  & \multicolumn{3}{c|}{Full}  & \multicolumn{4}{c|}{Diagonal} & \multicolumn{4}{c|}{Full} & \multicolumn{2}{c|}{Diagonal} & \multicolumn{2}{c}{Full}  \\ 
 & tll & acc & $t$ & tll & acc & $t$ & tll & acc & $\overline{M}$ & $t$ & tll & acc & $\overline{M}$ & $t$ & $\overline{M}$ & $t$ & $\overline{M}$ & $t$ \\
  \hline
Adam1 & -0.233 & 90.8 & 2.9 & -0.632 & 64.7 & 3.4 & -0.342 & 84.3 & 0.68 & 2.5 & -0.443 & 79.0 & 0.69 & 2.8 & 0.13 & 2.5 & 0.16 & 2.8 \\
  PGA1 & -0.214 & 90.9 & 2.7 & -0.221 & 90.8 & 3.4 & -0.333 & 84.6 & 1.85 & 2.2 & -0.341 & 84.4 & 1.82 & 2.8 & 0.54 & 2.2 & 0.62 & 2.8 \\
  nAdam1 & -0.276 & 88.7 & 2.8 & -0.664 & 66.1 & 4.3 & -0.331 & 84.7 & 0.73 & 2.2 & -0.409 & 81.4 & 0.70 & 3.6 & 0.45 & 2.2 & 0.64 & 3.6 \\
  Snngm1 & -0.212 & 90.9 & 2.6 & -0.290 & 87.0 & 4.2 & -0.337 & 84.3 & 2.12 & 2.2 & -0.344 & 84.0 & 0.67 & 3.5 & 2.20 & 2.2 & 1.75 & 3.5 \\
  Nagm1 & -0.212 & 90.9 & 2.7 & -0.234 & 90.1 & 4.2 & -0.340 & 84.3 & 1.73 & 2.3 & -0.331 & 84.7 & 0.68 & 3.4 & 2.68 & 2.3 & 0.85 & 3.4 \\ \hline
  Adam2 & -0.230 & 90.9 & 4.3 & -0.230 & 90.9 & 6.6 & -0.342 & 84.3 & 0.68 & 3.8 & -0.342 & 84.3 & 0.70 & 5.9 & 0.19 & 3.8 & 0.34 & 5.9 \\
  PGA2 & -0.214 & 90.9 & 4.3 & -0.214 & 90.9 & 6.6 & -0.333 & 84.6 & 1.21 & 3.9 & -0.333 & 84.5 & 1.28 & 5.8 & 0.55 & 3.9 & 0.90 & 5.8 \\
  nAdam2 & -0.219 & 90.9 & 4.3 & -0.670 & 74.1 & 7.5 & -0.334 & 84.6 & 0.79 & 3.8 & -0.514 & 76.1 & 0.94 & 6.7 & 0.46 & 3.8 & 1.51 & 6.7 \\
  Snngm2 & -0.212 & 90.9 & 4.3 & -0.212 & 90.9 & 7.6 & -0.337 & 84.3 & 2.02 & 3.8 & -0.337 & 84.4 & 4.84 & 7.3 & 2.61 & 3.8 & 7.89 & 7.3 \\
  Nagm2 & -0.212 & 90.9 & 4.3 & -0.212 & 90.9 & 7.5 & -0.339 & 84.3 & 1.76 & 3.9 & -0.346 & 84.3 & 2.04 & 7.3 & 2.80 & 3.9 & 6.46 & 7.3 \\ 
  VOGN & -0.212 & 90.9 & 4.3 & -0.212 & 90.9 & 9.9 & -0.332 & 84.6 & 0.70 & 3.8 & -0.336 & 84.4 & 0.77 & 9.5 & 2.22 & 3.8 & 5.61 & 9.5 \\ \hline
  NVP &  &  &  &  &  &  &  &  &  &  & -0.338 & 84.3 & 4.04 & 289 &  &  & 6.35 & 246 \\
  MCMC &  &  &  &  &  &  &  &  &  &  & -0.337 & 84.3 &  & 42170 &  &  &  & 511 
\end{tabular}}}
\end{table}

We generate a test set and training set each containing $10^5$ observations from the model by setting $d=100$, letting $\{\theta_i\}$ be evenly distributed in $[-1,1]$ and simulating the covariates from a standard normal. Two real datasets from the UCI Machine Learning Repository are considered. The German credit data classifies $1000$ people described by 20 attributes as good or bad credit risks, while the Adult data predicts whether the annual income of 48842 individuals exceeds \$50,000 per year using 14 features. There is no default test set for German while Adult has a 32,561/16,281 train-test split. Here we use the preprocessed a4a data from \url{www.csie.ntu.edu.tw/~cjlin/libsvmtools/datasets/binary.html}, which partitions the Adult training data into a 4781/27780 train-test split. The smaller training set makes MCMC feasible, while the full data is used in Section \ref{sec_deep glms}. After preprocessing and including an intercept, $d$ equals 49 and 124 for German and a4a respectively. For the variational algorithms, real datasets are processed in full batch, while minibatches of 5000 are used for the simulated data. MCMC and real NVP are performed for both real datasets, where 50,000 and 10,000 samples are obtained for German and a4a respectively.

In Table \ref{Tab 8}, Adam1 and Adam2 denote algorithms that use Adam and update $C$ based on the first and second derivative of $h(\theta)$ respectively (likewise for other methods). For the simulations, the highest test log-likelihood of $-0.212$ was achieved by Snngm, Nagm and VOGN using either the first or second derivative in the diagonal case, but only the second derivative in the full covariance case. This shows that second derivatives are useful in inferring full covariance matrices in high dimensions.

For a4a, the test log-likelihoods and accuracies do not seem to represent posterior approximation accuracy well. This phenomenon was also noted by \cite{Deshpande2024}. For example, in the diagonal case, nAdam1 has the highest test log-likelihood of -0.331 and accuracy of 84.7\%, but  a low $\overline{M}$ of 0.73. Focusing on the MMD, the highest $\overline{M}$ in the diagonal case is 2.12 by Snngm1, which is lower than 4.04 by real NVP. Thus, real NVP yields a better posterior approximation than mean-field Gaussian. However, Snngm2 achieved the overall highest $\overline{M}$  of 4.84, suggesting that a full covariance Gaussian can provide a better posterior approximation than real NVP, with shorter runtime. Of course, real NVP can be enhanced by increasing the flow length or number of layers in its neural networks, but these will incur higher computation costs. 

Observations for German data are similar to a4a. The highest $\overline{M}$ in the diagonal case is 2.80 by Nagm2, which is much lower than 6.35 by real NVP. However, Nagm2 and Snngm2 can achieve higher $\overline{M}$ than real NVP in the full covariance case, namely 6.46 and 7.89 respectively.

Generally, first order derivatives seem adequate for finding Gaussian approximations with diagonal covariance matrix, but second derivatives are very useful in the full covariance case, especially in high dimensions. Snngm and Nagm outperform Adam and nAdam in almost all cases and often achieve higher accuracies than VOGN and PGA for the same batchsize and epochs.
\begin{figure}
\centering
\includegraphics[width=\textwidth]{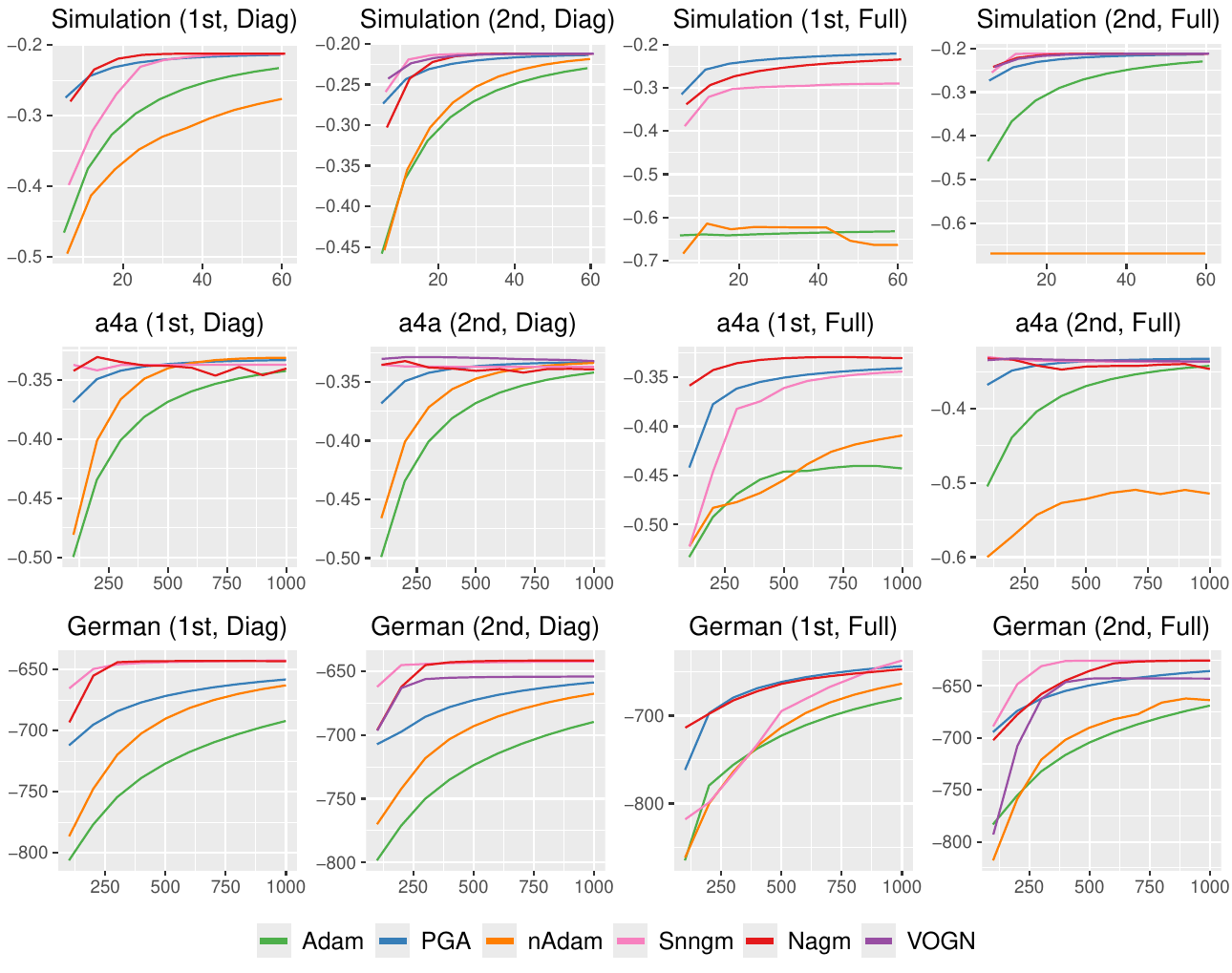}
\caption{Logistic regression: test log-likelihood (first two rows) and lower bound (third row) at each epoch.}
\label{Fig7}
\end{figure}

From Figure \ref{Fig7}, Adam and nAdam are always among the slowest to converge, while PGA, Snngm, Nagm and VOGN are faster and also achieve higher test log-likelihoods or lower bounds. Notably, PGA converges well both in speed and accuracy in the full covariance case by relying only on noisy first derivatives, while Snngm is weaker in this scenario.

\subsection{Convolutional neural networks} \label{sec CNN}
We train a convolutional neural network (CNN), LeNet-5 on MNIST and CIFAR-10, two computer vision datasets from the {\tt MLDatasets} Julia package. Sigmoid activations and average pooling layers in LeNet-5 are replaced by ReLUs and max-pooling layers. For training, MNIST contains 60,000 $28 \times 28$ pixel gray-scale images of handwritten digits (0--9), while CIFAR-10 contains 50,000 $32 \times 32$ pixel color images in 10 classes. Each has 10000 images for testing. The number of parameters is $d=61,706$ for MNIST and $d=83,126$ for CIFAR-10. Due to the high dimensions, only Gaussian variational approximations with diagonal covariances are considered. 

\begin{table}
\caption{\label{Tab 9} CNNs: test log-likelihood, accuracy and runtime.} 
\centering 
\fbox{
\begin{small}
\begin{tabular}{l|ccc|ccc}
& \multicolumn{3}{c|}{MNIST} & \multicolumn{3}{c}{CIFAR-10}\\
 & tll & acc & time & tll & acc & time \\   \hline
Adam & -0.057 & 98.7 & 76.3 & -1.347 & 52.6 & 160.6 \\
  PGA & -0.130 & 96.5 & 78.7 & -2.303 & 10.0 & 162.8 \\
  nAdam & -0.048 & 98.8 & 77.7 & -1.298 & 53.4 & 162.3 \\
  Snngm & -0.070 & 98.7 & 77.6 & -1.263 & 55.6 & 165.1 \\
  Nagm & -0.043 & 98.7 & 78.2 & -1.162 & 59.7 & 161.7 \\
  VOGN & -0.120 & 96.5 & 83.5 & -1.474 & 46.1 & 141.6 
\end{tabular}
\end{small}}
\end{table}

The CNNs are trained using the Julia {\tt Flux} library with Nvidia GPU support, and gradients are computed using automatic differentiation through {\tt Zygote}. Both datasets are loaded in minibatches of 128 for all algorithms except VOGN, which is more computationally intensive as it requires individual gradients (instead of only their sum) to compute the GGN approximation. As an online approach, VOGN is also more efficient with smaller batchsize when automatic differentiation is used. After experimenting with sizes 8, 16 and 32, VOGN is run with minibatches of 16 for 1 epoch (MNIST) and 2 epochs (CIFAR-10) so that its runtime is comparable to the rest.

\begin{figure}
\centering
\includegraphics[width=\textwidth]{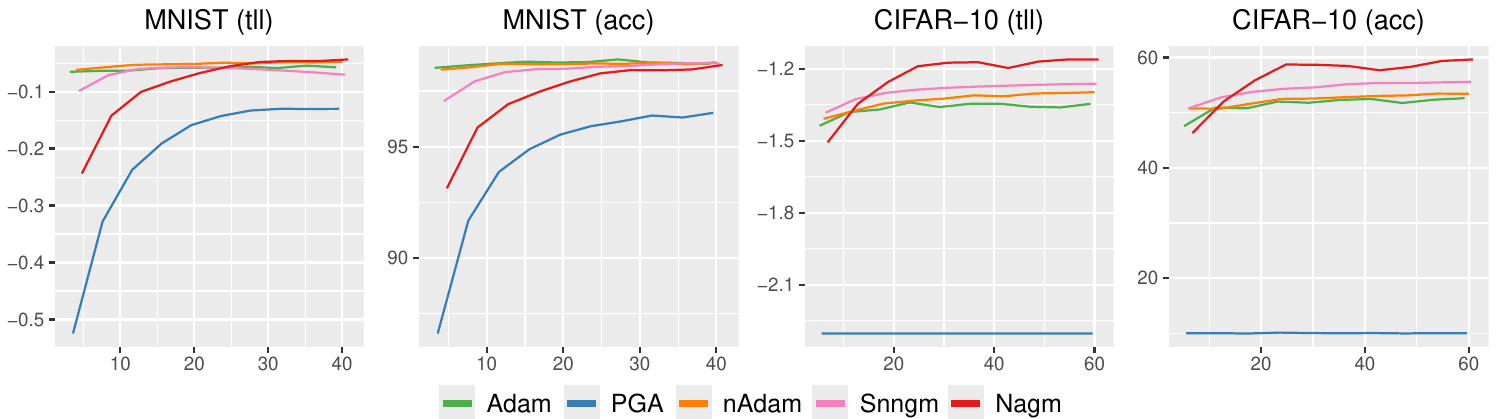}
\caption{CNNs: Test log-likelihoods (tll) and accuracies (acc) against epochs.}
\label{Fig8}
\end{figure}

From Table \ref{Tab 9}, Nagm achieved the highest test log-likelihoods for both datasets. The test accuracy for MNIST is much higher than CIFAR-10, with many algorithms achieving 98.7--98.8\%, while the highest test accuracy for CIFAR-10 was 59.7\% by Nagm. PGA has the lowest test log-likelihood and accuracy. From Figure \ref{Fig8}, its rate of convergence is slower than the rest for MNIST and it fails to make any progress over 60 epochs for CIFAR-10. Tuning of the default learning rates may be helpful in improving its performance. Nagm tends to be slower than other algorithms at first, but it is able to continue making steady progress even after other algorithms have reached a plateau.

\subsection{Deep net generalized linear models} \label{sec_deep glms}
Consider deep net GLMs \citep[deepGLMs, ][]{Tran2021}, where each response $y_i$ follows a distribution in the exponential family for $i=1, \dots, n$. Instead of connecting $\E(y_i)$ by a link function $g(\cdot)$ to a linear predictor ${\bf x}_i ^\top \theta$ as in a GLM, the covariates ${\bf x}_i \in \mathbb{R}^p$ are transformed via a deep feed forward neural network with $L$ layers, and $g(\E(y_i))$ is a linear transformation of units in the $L$th (output) layer. An ReLU activation function is applied to each layer after the affine transformation except the $L$th layer. For the deepGLM, $\theta$ consists of all weights and biases in the neural network. To perform variable selection, a normal-Gamma mixture shrinkage prior is placed on the weights $w_{x_j} \in \mathbbm{R}^m$ connecting input $x_j$ to the $m$ hidden units in the first layer: 
\[
w_{x_j}|\tau_j \sim \N(0, \tau_jI_m), \qquad 
\tau_j |\gamma_j \sim \text{Gamma}(\tfrac{m+1}{2}, \gamma_j^2/2), \quad 
j=1, \dots, p.
\]
A shrinkage prior $\N(0, I_{d_w}/\gamma_w)$ is assigned to the remaining weights $w$ of dimension $d_w$ to prevent overfitting, while a flat prior is used for the biases. The set of all variables is $\Theta = (\theta, v, \tau)$, where $v$ denotes any dispersion parameters in the exponential family, $\tau=(\tau_1, \dots, \tau_p)^\top$, and $\gamma = (\gamma_1, \dots, \gamma_p, \gamma_w)^\top$. A mean-field variational approximation $q(\Theta) = q_\lambda(\theta) q(v)\prod_{j=1}^p q(\tau_j)$ is considered, where $q_\lambda(\theta)$ is $\N(\mu, \Sigma)$ with $\Sigma$ being a diagonal matrix. Optimal densities of $q(v)$ and $q(\tau_j)$, and their parameter updates, are derived in the supplement S8. Updates of the shrinkage parameters $\gamma$, that maximize the evidence lower bound, are also derived. 

\begin{table}
\caption{\label{Tab 10} DeepGLMs: test log-likelihoods, accuracies and runtime.} 
\centering 
\small
\fbox{
\begin{tabular}{l|ccc|ccc|cc|cc}
& \multicolumn{3}{c|}{Binary simulation} & \multicolumn{3}{c|}{a9a} & \multicolumn{2}{c|}{Real simulation} & \multicolumn{2}{c}{Protein} \\ \cline{2-11}
& tll & acc & time & tll & acc & time & tll & time & tll & time \\  \hline
Adam & -0.0206 & 99.3 & 113.6 & -0.3225 & 85.1 & 20.3 & -1.9114 & 307.1 & -2.9425 & 86.3 \\
PGA & -0.0215 & 99.1 & 114.6 & -0.3260 & 85.0 & 20.6 & -1.8334 & 305.5 & -2.9537 & 84.3 \\
nAdam & -0.0221 & 99.1 & 115.3 & -0.3221 & 85.2 & 20.3 & -1.8184 & 307.3 & -2.9049 & 82.6 \\
Snngm & -0.0261 & 99.3 & 113.8 & -0.3218 & 85.1 & 20.4 & -1.6088 & 305.6 & -2.9184 & 82.7 \\
Nagm & -0.0145 & 99.4 & 113.1 & -0.3223 & 85.2 & 20.1 & -1.6832 & 304.9 & -2.8646 & 82.8 \\
VOGN & -0.0315 & 98.7 & 127.8 & -0.3265 & 84.9 & 21.8 & -1.8323 & 311.9 & -2.9491 & 84.5 \\
\end{tabular}}
\end{table}

Some data are simulated using the models in Sections 6.1.2 and 6.1.3 of \cite{Tran2020}. Let $n=10^5$ and $p=100$. In the first simulation, the binary response is generated as $y = \mathbbm{1} \{ 5 - 2(x_{1} + 2x_{2})^2 + 4 x_{3} x_{4} + 3 x_{5}  \geq 0 \}$, 
where $x_j \sim U(-1,1)$ for $j=1, \dots, p$. In the second simulation, a real-valued response is generated as 
\[
y = 5 + 10 x_1 + \frac{10}{x_2^2 + 1} + 5 x_3 x_4 + 2 x_4 + 5 x_4^2 + 5 x_5 + 2 x_6 + \frac{10}{x_7^2 + 1} + 5 x_8 x_9 + 5 x_9^2 + 5 x_{10} + \epsilon,
\]
where $\epsilon \sim \N(0,1)$ and ${\bf x}=(x_1, \dots, x_p)^\top$ is generated from a normal density with mean 0 and covariance matrix $[0.5^{|i-j|}]_{i,j}$. Two real datasets from the UCI machine learning repository are considered. The first is a real-valued response dataset of the physicochemical properties of protein tertiary structure, with $n_{\text{train}}=36,584$, $n_{\text{test}} = 9146$ and $d=9$ (80:20\% random train-test split). The second is the a9a data ($n_{\text{train}}=32,561$, $n_{\text{test}} = 16,281$ and $d=124$) processed from the Adult dataset described in Section \ref{sec_log_reg}. In each case, a 3-layer neural network (2 hidden layers each with 20 units and an output layer with a single unit) is used to transform the inputs.

DeepGLMs are also trained using {\tt Flux}. Minibatches of 5000, 4573 and 2036 are used for training the simulated, protein and a9a data respectively for all algorithms except VOGN. As VOGN is more efficient with smaller batchsize we used batchsize of 50 for the simulated data, 100 for a9a and 34 for protein data, after some tuning. The number of epochs for VOGN was adjusted so that runtimes are comparable to other algorithms.

\begin{figure}
\centering
\includegraphics[width=\textwidth]{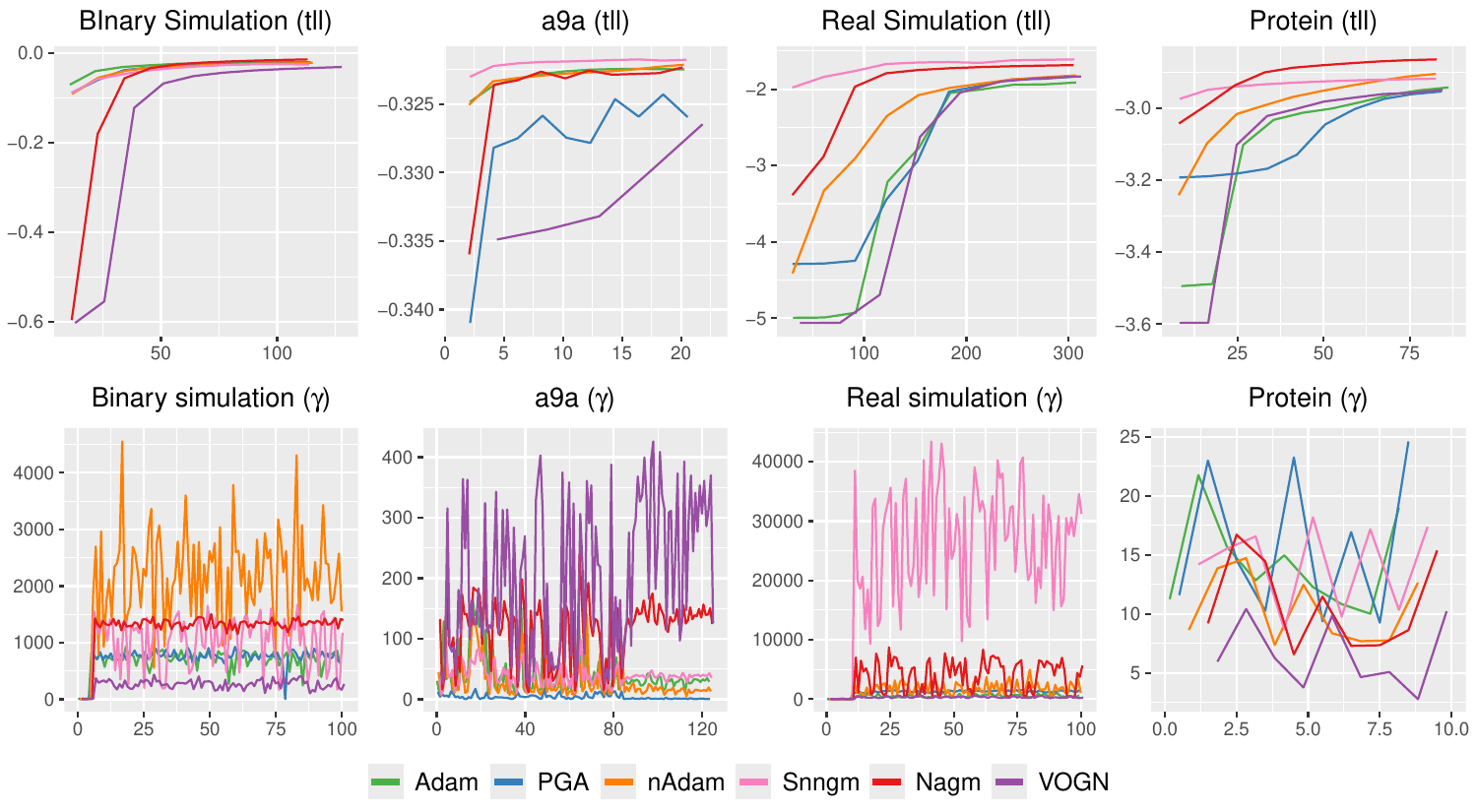}
\caption{DeepGLMs: Test log-likelihoods against time (first row) and values of shrinkage parameters $\gamma_j$ for each input variable $x_j$.}
\label{Fig9}
\end{figure}

Table \ref{Tab 10} shows that the highest test log-likelihood for (binary simulation, Protein) and (real simulation, a9a) are achieved by Nagm and Snngm respectively. The first row of Figure \ref{Fig9} reveals that Snngm usually has a high starting point and reaches its maximum quickly, while Nagm is slower initially but can often make steady progress even at a later stage. VOGN is more time-consuming than other algorithms as it incorporates second order derivative information via the GGN, but it offers greater stability and is useful when data must be processed online or in very small batches. The binary and real-valued simulated data only depend on the first five and ten inputs respectively, and the second row of Figure \ref{Fig9} shows that this information has been captured by $\{\gamma_j\}$, which are small for relevant inputs and large for irrelevant inputs (thus shrinking them to zero). Different algorithms react slightly differently to $\{\gamma_j\}$ although some common trends persist.

\subsection{Generalized linear mixed models}
Let $y_i = (y_{i1}, \dots, y_{in_i})^\top  $ denote the $i$th observation for $i=1, \dots, n$. In GLMMs, each $y_{ij}$ follows a distribution in the exponential family and $g(\E(y_{ij})) = \eta_{ij}$ for some link function $g(\cdot)$, where $\eta_{ij} = X_{ij}^\top \beta + Z_{ij}^\top \theta_i$ is the linear predictor. Here $X_{ij}$ and $Z_{ij}$ are covariates of length $p$ and $r$ respectively, $\beta$ is the fixed effects and $\theta_i \sim \N(0, B^{-1})$ is the random effects. To transform all variables onto $\mathbb{R}$, consider the Cholesky decomposition $B = WW^\top$ where $W$ is lower triangular with positive diagonal entries, and define $W^*$ such that $W_{ii}^* = \log(W_{ii})$ and $W_{ij}^* = W_{ij}$ if $i \neq j$. Then the joint distribution of the GLMM is of the form in \eqref{eq_joint density}, where $\theta_g = [\beta^\top, \omega^\top]^\top$ and $\omega = \vech(W^*)$. We assume the prior, $\theta_g  \sim \N(0, \sigma_0^2I_p)$, where $\sigma_0 = 10$.

We consider two variational approximations. The first is GVA \citep{Tan2018}, where posterior conditional independence structure is captured via a sparse precision matrix, with Cholesky factor $T$ of the form in \eqref{sparse T}. Thus GVA can be implemented using Algorithm 2S. The second is reparametrized variational Bayes \cite[RVB,][]{Tan2021}, where posterior dependence between local and global variables is first minimized by applying an invertible affine transformation on the local variables. Among the two transformations discussed in \cite{Tan2021}, RVB1 is more suited to Poisson and binomial models while RVB2 works better for Bernoulli models. Let $\tilde{\theta} = (\tilde{\theta}_1^\top , \dots, \tilde{\theta}_n^\top , \theta_g^\top )^\top$, where $\tilde{\theta}_1, \dots, \tilde{\theta}_n$ are the transformed local variables. Variational Bayes is then applied by assuming $q(\tilde{\theta}) = q(\theta_g) \prod_{i=1}^n q(\tilde{\theta}_i)$, and that all densities are Gaussian. Thus $q(\tilde{\theta}) = \N(\mu, \Sigma)$ where $\Sigma$ is a block diagonal matrix with $n+1$ blocks. If $CC^\top$ is a Cholesky decomposition of $\Sigma$, then RVB1 and RVB2 can be implemented using Algorithm 1S. In RVB, the local variables are transformed to be approximately Gaussian with mean 0 and variance 1. Hence, diagonal elements corresponding to local variables in $C$ are initialized at 1.

We consider four benchmark datasets. The first is the Epilepsy data \citep{Thall1990}, where $n=59$ epileptics are randomly assigned a new drug Progabide or a placebo, and $y_{ij}$ is the number of seizures of patient $i$ in the two weeks before clinic visit $j$ for $j=1, \dots, 4$. We fit the Poisson random intercept model,
\begin{align*}
\log \mu_{ij} = \beta_1+\beta_2 \text{Base}_i+\beta_3 \text{Trt}_i + \beta_4 \text{Base}_i \times \text{Trt}_i +\beta_5 \text{Age}_i +\beta_6 \text{Visit}_{ij} +b_{i1},
\end{align*}
where the covariates for patient $i$ are $\text{Base}_i$ (log(number of baseline seizures/4)), $\text{Trt}_i$  (1 for drug and 0 for placebo), $\text{Age}_i$ (log(age of patient at baseline) centered at zero) and $\text{Visit}_{ij}$ (coded as $-0.3$, $-0.1$, $0.1$, $0.3$ for $j=1, \dots, 4$).

The second is the Toenail data \citep{DeBacker1998}, where the binary response $y_{ij}$ of patient $i$ at the $j$th visit is 1 if degree of separation of nail plate from nail bed is moderate or severe and 0 if none or mild. Consider the random intercept model,
\begin{equation*}
\text{logit} (p_{ij}) = \beta_1 + \beta_2 \text{Trt}_i + \beta_3 t_{ij} + \beta_4 \text{Trt}_i \times t_{ij} + \theta_i, \quad  i=1, \dots, 294, \;1 \leq j \leq 7,
\end{equation*}
where for the $i$th patient, $\text{Trt}_i =1$ if 250mg of terbinafine is taken each day and 0 if 200mg of itraconazole is taken, and $t_{ij}$ is the time in months when the patient is evaluated at the $j$th visit. 

The third dataset comes from the Heart and Estrogen/Progestin Study \citep[HERS,][]{Hulley1998}. We examine 2031 women whose data for all covariates are available. The binary response $y_{ij}$ of patient $i$ at the $j$th visit indicates whether the systolic blood pressure is above 140. Consider the random intercept model,
\begin{equation*}
\begin{aligned}
\text{logit} (p_{ij}) &= \beta_1 + \beta_2 \text{age}_i +\beta_3 \text{BMI}_{ij} +\beta_4 \text{HTN}_{ij} + \beta_5 \text{visit}_{ij} + \theta_i, \quad 0 \leq j \leq 5,
\end{aligned}
\end{equation*}
where for patient $i$, $\text{age}_i$ is the normalized age at baseline, $\text{BMI}_{ij}$ is the normalized body mass index at $j$th visit, $\text{HTN}_{ij}$ indicates whether high blood pressure medication is taken at $j$th visit and $\text{visit}_{ij}$ is coded as $-1$, $-0.6$, $-0.2$, 0.2, 0.6, 1 for $j=0, 1, \dots, 5$ respectively.

\begin{figure}[h!]
\centering 
\includegraphics[width=\textwidth]{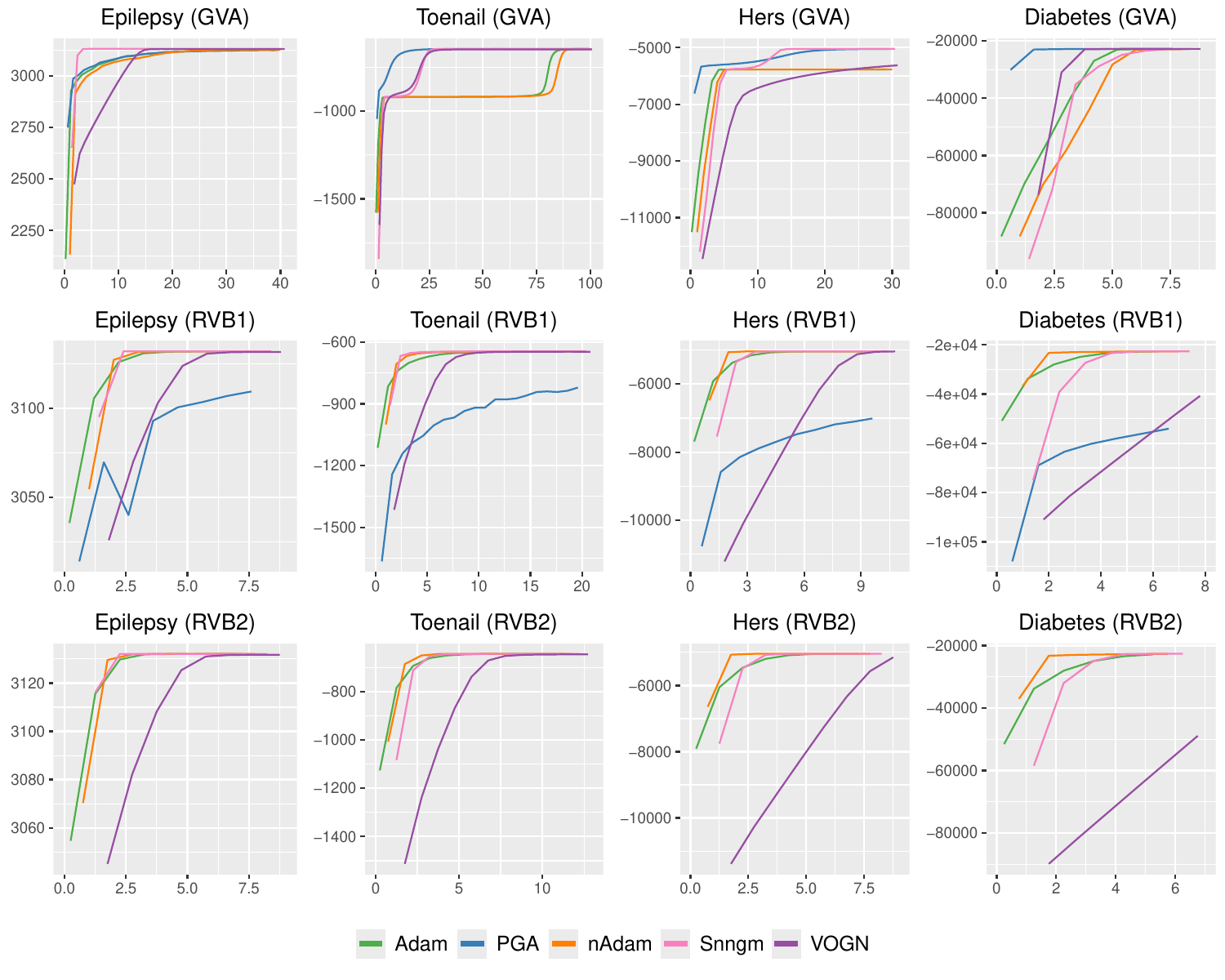}
\caption{GLMMs: Average lower bound attained against iterations (in thousands).}
\label{Fig10}
\end{figure}

\begin{table}
\caption{\label{Tab 11} GLMMs: maximum mean discrepancy measure $\overline{M}$ (first six rows) and runtime (last seven rows).} 
\centering \ra{1.05}
\fbox{\resizebox{\columnwidth}{!}{
\begin{tabular}{l|rrr|rrr|rrr|rrr}
& \multicolumn{3}{c|}{Epilepsy} & \multicolumn{3}{c|}{Toenail} & \multicolumn{3}{c|}{Hers}  & \multicolumn{3}{c}{Diabetes}  \\ 
 & GVA & RVB1 & RVB2 & GVA & RVB1 & RVB2 & GVA & RVB1 & RVB2 & GVA & RVB1 & RVB2 \\   \hline
Adam & 4.77 & {\bf 9.22} & 7.59 & 3.35 & 3.04 & 3.58 & 0.20 & 5.71 & 5.65 & 0.84 & 8.44 & 5.77 \\
PGA & 4.47 & 2.67 & 0.51 & 3.59 & 2.01 & 0.54 & 4.48 & 0.90 & 0.78 & {\bf 2.68} & 0.80 &  \\
nAdam & 4.85 & 9.02 & {\bf 9.51} & 3.48 & 3.23 & 3.59 & 0.19 & {\bf 5.75} & 5.68 & 0.84 & 8.62 & 6.45 \\
Snngm & {\bf 8.79} & 8.66 & 8.66 & {\bf 3.64} & {\bf 3.54} & {\bf 3.70} & {\bf 5.80} & 5.73 & {\bf 5.75} & 0.84 & {\bf 9.10} & {\bf 9.34}\\
VOGN & 4.47 & 5.88 & 5.90 & 2.37 & 2.43 & 3.15 & 0.49 & 4.77 & 3.94 & 0.84 & 1.41 & 1.03 \\ \cline{2-13}
NVP &  & 5.61 & &  & 3.67 & & & 5.52 & & & &     \\  \hline
Adam & 4.2 & 0.5 & 1.4 & 42.6 & 5.3 & 19.8 & 107.0 & 17.5 & 66.7 & 979.1 & 147.8 & 421.7 \\
PGA & 3.9 & 0.5 & 1.4 & 43.9 & 5.3 & 19.0 & 103.9 & 17.4 & 72.3 & 964.4 & 146.9 &     \\
nAdam & 7.6 & 0.6 & 1.5 & 72.4 & 5.7 & 20.4 & 146.0 & 19.1 & 68.9 & 2638.9 & 153.6 & 451.7 \\
Snngm & 7.3 & 0.6 & 1.5 & 71.4 & 5.7 & 20.3 & 156.3 & 19.6 & 69.4 & 2288.3 & 152.6 & 450.2 \\
VOGN & 6.0 & 1.3 & 1.8 & 52.4 & 11.3 & 21.1 & 104.3 & 41.1 & 77.6 & 1509.5 & 1787.3 & 1831.2 \\  \cline{2-13}
NVP & & 295.6 & & & 347.4 & & & 4109.8 & & & &     \\ 
MCMC & & 191.9 &  &  & 293.8 & & & 1431.9 & & & 13462.3 &     \\
\end{tabular}}}
\end{table}

Finally, we consider the diabetes dataset from the UCI machine learning repository, which contains 101766 hospital records of patients diagnosed with diabetes at 130 US hospitals from 1999--2008. We preprocess the data following \cite{Strack2014}, and study $n=16,341$ patients each with at least 2 hospital records. The binary response $y_{ij}$ of patient $i$ at the $j$th record indicates if the patient is readmitted in less than 30 days. We fit a random intercept model that includes as main terms, the HbA1c result (four levels) and diagnosis (nine levels), plus their interaction. 

For each model and approach (GVA, RVB1 and RVB2), we run the five algorithms, Adam, PGA, nAdam, Snngm and VOGN for the same number of iterations, and monitor convergence by averaging lower bound estimates over every 1000 iterations. Real NVP was performed for each dataset except Diabetes due to GPU memory constraints. 50,000 MCMC samples were obtained for Epilepsy, Toenail and Hers, while only 20,000 MCMC samples were obtained for Diabetes due to its large size. Figure \ref{Fig10} plots the average lower bounds against number of iterations, while Table \ref{Tab 11} shows the maximum mean discrepancy measure and runtime. These results are based on first order gradient estimates as $\nabla_\theta^2 h(\theta)$ is highly complex for GVA and RVB.

From Figure \ref{Fig10}, the performance of Snngm is stable across different approaches and datasets. Although it is not necessarily the fastest to converge, Snngm is always able to reach a good local mode, unlike Adam and nAdam, which were stuck at poor local modes for GVA (Toenail and Hers) for a long time. PGA performs well for GVA, but poorly for RVB1 and fails to converge for RVB2 (not shown). VOGN also does much better for GVA than for both RVB1 and RVB2.  

The maximum mean discrepancy in Table \ref{Tab 11} shows that Snngm achieves the highest or close to the highest $\overline{M}$ in many cases. The best Gaussian approximations based on GVA, RVB1 or RVB2 have higher  $\overline{M}$ than real NVP for the Epilepsy and Hers datasets. For Toenail, the $\overline{M}$ of real NVP is higher than GVA and RVB1 but slightly less than RVB2. Among the three variational approaches, RVB2 achieved the highest $\overline{M}$ across all datasets and has the best posterior approximation accuracy.

For GVA, natural gradient based algorithms (nAdam, Snngm, VOGN) are more time-consuming than those based on Euclidean gradient (Adam, PGA), as more matrix inversions are required to compute the natural gradients. For RVB1 and RVB2, runtimes across all algorithms are similar, except for VOGN whose computation cost increases drastically with the variable dimension. This is because RVB considers the Cholesky factor of the covariance matrix instead of the precision matrix, whose natural gradients are more efficient to compute as they do not require any matrix inversions. On the other hand, VOGN still requires matrix inversions as it updates the precision matrix, and is more computationally intensive.

\section{Conclusion} \label{sec_Conclusion}
Gaussian variational approximation is widely used and natural gradients provide a means of improving the convergence in stochastic gradient ascent, which is especially important when suboptimal local modes are present. However, the natural gradient update of the precision matrix does not ensure positive definiteness and expensive matrix inversion may be required. To tackle this issue, we consider Cholesky decomposition of the covariance or precision matrix, show that the inverse Fisher information can be found analytically and present natural gradient updates of the Cholesky factors in closed form. We also derive unbiased estimates of the gradient of the lower bound in terms of either the first or second derivative of the log posterior. First order gradient estimates are efficient, but second order gradient estimates are more stable, and can yield more accurate Gaussian variational approximations when large dense covariance matrices are used. 

For high-dimensional models, we show that efficient natural gradient updates can be derived when sparsity constraints are imposed on the covariance or precision matrix, to incorporate variational Bayes assumptions or posterior conditional independence. Finally, we observe that Adam does not always perform well with natural gradients, and propose a basic approach that implements natural gradient with momentum using a constant stepsize (Nagm), and stochastic normalized natural gradient ascent with momentum (Snngm) as alternatives. The convergence of Snngm is proven for $L$-Lipschitz smooth functions with bounded gradients. We compare these alternatives with some popular machine learning approaches such as VOGN and PGA.

Through extensive experiments with statistical and deep learning models, we observe that Snngm performs better when gradient estimates are less noisy, such as when larger batch sizes or second order gradient estimates are used. The main advantages of Snngm are its stability and resilience to poor initializations and larger stepsizes (due to the normalization applied), rapid convergence in the initial stages and ability to escape poor local modes. In the later stages of optimization, it may be useful to reduce the stepsize instead of keeping it constant, so that Snngm can converge more precisely to the mode. 

Unlike Snngm, Nagm requires a smaller stepsize and gradient clipping for stability as it does not employ any normalization. While Nagm tends to converge slowly at first, it gains momentum as optimization proceeds and is often able to converge steadily to a better local mode even when other methods have reached a plateau. It often performs better than Snngm in deep learning where gradient estimates tend to be noisier due to the use of small minibatches.

As an online approach, VOGN is designed to work efficiently with small minibatches. It uses the GGN approximation to compute second order gradient estimates, which are more stable than first order ones, and is thus able to converge rapidly despite the noise inherent in small minibatches. However, the GGN approximation is computationally intensive, especially when the Gaussian covariance matrix is non-diagonal, and hence VOGN will require longer runtimes in such scenarios. 

The performance of PGA and nAdam are quite erratic and vary widely across different models and datasets. They converge rapidly in some settings, but may be slow or diverge in other settings, even after tuning of the stepsize. Adam works well generally, but can be slow to converge sometimes or may get stuck in poor local modes in some scenarios.

While Snngm and Nagm may not necessarily be the fastest to converge in any setting, their performance are more consistent than PGA or nAdam, and they are able to escape the local modes that Adam can get trapped in. Unlike VOGN which thrives in the online setting, they perform better with larger batchsize. Our experiments indicate that Nagm and Snngm each have their unique characteristics and advantages. They are competitive with existing methods in terms of test log-likelikhood and posterior approximation accuracy, and help to fill in some gaps that have been left behind.

\section*{Data Availability}
All datasets are available online and weblinks are provided in the {\tt readme.txt} file in the supplement. 

\section*{Conflict of interest}
There is no conflict of interest.

\section*{Funding}
This research is supported by the Ministry of Education, Singapore, under its Academic Research Fund Tier 2 (Award MOE-T2EP20222-0002).

\vskip 0.2in
\bibliographystyle{chicago}
\bibliography{refnatgrad}

\newpage

\setcounter{section}{0} \renewcommand{\thesection}{S\arabic{section}}
\setcounter{figure}{0} \renewcommand{\thefigure}{S\arabic{figure}}
\setcounter{table}{0} \renewcommand{\thetable}{S\arabic{table}}
\setcounter{equation}{0} \renewcommand{\theequation}{S\arabic{equation}}
\setcounter{lemma}{0} \renewcommand{\thelemma}{S\arabic{lemma}}

\noindent
{\bf \Large Supplementary material}

\section{Natural gradient updates in terms of mean and covariance/precision matrix} \label{S1}

First we derive the natural gradient of $\mL$ with respect to the natural parameter $\lambda$ using $\widetilde{\nabla}_\lambda \mL = \nabla_m \mL$. For the Gaussian, $m = \E[s(\theta)] = (m_1^\top , m_2^\top  )^\top $, where $m_1 = \mu$, $m_2 = \vech(\Sigma + \mu \mu^\top ) $. We introduce $\zeta = (\zeta_1^\top , \zeta_2^\top )^\top $, where
\[
\zeta_1 = \mu = m_1, \quad \zeta_2 = \vech(\Sigma) = m_2 - \vech(m_1 m_1^\top ).
\]
Then
\[
\nabla_m \zeta 
= \begin{bmatrix} I_d & -2(I_d \otimes \mu^\top ){D^+}^\top  \\ 0_{d(d+1)/2 \times d} & I_{d(d+1)/2}  \end{bmatrix},
\]
where $D^+ = (D^\top  D)^{-1} D^\top $ is the Moore-Penrose inverse of $D$, and $D^+ \vc(A) = \vech(A)$ if $A$ is symmetric. Applying chain rule, the natural gradient is 
\[
\begin{aligned}
\widetilde{\nabla}_\lambda \mL 
&= \nabla_m \mL =\nabla_m \zeta \nabla_\zeta \mL \\
&= \begin{bmatrix} I_d & -2(I_d \otimes \mu^\top ){D^+}^\top  \\ 0_{d(d+1)/2 \times d} & I_{d(d+1)/2}  \end{bmatrix} \begin{bmatrix} \nabla_\mu \mL \\ \nabla_{\vech(\Sigma)} \mL  \end{bmatrix} \\
&= \begin{bmatrix} \nabla_\mu \mL -2(I_d \otimes \mu^\top ) \vc(\nabla_{\Sigma} \mL) \\  D^\top  \vc(\nabla_{\Sigma} \mL) \end{bmatrix} \\
&= \begin{bmatrix} \nabla_\mu \mL - 2(\nabla_{\Sigma}\mL) \mu \\  D^\top  \vc(\nabla_{\Sigma} \mL) \end{bmatrix}.
\end{aligned}
\]
Note that $\nabla_{\vech(\Sigma)} \mL = D^\top \nabla_{\vc(\Sigma)} \mL = D^\top \vc(\nabla_{\Sigma} \mL$). The update for the natural parameter is thus given by 
\[
\begin{aligned}
\lambda^{(t+1)} &= \lambda^{(t)} + \rho_t \widetilde{\nabla}_\lambda \mL^{(t)}, \\
\implies \begin{bmatrix}
\Sigma^{-1(t+1)} \mu^{(t+1)} \\ -\frac{1}{2}D^\top  \vc(\Sigma^{-1(t+1)}) 
\end{bmatrix} &= 
\begin{bmatrix}
\Sigma^{-1(t)} \mu^{(t)} \\ -\frac{1}{2}D^\top  \vc(\Sigma^{-1(t)})
\end{bmatrix} + 
\rho_t  \begin{bmatrix} \nabla_\mu \mL^{(t)} - 2(\nabla_{\Sigma}\mL^{(t)}) \mu^{(t)} \\  D^\top  \vc(\nabla_{\Sigma} \mL^{(t)}) \end{bmatrix}.
\end{aligned}
\]
The second line simplifies to 
\[
\Sigma^{-1(t+1)} = \Sigma^{-1(t)} -2 \rho_t \nabla_{\Sigma} \mL^{(t)}.
\]
For the first line, 
\[
\begin{aligned}
\Sigma^{-1(t+1)} \mu^{(t+1)}   &= \Sigma^{-1(t)} \mu^{(t)} + \rho_t  \nabla_\mu \mL^{(t)} - 2 \rho_t (\nabla_{\Sigma}\mL^{(t)}) \mu^{(t)} \\
&=\{ \Sigma^{-1(t)} - 2 \rho_t \nabla_{\Sigma}\mL^{(t)}  \} \mu^{(t)} + \rho_t  \nabla_\mu \mL^{(t)} \\
& = \Sigma^{-1(t+1)}  \mu^{(t)} + \rho_t  \nabla_\mu \mL^{(t)} \\
\implies  \mu^{(t+1)} & = \mu^{(t)} + \rho_t \Sigma^{(t+1)}  \nabla_\mu \mL^{(t)}.
\end{aligned}
\]
As the update for $\mu^{(t+1)}$ depends on $\Sigma^{(t+1)}$, we need to update $\Sigma$ first before updating $\mu$.

Next, consider the parametrization, $\kappa = (\mu^\top , \vech(\Sigma)^\top )^\top $. The Fisher information matrix and its inverse are respectively, 
\begin{equation*}
F_\kappa = \begin{bmatrix}
\Sigma^{-1} & 0 \\
0 & \frac{1}{2}D^\top  (\Sigma^{-1} \otimes \Sigma^{-1}) D
\end{bmatrix}, 
\quad
F_\kappa^{-1} = \begin{bmatrix}
\Sigma & 0 \\
0 & 2D^+ (\Sigma \otimes \Sigma) {D^+}^\top 
\end{bmatrix}.
\end{equation*}
Hence the natural gradient is 
\[
\widetilde{\nabla}_\kappa \mL = \begin{bmatrix}
\Sigma & 0 \\
0 & 2D^+ (\Sigma \otimes \Sigma) {D^+}^\top 
\end{bmatrix}
\begin{bmatrix}
\nabla_\mu \mL \\ \nabla_{\vech(\Sigma)} \mL
\end{bmatrix} 
= \begin{bmatrix}
\Sigma \nabla_\mu \mL \\ 2 \vech(\Sigma \nabla_\Sigma \mL \Sigma)
\end{bmatrix}.
\]
Alternatively, $\widetilde{\nabla}_\kappa \mL =  (\nabla_\lambda \kappa)^\top  \widetilde{\nabla}_\lambda \mL$, which is equal to
\[
\begin{bmatrix}
\Sigma & 2 \Sigma (\mu^\top  \Sigma^{-1} \otimes \Sigma^{-1}) D D^+ (\Sigma \otimes \Sigma) {D^+}^\top  \\
0 & 2 D^+(\Sigma \otimes \Sigma) {D^+}^\top  \end{bmatrix}
\begin{bmatrix} \nabla_\mu \mL - 2(\nabla_{\Sigma}\mL) \mu \\ D^\top  \vc(\nabla_{\Sigma} \mL) \end{bmatrix}  
= \begin{bmatrix} \Sigma \nabla_\mu \mL \\ 2 \vech (\Sigma \nabla_{\Sigma} \mL \Sigma ) \end{bmatrix}.
\]

For the parametrization $\xi = (\mu^\top , \vech(\Sigma^{-1})^\top )^\top $, the Fisher information and its inverse are respectively,
\begin{equation*}
F_\xi = \begin{bmatrix}
\Sigma^{-1} & 0 \\
0 & \frac{1}{2}D^\top  (\Sigma \otimes \Sigma) D
\end{bmatrix}, 
\quad 
F_\xi^{-1} = \begin{bmatrix}
\Sigma & 0 \\
0 & 2D^+ (\Sigma^{-1} \otimes \Sigma^{-1}) {D^+}^\top 
\end{bmatrix}.
\end{equation*}
Note that $\nabla_{\vech(\Sigma^{-1})} \mL = - D^\top (\Sigma \otimes \Sigma) \nabla_{\vc(\Sigma)} \mL$. Hence the natural gradient is 
\begin{equation*}
\widetilde{\nabla}_\xi \mL= \begin{bmatrix}
\Sigma & 0 \\
0 & 2D^+ (\Sigma^{-1} \otimes \Sigma^{-1}) {D^+}^\top 
\end{bmatrix}\begin{bmatrix}
\nabla_\mu \mL \\ - D^\top (\Sigma \otimes \Sigma) \nabla_{\vc(\Sigma)} \mL
\end{bmatrix} = \begin{bmatrix}
\Sigma \nabla_\mu \mL \\ - 2 \vech(\nabla_\Sigma \mL)
\end{bmatrix},
\end{equation*}
Alternatively, $\widetilde{\nabla}_\xi \mL =  (\nabla_\lambda \xi)^\top  \widetilde{\nabla}_\lambda \mL $, which is equal to
\begin{equation*}
= \begin{bmatrix}
\Sigma & 2\Sigma (\mu^\top  \otimes I){D^+}^\top  \\
0 & -2 (D^\top  D)^{-1} \end{bmatrix}
\begin{bmatrix} \nabla_\mu \mL - 2(\nabla_{\Sigma}\mL) \mu \\ D^\top  \vc(\nabla_{\Sigma} \mL) \end{bmatrix} = \begin{bmatrix} \Sigma \nabla_\mu \mL \\ -2 \vech (\nabla_{\Sigma} \mL) \end{bmatrix}.
\end{equation*}

\section{Loglinear model}  \label{S2}
Let $y = (y_1, \dots, y_n)^\top $ and $X = (x_1, \dots, x_n)^\top $. We have $\delta_i = \exp(x_i^\top  \theta)$ for $i=1, \dots, n$. The lower bound can be evaluated analytically and it is given by 
\[
\begin{aligned}
\mL(\lambda) &= \E_q \{\log p(y, \theta) - \log q_\lambda(\theta)\} \\
&= \E_q \bigg[  y^\top  X \theta - \sum_{i=1}^n \{\exp(x_i^\top  \theta)  + \log(y_i!)\} - \frac{d}{2} \log(2\pi) - \frac{d}{2} \log(\sigma_0^2)  - \frac{\theta^\top  \theta}{2\sigma_0^2} \bigg] \\
& \quad - \E_q \bigg[ -\frac{d}{2} \log(2\pi) - \frac{1}{2} \log|\Sigma|- \frac{1}{2} (\theta - \mu)^\top  \Sigma^{-1} (\theta - \mu) \bigg] \\
&= y^\top  X \mu - \sum_{i=1}^n \{w_i + \log(y_i!) \} - \frac{\mu^\top  \mu + \tr(\Sigma)}{2\sigma_0^2} + \frac{1}{2} \log|\Sigma| + \frac{d}{2} \{1 -  \log(\sigma_0^2)\},
\end{aligned}
\]
where $w_i =  \exp (x_i^\top  \mu + \frac{1}{2} x_i^\top  \Sigma x_i )$. Let $w=(w_1, \dots, w_n)^\top $ and $W = \diag(w)$. Then the Euclidean gradients are given by 
\[
\begin{aligned}
\nabla_\mu \mL &= X^\top  (y - w)  - \mu/\sigma_0^2, \\
\nabla_{\vc(\Sigma)} \mL &= \tfrac{1}{2} \vc ( \Sigma^{-1} - I/\sigma_0^2 - X^\top W X ), \\
\nabla_{\vech(C)} \mL &=2 \vech(\nabla_{\Sigma} \mL  \; C) 
= \vech (C^{-\top} - C/\sigma_0^2 - X^\top  W X C ) , \\
\nabla_{\vech(T)} \mL &= -2\vech(\Sigma \,\nabla_\Sigma \mL \,T^{-\top} ) 
= \vech ( \{ \Sigma X^\top  W X  + \Sigma/\sigma_0^2 - I \} T^{-\top}  ).
\end{aligned}
\]

\section{Natural gradient updates in terms of mean and Cholesky factor} \label{S3}
First we present the proof of Lemma 1. As this proof requires several results from \cite{Magnus1980} concerning the elimination matrix $L$, these are collected here in Lemma S1 for ease of reference.  
\begin{lemma} \label{Lem_L}
If $P$ and $Q$ are lower triangular $d \times d$ matrices and $N = (I+K)/2$, then
\begin{enumerate}[(i)]
\item $L L^\top  = I_{d(d+1)/2}$,
\item $(L N L^\top )^{-1} = 2I_{d(d+1)/2} -LKL^\top $,
\item $N = D L N$,
\item $L^\top  L (P^\top  \otimes Q)L^\top  = (P^\top  \otimes Q) L^\top $ and its transpose, $L (P \otimes Q^\top )L^\top  L = L (P \otimes Q^\top )$,
\item $L (P^\top  \otimes Q) L^\top  = D^\top  (P^\top  \otimes Q) L^\top $ and its transpose, $L (P \otimes Q^\top ) L^\top  = L (P \otimes Q^\top ) D$.
\end{enumerate}
\begin{proof}[Lemma S1] 
The proofs can be found in Lemma 3.2 (ii), Lemma 3.4 (ii), Lemma 3.5 (ii) and Lemma 4.2 (i) and (iii) of \cite{Magnus1980} respectively.
\end{proof}
\end{lemma}

\subsection{Proof of Lemma 1}
First, we prove (i):
\begin{align*}
\mathfrak{I}(C) 
&= L \{ K(C^{-\top}  \otimes C^{-1} ) + I_d \otimes C^{-\top} C^{-1} \} L^\top  \\ 
&= L \{ K(C^{-\top}  \otimes I_d)(I_d \otimes C^{-1} ) + ( I_d \otimes C^{-\top} ) ( I_d \otimes C^{-1}) \}L^\top \\ 
&= L \{ ( I_d \otimes C^{-\top} ) K+ ( I_d \otimes C^{-\top} ) \} ( I_d \otimes C^{-1})L^\top \\ 
&= L ( I_d \otimes C^{-\top} )(K+ I_{d^2}) ( I_d \otimes C^{-1})L^\top \\ 
&= 2 L ( I_d \otimes C^{-\top} )N ( I_d \otimes C^{-1})L^\top .
\end{align*}
Next we prove (ii) and (iii) by using the results in Lemma S1. The roman letters in square brackets on the right indicate which parts of Lemma S1 are used. For (ii),
\begin{align*}
&\{ 2 L  ( I_d \otimes C^{-\top} )N ( I_d \otimes C^{-1})L^\top  \} \left\{\tfrac{1}{2} L ( I_d \otimes C) L^\top  (L N L^\top )^{-1} L (I_d \otimes C^\top )  L^\top  \right\} \\
&= L  ( I_d \otimes C^{-\top} )(DL N) ( I_d \otimes C^{-1}) ( I_d \otimes C) L^\top  (L N L^\top )^{-1} L (I_d \otimes C^\top )  L^\top  && \text{[(iii) \& (iv)]} \\
&= L  ( I_d \otimes C^{-\top} )L^\top  (L N L^\top )  (L N L^\top )^{-1} L (I_d \otimes C^\top )  L^\top  &&  \text{[(v)]}\\
&= L  ( I_d \otimes C^{-\top} )L^\top  L (I_d \otimes C^\top ) L^\top  \\
&= L  ( I_d \otimes C^{-\top} ) (I_d \otimes C^\top ) L^\top  && \text{[(iv)]}\\
&= L  L^\top  = I_{d(d+1)/2}. && \text{[(i)]}
\end{align*}
For (iii),
\begin{align*}
\mathfrak{I}(C)^{-1} \vech(G)  &= \tfrac{1}{2}L ( I_d \otimes C) L^\top (L N L^\top )^{-1}L (I_d \otimes C^\top ) L^\top \vech( \bar{G})  \\
&= \tfrac{1}{2}L ( I_d \otimes C) L^\top (2I_{d(d+1)/2} -LKL^\top ) L (I_d \otimes C^\top ) \vc(\bar{G})  && \text{[(ii)]} \\
&= \tfrac{1}{2}L ( I_d \otimes C)  (2I_{d^2} -L^\top LK) L^\top  L \vc( C^\top  \bar{G}) \\
&= \tfrac{1}{2}L ( I_d \otimes C) (2I_{d^2} -L^\top LK) L^\top \vech(\bar{H})  \\
& = \tfrac{1}{2}L ( I_d \otimes C) (2I_{d^2} -L^\top LK) \vc(\bar{H})  \\
& = L ( I_d \otimes C) \vc(\bar{H})  -  \tfrac{1}{2}L ( I_d \otimes C)  L^\top LK \vc(\bar{H})  \\
& = L\vc(C\bar{H}) - \tfrac{1}{2}L ( I_d \otimes C)  L^\top  \vech(\bar{H}^\top )  \\
& = \vech(C\bar{H}) - \tfrac{1}{2}L ( I_d \otimes C) \vc(\dg (\bar{H}))  \\
& = \vech(C\bar{H}) - \tfrac{1}{2} \vech(C\dg (\bar{H}))  
= \vech(C \dH).
\end{align*}

\subsection{Proof of Theorem 1} 
First we derive the Fisher information and its inverse for each of the two parametrizations. We have 
\[
\ell_q = \log q_\lambda(\theta) = -\tfrac{d}{2} \log(2\pi) - \tfrac{1}{2} \log|\Sigma| - \tfrac{1}{2}(\theta-\mu)^\top  \Sigma^{-1} (\theta-\mu).
\]

For the first parametrization, $\lambda = (\mu^\top  , \vech(C)^\top  )^\top $,  let $z = C^{-1} (\theta -\mu)$ to simplify expressions. The first order derivatives are
\[
\nabla_\mu \ell_q = \Sigma^{-1} (\theta - \mu), \quad 
\nabla_{\vech(C)} \ell_q = \vech(C^{-\top}  z z^\top  - C^{-\top} ),
\] 
and $\nabla_\lambda^2 \ell_q$ is given by
\[
 -\begin{bmatrix}
\Sigma^{-1} & \{(C^{-\top}  \otimes z^\top  C^{-1}) + (z^\top  \otimes \Sigma^{-1})\}L^\top  \\
\cdot  &  L[\{ (C^{-1} \otimes C^{-\top}  zz^\top ) + (zz^\top  C^{-1} \otimes C^{-\top} ) - (C^{-1} \otimes C^{-\top} )  \}K + zz^\top  \otimes \Sigma^{-1}]L^\top 
\end{bmatrix}.
\]
Taking the negative expectation of $\nabla_\lambda^2 \ell_q$ and applying the fact that $\E(z) = 0$ and $\E(z z^\top ) = I_d$, we obtain 
\begin{equation*} 
F_\lambda = \begin{bmatrix}
\Sigma^{-1} & 0 \\
0 & L \{  (C^{-1} \otimes C^{-\top}) K +  (I_d \otimes \Sigma^{-1}) \}L^\top  
\end{bmatrix} =
\begin{bmatrix}
\Sigma^{-1} & 0 \\ 0 & \mathfrak{I}(C)
\end{bmatrix}.
\end{equation*}
Thus $F_\lambda^{-1} = \blockdiag(\Sigma, \mathfrak{I}(C)^{-1})$.

For the second parametrization, $\lambda = (\mu^\top , \vech(T)^\top )^\top $, the first order derivative, 
\[
\nabla_{\vech(T)} \ell_q =\vech(T^{-\top}  - (\theta -\mu) (\theta -\mu)^\top  T) 
\]
and 
\[
\nabla_\lambda^2 \ell_q= -
\begin{bmatrix}
\Sigma^{-1} & - \{(\theta -\mu)^\top  T \otimes I_d + T \otimes (\theta -\mu)^\top \}L^\top  \\
\cdot  &   L\{ (T^{-1} \otimes T^{-\top}  ) K + I_d \otimes (\theta -\mu)(\theta -\mu)^\top  \}L^\top 
\end{bmatrix}.
\]
Taking the negative expectation of $\nabla_\lambda^2 \ell_q$ and applying the fact that $\E(\theta) = \mu$ and $\E[(\theta -\mu)(\theta -\mu)^\top ] = \Sigma$, we obtain 
\begin{equation*} 
F_\lambda = \begin{bmatrix}
\Sigma^{-1} & 0 \\
0 & L\{ (T^{-1} \otimes T^{-\top} ) K + I_d \otimes \Sigma \}L^\top  
\end{bmatrix} =
\begin{bmatrix}
\Sigma^{-1} & 0 \\ 0 & \mathfrak{I}(T)
\end{bmatrix}.
\end{equation*}
Thus $F_\lambda^{-1} = \blockdiag(\Sigma, \mathfrak{I}(T)^{-1})$.

Suppose $\nabla_{\vech(C)} \mL = \vech(G)$. Then, the natural gradient is 
\[
\widetilde{\nabla} _{\lambda} \mL = F_{\lambda}^{-1} \nabla_{\lambda} \mL 
= \begin{bmatrix} \Sigma & 0 \\ 0 & \mathfrak{I}(C)^{-1} \end{bmatrix}
\begin{bmatrix} \nabla_\mu \mL \\ \vech(G) \end{bmatrix} 
= \begin{bmatrix} \Sigma \nabla_\mu \mL  \\  \vech(C \dH) \end{bmatrix},
\]
where we have applied Lemma 1 (iii) in the last step. The proof is similar for the second parametrization.

\subsection{Proof of Corollary 1}
If $\xi = ((T^\top  \mu)^\top , \vech(T)^\top )^\top $, then 
\begin{equation*}
\begin{aligned}
\widetilde{\nabla}_\xi \mL =  (\nabla_\lambda \xi)^\top  \widetilde{\nabla}_\lambda \mL 
= \begin{bmatrix} T^\top  & (I \otimes \mu^\top ) L^\top  \\ 0 & I \end{bmatrix}
\begin{bmatrix} \Sigma \nabla_\mu \mL \\  \vech(T \dH) \end{bmatrix} 
= \begin{bmatrix}
T^{-1} \nabla_\mu \mL + \dH^\top  T^\top  \mu \\  \vech(T \dH) \end{bmatrix}.
\end{aligned}
\end{equation*}
The natural gradient ascent update is 
\[
\begin{aligned}
{T^{(t+1)}}^\top \mu^{(t+1)} &= {T^{(t)}}^\top \mu^{(t)} + \rho_t \{ {T^{(t)}}^{-1} \nabla_\mu \mL  + {\dH^{(t)}}^\top {T^{(t)}}^\top \mu^{(t)}  \},  \\
T^{(t+1)} &= T^{(t)} + \rho_t T^{(t)}  \dH^{(t)},
\end{aligned}
\]
The first line simplifies to 
\[
\begin{aligned}
{T^{(t+1)}}^\top \mu^{(t+1)} &=\{T^{(t)} + \rho_t T^{(t)}\dH^{(t)} \} ^\top  \mu^{(t)} + \rho_t {T^{(t)}}^{-1} \nabla_\mu \mL  \\
\implies \mu^{(t+1)} &= \mu^{(t)} + \rho_t {T^{(t+1)}}^{-\top} {T^{(t)}}^{-1} \nabla_\mu \mL.
\end{aligned}
\]

\subsection{Proof of Lemma 2}
Define $g(\theta) = (\theta - \mu) ^\top  e_j h(\theta)$ to be a function from $\mathbbm{R}^d$ to $\mathbbm{R}$ and $e_j$ to be a $d \times 1$ vector with the $j$th element equal to one and zero elsewhere. Then
\[
\nabla_\theta g(\theta) = h(\theta) e_j  +  (\theta - \mu) ^\top  e_j \nabla_\theta h(\theta).
\] 
Replacing $h(\theta)$ by $g(\theta)$ in Stein's Lemma, we obtain
\[
\E_q[\Sigma^{-1} (\theta-\mu)  (\theta - \mu) ^\top  e_j h(\theta)] =  \E_q[h(\theta) e_j  +  (\theta - \mu) ^\top  e_j \nabla_\theta h(\theta)].
\]
This implies that for any $1 \leq i,j \leq d$, 
\[
e_i^\top  \E_q[\{\Sigma^{-1} (\theta-\mu)  (\theta - \mu) ^\top  - I_d\}  h(\theta)]e_j =  e_i^\top  \E_q[ \nabla_\theta h(\theta)(\theta - \mu) ^\top ] e_j.
\]
Thus $\E_q[\{\Sigma^{-1} (\theta-\mu)  (\theta - \mu) ^\top  - I_d\}  h(\theta)] =  \E_q[ \nabla_\theta h(\theta)(\theta - \mu) ^\top ]$ since every $(i,j)$ element of these two matrices agree with each other.

\subsection{Proof of Theorem 2}
Post-multiplying the identity in Lemma 2 by $C^{-\top} $, we obtain
\[
\E_q[\{\Sigma^{-1} (\theta-\mu)  (\theta - \mu) ^\top  C^{-\top}  - C^{-\top} \}  h(\theta)] =  \E_q[ \nabla_\theta h(\theta)(\theta - \mu) ^\top  C^{-\top} ] = \E_q(\G_1).
\]
\[
\begin{aligned}
\therefore \;\;  \nabla_{\vech(C)} \mL &= \int \nabla_{\vech(C)} q_\lambda (\theta)  h(\theta) \df \theta \\
& =  \int q_\lambda (\theta) \vech \left\{\Sigma^{-1}(\theta -\mu)(\theta-\mu)^\top C^{-\top}  -  C^{-\top}  \right\} h(\theta) \df \theta \\
& = \E_q \left[ \vech \{\Sigma^{-1}(\theta -\mu)(\theta-\mu)^\top C^{-\top}  -  C^{-\top} \} h(\theta) \right] 
= \E_q \vech(\G_1),
\end{aligned}
\]
and the first part of the first identity in Theorem 2 is shown. For the second part of the identity, we have $\E_q[\Sigma^{-1} (\theta-\mu) \nabla_\theta h(\theta)^\top ]= \E_q[\nabla_\theta^2 h(\theta) ]$ from Price's Theorem. Taking the transpose and post-multiplying by $C$, we obtain 
\[
\E_q(\G_1) = \E_q[ \nabla_\theta h(\theta)  (\theta-\mu)^\top  C^{-\top} ]= \E_q[\nabla_\theta^2 h(\theta)C ] = \E_q(\F_1).
\]

Taking the transpose of the identity in Lemma 2 and post-multiplying by $T^{-\top} $,
\[
\E_q[\{ (\theta-\mu)  (\theta - \mu) ^\top  T - T^{-\top} \}  h(\theta)] = \E_q[ (\theta - \mu) \nabla_\theta h(\theta)^\top  T^{-\top} ] = -\E_q(\G_2).
\] 

\[
\begin{aligned}
\therefore \;\; \nabla_{\vech(T)} \mL &= \int \nabla_{\vech(T)} q_\lambda (\theta)  h(\theta) \df \theta \\
& = \int q_\lambda (\theta) \vech \{T^{-\top}  -  (\theta-\mu)(\theta - \mu)^\top  T\}  h(\theta) \df \theta \\
& = \E_q[ \vech \{T^{-\top}  -  (\theta-\mu)(\theta - \mu)^\top  T\} h(\theta)] 
= \E_q \vech (\G_2) ,
\end{aligned}
\]
and the first part of the second identity in Theorem 2 is shown. For the second part of the identity, we have $\E_q[\Sigma^{-1} (\theta-\mu) \nabla_\theta h(\theta)^\top ]= \E_q[\nabla_\theta^2 h(\theta) ]$ from Price's Theorem. Pre-multiplying by $\Sigma$ and post-multiplying by $T^{-\top} $, we obtain 
\[
- \E_q (\G_2) = \E_q[(\theta-\mu) \nabla_\theta h(\theta)^\top T^{-\top} ]= \E_q[\Sigma \nabla_\theta^2 h(\theta) T^{-\top} ] = - \E_q(\F_2).
\]

To obtain unbiased estimates in terms of the first order derivative, we can also apply the reparametrization trick directly, which will lead to the same results. For $\lambda = (\mu^\top, \vech(C)^\top)^\top$ where $\Sigma = CC^\top$, let $\theta = \mu + Cz$. For $\lambda = (\mu^\top, \vech(T)^\top)^\top$ where $\Sigma^{-1} = T T^\top$, let $\theta = \mu + T^{-\top}z$.
\[
\begin{aligned}
\nabla_{\vech(C)} \mL &= \nabla_{\vech(C)} \theta \nabla_\theta h(\theta)  \\
&= L (z \otimes I) \nabla_\theta h(\theta) \\
&= L \vc(\nabla_\theta h(\theta) z^\top) \\
&= \vech(\nabla_\theta h(\theta) z^\top),
\end{aligned}
\qquad 
\begin{aligned}
\nabla_{\vech(T)} \mL &= \nabla_{\vech(T)} \theta \nabla_\theta h(\theta)  \\
&= -L (T^{-1} \otimes T^{-\top}  z) \nabla_\theta h(\theta) \\
&= -L \vc(T^{-\top}  z \nabla_\theta h(\theta) T^{-\top}) \\
&= - \vech(T^{-\top}  z \nabla_\theta h(\theta) T^{-\top}).
\end{aligned}
\]

\section{Convergence of Snngm} \label{S5}
First we derive some intermediate results that are needed for the proof of Theorem 3. Let $\langle \cdot, \cdot\rangle$ denote the inner product and $\bar{g}_{t}  = \widetilde{g}_t/\|\widetilde{g}_t\|$ so that $\|\bar{g}_t\| = 1$.

\begin{lemma} \label{lem_beta}
\[
\sum_{i=0}^{t-1} i \beta^i \leq \frac{\beta(1-\beta^t)}{(1-\beta)^2}.
\]
\begin{proof}
\[
\begin{aligned}
\sum_{i=0}^{t-1} i \beta^i &= \beta \{1 + 2\beta + \dots (t-1) \beta^{t-2}\}
= \beta \frac{\df}{\df \beta} (\beta + \beta^2 + \dots + \beta^{t-1}) \\
&= \beta \frac{\df}{\df \beta} \left\{ \frac{\beta(1-\beta^{t-1})}{1-\beta} \right\}
= \frac{\beta \{1 - \beta^t - t \beta^{t-1} (1-\beta)\} }{(1-\beta)^2} \leq \frac{\beta(1-\beta^t)}{(1-\beta)^2}.
\end{aligned}
\]
\end{proof}
\end{lemma}

\subsection{Bounds for norm of natural gradient} 
We have $\| \widetilde{g}_t \|^2 = \widehat{g}_t^\top   F_t^{-2}\widehat{g}_t$.  By (A2), $\|\widehat{g}_t\| \leq R$ and by (A4), $R_1 \leq \ev(F_t) \leq R_2$. This implies that $1/R_2 \leq \ev(F_t^{-1}) \leq 1/R_1$. Using the result in Pg. 18 of \cite{Magnus2019},
\[
\begin{aligned}
\frac{1}{R_2} &\leq \frac{g_t^\top   F_t^{-1}g_t}{g_t^\top  g_t} \leq \frac{1}{R_1}  
\implies \frac{\|g_t\|^2}{R_2} \leq \langle g_t, F_t^{-1}g_t \rangle  \leq \frac{\|g_t\|^2}{R_1}, \\
\frac{1}{R_2^2} &\leq \frac{\widehat{g}_t^\top   F_t^{-2}\widehat{g}_t}{\widehat{g}_t^\top   \widehat{g}_t} \leq \frac{1}{R_1^2}  
\implies \frac{\|\widehat{g}_t\|}{R_2} \leq \| \widetilde{g}_t \| \leq \frac{\|\widehat{g}_t\|}{R_1} \leq \frac{R}{R_1}. 
\end{aligned}
\]

\subsection{Bound on momentum} \label{bound on momentum}
Since $m_0 = 0$, $m_t = \beta m_{t-1} + (1-\beta) \bar{g}_{t} = (1-\beta) \sum_{i=0}^{t-1} \beta^i \bar{g}_{t-i}$. Thus,
\begin{equation} \label{bd on m_t}
\|m_t\|  \leq (1-\beta) \sum_{i=0}^{t-1} \beta^{i} \|\bar{g}_{t-i}\|
= (1-\beta) \sum_{i=0}^{t-1} \beta^i 
= 1-\beta^t. 
\end{equation}

\subsection{Inequality from $L$-Lipschitz smooth assumption}
Define $G(t) = \mL(\lambda + t(\lambda'-\lambda))$. Then $G'(t) = \nabla_\lambda \mL(\lambda + t(\lambda'-\lambda))^\top  (\lambda'-\lambda)$. Now,
\[
G(1)  = G(0) + G'(0) + \int_0^1 G'(t) - G'(0) \; \df t.
\]
Therefore, by Cauchy-Schwarz inequality, 
\begin{align} \label{Lsmooth}
|G(1) - G(0) -G'(0) | &\leq \int_0^1 | \langle \nabla_\lambda \mL(\lambda + t(\lambda'-\lambda)) - \nabla_\lambda \mL(\lambda), \lambda' - \lambda \rangle |\, \df t \nonumber \\
& \leq  L \|\lambda'-\lambda \|^2 \int_0^1 t \, \df t 
= L \|\lambda'-\lambda \|^2/2.\nonumber \\
\therefore  |\mL(\lambda') - \mL(\lambda) &-\langle \nabla_\lambda \mL(\lambda), \lambda' - \lambda \rangle |  \leq L \|\lambda'-\lambda \|^2/2 \nonumber \\ 
\implies  -\mL(\lambda') + \mL(\lambda) &+ \langle \nabla_\lambda \mL(\lambda), \lambda' - \lambda \rangle \leq L \|\lambda'-\lambda \|^2/2.
\end{align}

\subsection{Proof of Theorem 3}
Set $\lambda=\lambda^{(t)}$ and $\lambda'=\lambda^{(t+1)}$ in \eqref{Lsmooth}. Since $\lambda^{(t+1)} = \lambda^{(t)} + \alpha m_t/(1-\beta^t)$, $m_t = (1-\beta) \sum_{i=0}^{t-1} \beta^i \bar{g}_{t-i}$ and $\|m_t\| \leq 1 -\beta^t$ from \eqref{bd on m_t},
\begin{align} \label{E1}
\mL(\lambda^{(t)}) & \leq \mL(\lambda^{(t+1)}) - \langle \nabla_\lambda \mL(\lambda^{(t)}), \lambda^{(t+1)}- \lambda^{(t)} \rangle  + L \| \lambda^{(t+1)}- \lambda^{(t)} \|^2/2 \nonumber \\
&= \mL(\lambda^{(t+1)}) - \frac{\alpha}{1-\beta^t} \langle \nabla \mL(\lambda^{(t)}),m_t \rangle  + \frac{L \alpha^2}{2(1-\beta^t)^2} \| m_t \|^2 \nonumber \\
& \leq \mL(\lambda^{(t+1)}) - \frac{\alpha (1-\beta)}{1-\beta^t} \sum_{i=0}^{t-1} \beta^i \langle \nabla_\lambda \mL(\lambda^{(t)}), \bar{g}_{t-i} \rangle + \frac{L \alpha^2}{2}.
\end{align}

Write $\langle \nabla_\lambda \mL(\lambda^{(t)}), \bar{g}_{t-i} \rangle =  \langle \nabla_\lambda \mL(\lambda^{(t)}) - \nabla_\lambda \mL(\lambda^{(t-i)}), \bar{g}_{t-i} \rangle  +  \langle \nabla_\lambda \mL(\lambda^{(t-i)}), \bar{g}_{t-i} \rangle$. For the first term, applying the Cauchy–Schwarz inequality and $L$-Lipschitz smooth assumption (A3),
\[
|\langle \nabla_\lambda \mL(\lambda^{(t)}) - \nabla_\lambda \mL(\lambda^{(t-i)}), \bar{g}_{t-i} \rangle| \leq \|\nabla_\lambda \mL(\lambda^{(t)}) - \nabla_\lambda \mL(\lambda^{(t-i)}) \| \|\bar{g}_{t-i}\| \leq L \|\lambda^{(t)} - \lambda^{(t-i)} \|,
\]
where $\lambda^{(t)} - \lambda^{(t-i)} = \sum_{j=t-i}^{t-1} \{\lambda^{(j+1)} - \lambda^{(j)}\} = \alpha \sum_{j=t-i}^{t-1} m_j/(1-\beta^j)$. Hence
\begin{align*}
\|\lambda^{(t)} - \lambda^{(t-i)} \|  \leq \alpha \sum_{j=t-i}^{t-1} \frac{\| m_j \|}{1-\beta^j} \leq \alpha i.
\end{align*}

For the second term,
\[
\langle \nabla_\lambda \mL(\lambda^{(t-i)}), \bar{g}_{t-i} \rangle 
= \langle g_{t-i}, F_{t-i}^{-1}\widehat{g}_{t-i} \rangle/\|\widetilde{g}_{t-i} \|
\geq \frac{R_1}{R}  \langle g_{t-i}, F_{t-i}^{-1}\widehat{g}_{t-i} \rangle.
\]
Substituting these back into \eqref{E1} and applying Lemma \ref{lem_beta},   
\begin{align*}
\mL(\lambda^{(t)}) & \leq \mL(\lambda^{(t+1)}) + \frac{\alpha (1-\beta)}{1-\beta^t} \sum_{i=0}^{t-1} \beta^i \left( L\alpha i - \frac{R_1}{R}\langle g_{t-i}, F_{t-i}^{-1}\widehat{g}_{t-i} \rangle \right) + \frac{L \alpha^2}{2} \\ 
& \leq \mL(\lambda^{(t+1)}) + \frac{L\alpha ^2 \beta }{(1-\beta)} - \frac{R_1 \alpha (1-\beta)}{R(1-\beta^t)} \sum_{i=0}^{t-1} \beta^i\langle g_{t-i}, F_{t-i}^{-1}\widehat{g}_{t-i} \rangle + \frac{L \alpha^2}{2}.
\end{align*}
Taking expectation,
\[
\begin{aligned}
 \frac{R_1 \alpha (1-\beta)}{R} \sum_{i=0}^{t-1} \beta^i \langle g_{t-i}, F_{t-i}^{-1} g_{t-i} \rangle 
&\leq  \frac{R_1 \alpha (1-\beta)}{R(1-\beta^t)} \sum_{i=0}^{t-1} \beta^i \langle g_{t-i}, F_{t-i}^{-1} g_{t-i} \rangle \\
&\leq \mL(\lambda^{(t+1)}) - \mL(\lambda^{(t)}) + \frac{L\alpha ^2 \beta }{(1-\beta)}  + \frac{L \alpha^2}{2}.
\end{aligned}
\]
Summing over $t=1$ to $t=N$ and applying $\mL(\lambda^{N+1}) \leq \mL^*$ by (A1),
\[
\begin{aligned}
 \frac{R_1 \alpha (1-\beta)}{R} \sum_{t=1}^N  \sum_{i=0}^{t-1} \beta^i \langle g_{t-i}, F_{t-i}^{-1} g_{t-i} \rangle 
&\leq \mL(\lambda^{(N+1)}) - \mL(\lambda^{(1)}) + \frac{N L\alpha ^2 \beta }{(1-\beta)}  + \frac{N L \alpha^2}{2} \\
&\leq \mL^* - \mL(\lambda^{(1)}) + \frac{N L\alpha ^2 \beta }{(1-\beta)}  + \frac{N L \alpha^2}{2}.
\end{aligned}
\]
Since $\langle g_{t-i}, F_{t-i}^{-1} g_{t-i} \rangle\geq \|g_{t-i}\|^2/R_2$,
\[
\begin{aligned}
 \frac{R_1 \alpha (1-\beta)}{R_2 R} \sum_{t=1}^N  \sum_{i=0}^{t-1} \beta^i \|g_{t-i}\|^2 
&\leq \mL^* - \mL(\lambda^{(1)}) + \frac{N L\alpha ^2 \beta }{(1-\beta)}  + \frac{N L \alpha^2}{2}.
\end{aligned}
\]
Let $j = t-i$ and interchanging the summation,
\begin{align*}
 \sum_{t=1}^N  \sum_{i=0}^{t-1} \beta^i \| g_{t-i} \|^2
&= \sum_{t=1}^N  \sum_{j=1}^t \beta^{t-j} \| g_{j} \|^2  
= \sum_{j=1}^N  \sum_{t=j}^N  \beta^{t-j} \| g_{j} \|^2  
= \frac{\sum_{j=1}^N  (1-\beta^{N-j+1})\| g_{j} \|^2}{1-\beta}.
\end{align*}
Therefore,
\begin{align*}
\E\|g_\tau \|^2 &= \sum_{j=1}^N  \frac{1 - \beta^{N-j+1}}{C} \|g_j \|^2  \leq  \frac{1}{\widetilde{N}}\sum_{j=1}^N  (1-\beta^{N-j+1})\| g_{j} \|^2\\
&= \frac{1-\beta}{\widetilde{N}}  \sum_{t=1}^N  \sum_{i=0}^{t-1} \beta^i \| g_{t-i} \|^2 \\
& \leq \frac{R R_2}{\widetilde{N} R_1 \alpha} \left\{ \mL^* - \mL(\lambda^{(1)}) + \frac{N L\alpha ^2 \beta }{(1-\beta)}  + \frac{N L \alpha^2}{2} \right\}.
\end{align*}

\section{Methods for comparison}
In the experiments, we compare proposed methods with Adam \citep{Kingma2015}, variational online Gauss-Newton \citep[VOGN,][]{Khan2018} and ProxGenAdam \citep{Kim2023}. The updates used in VOGN and ProxGenAdam are summarized below.

\subsection{Variational online Gauss-Newton (VOGN)}
\cite{Khan2018} used the generalized Gauss-Newton approximation for the Hessian of the log-likelihood in the natural gradient update of the natural parameter for the Gaussian density. They assumed a normal prior, $\theta \sim \N(0, \sigma_0^2 I)$, and a log-likelihood of the form $\log p(y|\theta) = \sum_{i=1}^n \log p(y_i|\theta)$. Suppose $B_t$ contains the indices  for a minibatch of observations sampled uniformly at random from $\{y_1, \dots, y_n\}$ at iteration $t$. Let
\[
\hat{g}_t^L = \frac{n}{|B_t|} \sum_{i \in B_t} \nabla_\theta \log p(y_i|\theta)
\]
be the unbiased estimate of $\nabla_\theta \log p(y|\theta)$ and $\hat{H}_t^L$ be the generalized Gauss-Newton estimate of $-\nabla_\theta^2 \log p(y|\theta)$ at iteration $t$. Then the $(j,k)$ element of $\hat{H}_t^L$ is given by
\begin{equation*} \label{GGN}
\frac{n}{|B_t|} \sum_{i \in B_t} \{ \nabla_{\theta_j} \log p(y_i|\theta) \} \{ \nabla_{\theta_k} \log p(y_i|\theta) \}.
\end{equation*}
The remaining terms in the lower bound and their gradients are evaluated analytically:
\[
\begin{aligned}
\nabla_\mu \E_q[\log p(\theta) + \log q_\lambda (\theta)] &= - \mu/\sigma_0^2, \\
\nabla_{\Sigma} \E_q[\log p(\theta) + \log q_\lambda (\theta)] &= \tfrac{1}{2} ( \Sigma^{-1} - I_d/\sigma_0^2).
\end{aligned}
\]
  Let $S = \Sigma^{-1}$ denote the precision matrix. Applying the theorems of Bonnet and Price for the log-likelihood term, the natural gradient update at iteration $t$ is
\begin{equation*} 
\begin{aligned}
{S^{(t+1)}} &= {S^{(t)}} - 2 \rho_t (- \tfrac{1}{2}\hat{H}_t^L + \tfrac{1}{2}{S^{(t)}} - \tfrac{1}{2} I_d/\sigma_0^2 ) = (1- \rho_t){S^{(t)}} + \rho_t (\hat{H}_t^L + I_d/\sigma_0^2), \\
\mu^{(t+1)} &= \mu^{(t)} + \rho_t \Sigma^{(t+1)} (\hat{g}_t^L - \mu^{(t)}/\sigma_0^2).
\end{aligned}
\end{equation*}
\cite{Khan2018} further incorporated momentum and introduced bias correction to arrive at Adam-like updates. Let $TT^\top$ be the Cholesky factorization of $S$.
\begin{itemize}
\item Initialize $m_0 = 0$, $\mu_1$ and $T^{(1)}$. Set $S^{(1)} = T^{(1)} {T^{(1)}}^\top$. For $t=1, \dots, T$, 
\begin{enumerate}[1.]
\item Generate $z \sim N(0, I_d)$ and compute $\theta^{(t)} = {T^{(t)}}^{-\top} z + \mu^{(t)}$, $\hat{g}_t^L$ and $\hat{H}_t^L$.
\item $m_t = \beta_1 m_{t-1}+ (1 - \beta_1) (\hat{g}_t^L - \mu^{(t)}/\sigma_0^2)$,
\item $\hat{m}_t = m_t /(1-\beta_1^t)$,
\item $S^{(t+1)} = \beta_2 S^{(t)} + (1 - \beta_2) (\hat{H}_t^L  + I_d/\sigma_0^2)$,
\item $\mu^{(t+1)} = \mu^{(t)} + \alpha {S^{(t+1)}}^{-1} \hat{m}_t$.
\item Find the Cholesky factor, $T^{(t+1)}$, of ${S^{(t+1)}}$.
\end{enumerate}
\end{itemize}
Using the default values in Adam as reference, we set $\beta_1 = 0.9$, $\beta_2 = 0.999$ and $\alpha = 0.001$. We adjust $\alpha$ upwards by a factor of 10 each time if necessary. 

For the loglinear model where $\mL$ is tractable and $\nabla_\Sigma \E_q[\log p(y|\theta)$ is positive definite, we replace the estimates $\hat{g}_t^L$ and $\hat{H}_t^L$ respectively by $\frac{n}{|B_t|} \sum_{i \in B_t} \nabla_\mu \E_q[\log p(y_i|\theta)]$ and $- \frac{2n}{|B_t|} \sum_{i \in B_t} \nabla_\Sigma \E_q[\log p(y_i|\theta)]$. 

For GLMMs, $\theta = (\theta_1^\top, \dots, \theta_n^\top, \theta_g)^\top$ and the prior is not of the form $\theta \sim \N(0, \sigma_0^2 I)$. Moreover, $\E_q[\log p(\theta_i|\theta_g)]$ cannot be evaluated analytically. Hence, we consider a generalized Gauss-Newton approximation of
\[
\log p(y, \theta) = \sum_{i=1}^n f_i(\theta), \;\; \text{where} \;\;
f_i(\theta) = \log p(y_i|\theta_i, \theta_g) + \log p(\theta_i|\theta_G) + \frac{1}{n}\log p(\theta_G), 
\]
where $- \nabla_\theta^2 \log p(y, \theta) \approx J^\top J$ and
\[
J = \begin{bmatrix}
\frac{\partial f_1}{\partial \theta_1} & 0 & \dots & 0 & \frac{\partial f_1}{\partial \theta_g} \\
0 & \frac{\partial f_2}{\partial \theta_2} & \dots & 0 & \frac{\partial f_2}{\partial \theta_g} \\ 
\vdots & \vdots & \ddots & \vdots & \vdots \\
0 & 0 & \dots & \frac{\partial f_n}{\partial \theta_1} & \frac{\partial f_n}{\partial \theta_g}
\end{bmatrix}.
\]
Correspondingly, we replace $(\hat{g}_t^L - \mu^{(t)}/\sigma_0^2)$ in step (1) by an estimate of $\nabla \log p(y, \theta)$ at iteration $t$ and 
$(\hat{H}_t^L  + I_d/\sigma_0^2)$ in step (4) by $J^\top J$.

\subsection{ProxGenAdam}
Let 
\[
-\mL(\lambda) = \underbrace{-\E_q[\log p(y, \theta)]}_{\mL_1(\lambda)} + \underbrace{\E_q[\log q_\lambda (\theta)]}_{\mL_2(\lambda)}.
\]
The stochastic gradient descent (SGD) update, $\lambda \leftarrow \lambda - \gamma (\nabla_\lambda \mL_1(\lambda) + \nabla_\lambda \mL_2(\lambda))$, where $\gamma$ is the stepsize,  can be interpreted as minimizing a linear approximation of the negative lower bound with a quadratic penalty:
\[
s(v) = \mL_1(\lambda) + \mL_2(\lambda) + \nabla_\lambda \mL_1(\lambda)^\top (v - \lambda) +
\nabla_\lambda \mL_2(\lambda)^\top (v - \lambda) + \frac{1}{2\gamma} \| v - \lambda \|^2.
\]
\cite{Domke2020} considered a Cholesky decomposition, $\Sigma = CC^\top$, and noted that $\mL_2(\lambda)$ is non-smooth. He proposed proximal optimization by using this term in the objective instead of its linear approximation. This led to minimization of 
\[
\begin{aligned}
s(v) &= \mL_1(\lambda) + \nabla_\lambda \mL_1(\lambda)^\top (v - \lambda) + \mL_2(v) + \frac{1}{2\gamma} \| v - \lambda \|^2 \\
&= \mL_2(v) + \frac{1}{2\gamma} \| v - \lambda' \|^2 + \text{terms not inolving $v$},
\end{aligned}
\]
where $\lambda' = \lambda - \gamma \nabla_\lambda \mL_1(\lambda)$. Since $\mL_2(\lambda) = -\frac{d}{2} \log(2\pi) - \sum_{i=1}^d \log |C_{ii}| - \frac{d}{2}$ involves only the diagonal elements of $C$, the updates of all other elements in $\lambda$ remain the same, while we seek $C_{ii}$ that minimizes
\[
L_i = -\log |C_{ii}| + \frac{1}{2\gamma} (C_{ii} - C_{ii}')^2, \qquad (i =1, \dots, d),
\]
where $C_{ii}'$ is the SGD update for $C_{ii}$ based on $\mL_1(\lambda)$ only. We have 
\[
\begin{aligned}
\frac{\partial L_i}{\partial C_{ii}} &= \frac{(C_{ii} - C_{ii}')}{\gamma} - \frac{1}{C_{ii}} = 0 \implies C_{ii}  = \frac{C_{ii}' \pm \sqrt{{C_{ii}'}^2 + 4\gamma}}{2}, \\
\frac{\partial^2 L_i}{\partial C_{ii}^2} &= \frac{1}{\gamma} + \frac{1}{C_{ii}^2} > 0.
\end{aligned}
\]
Hence, both solutions lead to minimum points. However, if $C_{ii}' > 0$, then the global minimum occurs at $C_{ii} = \{ C_{ii}' + ({C_{ii}'}^2 + 4\gamma)^{1/2} \}/2$. On the other hand, if $C_{ii}' < 0$, then the global minimum occurs at $C_{ii} = \{ C_{ii}' - ({C_{ii}'}^2 + 4\gamma)^{1/2} \}/2$. Thus, proximal gradient descent effectively keeps the diagonal entries of $C$ further away from 0.  

We also consider Cholesky decomposition of the precision matrix, $\Sigma^{-1} = TT^\top$, where we minimize 
\[
L_i = \log |T_{ii}| + \frac{1}{2\gamma} (T_{ii} - T_{ii}')^2, \qquad (i =1, \dots, d)
\]
with respect to $T_{ii}$ instead, and $T_{ii}'$ is the SGD update for $T_{ii}$ based on $\mL_1(\lambda)$ only. We have
\[
\begin{aligned}
\frac{\partial L_i}{\partial T_{ii}} &= \frac{(T_{ii} - T_{ii}')}{\gamma} + \frac{1}{T_{ii}} = 0 \implies T_{ii}  = \frac{T_{ii}' \pm \sqrt{{T_{ii}'}^2 - 4\gamma}}{2}, \\
\frac{\partial^2 L_i}{\partial T_{ii}^2} &= \frac{1}{\gamma} - \frac{1}{T_{ii}^2} = \frac{T_{ii}^2 - \gamma }{\gamma T_{ii}^2}.
\end{aligned}
\]
We require $T_{ii}'^2 > 4\gamma$ for the solution to be well-defined. If $T_{ii}' > 0$, then we have a minimum point at $T_{ii} = \{ T_{ii}' + ({T_{ii}'}^2 - 4\gamma)^{1/2} \}/2$ and a maximum point at the other solution. On the other hand, if  $T_{ii}' < 0$, then we have a minimum point at $T_{ii} = \{ T_{ii}' - ({T_{ii}'}^2 - 4\gamma)^{1/2} \}/2$ and a maximum point at the other solution. Thus proximal gradient descent effectively moves the diagonal entries of $T$ closer to 0.

In ProxGenAdam \citep{Yun2021, Kim2023}, momentum and adaptive stepsizes are combined with the proximal steps to yield Adam-like updates outlined below, although bias correction of estimated moments is not performed. 
\begin{itemize}
\item Initialize $m_0 = 0$, $v_0 = 0$ and $\lambda^{(1)}$. For $t=1, 2, \dots$, 
\begin{itemize}
\item Compute $\hat{g}_t^\ell$, an unbiased estimate of $\E_q[\log p(y, \theta)]$.
\item $m_t = \beta_1 m_{t-1} + (1 - \beta_1) \hat{g}_t^\ell$,
\item $v_t = \beta_2 v_{t-1} + (1 - \beta_2)  \hat{g}_t^{\ell \, 2}$,
\item $\lambda^{(t+1)} = \lambda^{(t)} +\frac{ \alpha}{\sqrt{v_t} + \epsilon} m_t$.
\item For $i=1, \dots, d$, update
\[
\begin{aligned}
C_{ii}^{(t+1)} &\leftarrow \frac{ C_{ii}^{(t+1)} + \text{sgn}(C_{ii}^{(t+1)}) \sqrt{{C_{ii}^{(t+1)}} ^2 + 4 \gamma_{t}} }{2}, \\
\text{or} \quad 
T_{ii}^{(t+1)} &\leftarrow \frac{ T_{ii}^{(t+1)} + \text{sgn}(T_{ii}^{(t+1)}) \sqrt{{T_{ii}^{(t+1)}} ^2 - 4 \gamma_{t}} }{2} 
\;\; \text{if} \;\; {T_{ii}^{(t+1)}} ^2 > 4 \gamma_{t}.
\end{aligned}
\]
\end{itemize}
\end{itemize}
We set $\alpha = 0.001$ , $\beta_1 = 0.9$, $\beta_2 = 0.999$ and $\epsilon = 10^{-8}$ similar to Adam and $\gamma_{t} = \frac{ \alpha}{\sqrt{v_{t}} + \epsilon}$.

\section{Natural gradient for sparse precision matrix} \label{S4}
We derive the natural gradient where the precision matrix is sparse.  In this case,
\[
\begin{aligned}
\ell_q &= \log q_\lambda (\theta) = -\frac{d}{2} \log (2\pi) + \log|T_g| + \sum_{i=1}^N \log |T_i| - \frac{1}{2} \sum_{i=1}^n  (\theta_i - \mu_i)^\top  T_i T_i^\top  (\theta_i - \mu_i) \\
&- \frac{1}{2} (\theta_g - \mu_g)^\top  \left( \sum_{i=1}^n T_{gi} T_{g_i}^\top  + T_g T_g^\top  \right)  (\theta_g - \mu_g) - (\theta_g - \mu_g)^\top  \sum_{i=1}^n T_{gi} T_i^\top   (\theta_i - \mu_i).
\end{aligned}
\]

\subsection{Proof of Theorem 4}
If we integrate out all other variables from $q_\lambda(\theta) =  q(\theta_g) \prod_{i=1}^n q(\theta_i|\theta_g)$ except $\theta_i$ and $\theta_g$, then $q(\theta_i, \theta_g) = q(\theta_i|\theta_g) q(\theta_g)$, whose covariance matrix is 
\[
\begin{aligned}
\begin{bmatrix} T_i & 0 \\ T_{gi} & T_g \end{bmatrix}^{-\top}  
&\begin{bmatrix} T_i & 0 \\ T_{gi} & T_g \end{bmatrix}^{-1} 
= \begin{bmatrix} T_i^{-\top}  & -T_i^{-\top} T_{gi}^\top  T_g^{-\top}    \\ 0 & T_g^{-\top}  \end{bmatrix}
\begin{bmatrix} T_i^{-1} & 0 \\ -T_g^{-1} T_{gi} T_i^{-1} & T_g^{-1} \end{bmatrix} \\
&= \begin{bmatrix} T_i^{-\top}   T_i^{-1} + T_i^{-\top} T_{gi}^\top  T_g^{-\top}  T_g^{-1} T_{gi} T_i^{-1} & -T_i^{-\top} T_{gi}^\top  T_g^{-\top}   T_g^{-1} \\[1mm]
 -T_g^{-\top}  T_g^{-1} T_{gi} T_i^{-1} & T_g^{-\top}  T_g^{-1} \end{bmatrix}.
\end{aligned}
\]
Hence $\Cov(\theta_g) = \E\{ (\theta_g - \mu_g)(\theta_g - \mu_g)^\top \} = T_g^{-\top}  T_g^{-1}$, 
\[
\begin{aligned}
\Cov(\theta_i) &= \E\{(\theta_i - \mu_i) (\theta_i - \mu_i)^\top \} = T_i^{-\top}   T_i^{-1} + T_i^{-\top} T_{gi}^\top  T_g^{-\top}  T_g^{-1} T_{gi} T_i^{-1}, \\
\Cov(\theta_i, \theta_g) &= \E\{(\theta_i - \mu_i)(\theta_g - \mu_g)^\top  \}
= -T_i^{-\top} T_{gi}^\top  T_g^{-\top}   T_g^{-1}.
\end{aligned}
\]

First, we find the elements in the Fisher information matrix. Differentiating $\ell_q$ with respect to $T_i$ and taking expectation with respect to $q_\lambda (\theta)$, 
\[
\begin{aligned}
\nabla_{\vech(T_i)} \ell_q &= \vech\{T_i^{-\top}  - (\theta_i - \mu_i) (\theta_i - \mu_i)^\top  T_i - (\theta_i - \mu_i)(\theta_g - \mu_g)^\top  T_{gi} \}, \\
\nabla_{\vech(T_i)}^2 \ell_q &= - L \{ (T_i^{-1} \otimes T_i^{-\top} )K +  I \otimes (\theta_i - \mu_i)  (\theta_i - \mu_i)^\top  \}L^\top , \\
\E[\nabla_{\vech(T_i)}^2 \ell_q] &= -L \{  (T_i^{-1} \otimes T_i^{-\top} )K + I \otimes T_i^{-\top}  (I + T_{gi}^\top  T_g^{-\top}  T_g^{-1} T_{gi}) T_i^{-1} \}L^\top  \\
&= - \mathfrak{I}(T_i) -  L ( I \otimes T_i^{-\top}  T_{gi}^\top  T_g^{-\top}  T_g^{-1} T_{gi} T_i^{-1} ) L^\top .
\end{aligned}
\]
Differentiating $\nabla_{\vech(T_i)} \ell_q$ with respect to $T_{gi}$ and taking expectation with respect to $q_\lambda (\theta)$, 
\[
\begin{aligned}
\nabla_{\vech(T_i), \vc(T_{gi})}^2 \ell_q &= - L \{I \otimes (\theta_i - \mu_i)(\theta_g - \mu_g)^\top   \}, \\
\E [\nabla_{\vech(T_i), \vc(T_{gi})}^2 \ell_q]&= L (I \otimes T_i^{-\top}  T_{gi}^\top  T_g^{-\top}  T_g^{-1} ).
\end{aligned}
\]
Differentiating $\ell_q$ with respect to $T_{gi}$ and taking expectation with respect to $q_\lambda (\theta)$, 
\[
\begin{aligned}
\nabla_{\vc(T_{gi})} \ell_q &= - \vc[ (\theta_g - \mu_g)(\theta_g - \mu_g)^\top   T_{gi} + (\theta_g - \mu_g)(\theta_i - \mu_i)^\top  T_i]. \\
\nabla_{\vc(T_{gi})}^2 \ell_q &= - (I \otimes  (\theta_g - \mu_g)(\theta_g - \mu_g)^\top  ) . \\
\E[\nabla_{\vc(T_{gi})}^2 \ell_q] &= - (I \otimes T_g^{-\top} T_g^{-1}).
\end{aligned}
\]
Differentiating $\ell_q$ with respect to $T_g$ and taking expectation with respect to $q_\lambda (\theta)$, 
\begin{align*}
\nabla_{\vech (T_g)} \ell_q &= \vech[ T_g^{-\top}  - (\theta_g - \mu_g)(\theta_g - \mu_g)^\top  T_g]. \\
\nabla_{\vech (T_g)}^2  \ell_q &= - L [ (T_g^{-1} \otimes T_g^{-\top} )K + I \otimes (\theta_g - \mu_g)(\theta_g - \mu_g)^\top  ]  L^\top.  \\
\E[\nabla_{\vech (T_g)}^2  \ell_q ]&= - L [ (T_g^{-1} \otimes T_g^{-\top} )K + I \otimes T_g^{-\top}  T_g^{-1} ]  L^\top  = - \mathfrak{I}(T_g).
\end{align*}
Thus the Fisher information matrix is $F_\lambda = \blockdiag(\Sigma^{-1}, F_1, \dots, F_n, \mathfrak{I}(T_g) )$, where 
\begin{equation*}
\begin{aligned}
F_i = \begin{bmatrix}
F_{11i} & F_{12i} \\
F_{12i}^\top  & F_{22i}
\end{bmatrix}
\end{aligned}
\qquad  \text{and} \qquad
\begin{aligned}
F_{11i} &= \mathfrak{I}(T_i) +  L ( I \otimes T_i^{-\top}  T_{gi}^\top  T_g^{-\top}  T_g^{-1} T_{gi} T_i^{-1} ) L^\top , \\
F_{12i} &= - L (I \otimes T_i^{-\top}  T_{gi}^\top  T_g^{-\top}  T_g^{-1}) , \\
F_{22i} &= (I \otimes T_g^{-\top} T_g^{-1}). 
\end{aligned}
\end{equation*}
Since $F_{22i}^{-1} = I \otimes T_gT_g^\top $, and $F_{12i}F_{22i}^{-1} = - L (I \otimes T_i^{-\top}  T_{gi}^\top )$, $F_{11i} - F_{12i} F_{22i}^{-1} F_{12i}^\top  = \mathfrak{I}(T_i)$. Hence using block matrix inversion,
\[
F_i^{-1} = \begin{bmatrix}
\mathfrak{I}(T_i)^{-1} & \mathfrak{I}(T_i)^{-1} L (I \otimes T_i^{-\top}  T_{gi}^\top )  \\
\cdot &  (I \otimes T_g T_g^\top ) +  (I \otimes T_{gi} T_i^{-1}) L^\top  \mathfrak{I}(T_i)^{-1} L (I \otimes T_i^{-\top}  T_{gi}^\top )
\end{bmatrix}.
\]

Next, we derive the expression for the natural gradient. Let $A_i, G_{gi}, G_i$ for $i=1, \dots, n$ and $G_g$ be as defined in theorem 4. In addition, let $H_i = T_i^\top \bar{G}_i$ for $i=1, \dots, n$ and $H_g = T_g^\top \bar{G}_{g}$. For $i=1, \dots, n$, 
\[
\begin{aligned}
\begin{bmatrix}
\widetilde{\nabla}_{\vech(T_i)} \mL \\ \widetilde{\nabla}_{\vc(T_{gi})} \mL
\end{bmatrix} 
&= F_i^{-1} \begin{bmatrix}  \vech(A_i)  \\  \vc(G_{gi})  \end{bmatrix}.
\end{aligned}
\]
Applying Lemma 1, 
\begin{align*}
\widetilde{\nabla}_{\vech(T_i)} \mL &= \mathfrak{I}(T_i)^{-1}  \vech(A_i) +  \mathfrak{I}(T_i)^{-1}  L (I \otimes T_i^{-\top}  T_{gi}^\top ) \vc(G_{gi}) \\
&= \mathfrak{I}(T_i)^{-1} \vech(A_i + T_i^{-\top}  T_{gi}^\top  G_{gi})  \\
&= \mathfrak{I}(T_i)^{-1} \vech(G_i )  \\
&= \vech(T_i \dH_i). \\
\widetilde{\nabla}_{\vc(T_{gi})} \mL 
&= (I \otimes T_gT_g^\top )  \vc(G_{gi})+  (I \otimes T_{gi} T_i^{-1}) L^\top  \widetilde{\nabla}_{\vech(T_i)} \mL \\
&= \vc(T_g T_g^\top  G_{gi}) +  (I \otimes T_{gi} T_i^{-1}) L^\top  \vech (T_i \dH_i) \\
&= \vc( T_gT_g^\top G_{gi} + T_{gi}\dH_i).
\end{align*} 
Finally, $\widetilde{\nabla}_{\vech(T_g)} \mL = \mathfrak{I}(T_g)^{-1} \vech(G_g) = \vech(T_g \dH_g)$ from Lemma 1. Hence the full natural gradient is 
\[
\widetilde{\nabla}_\lambda \mL = F_\lambda^{-1}  \begin{bmatrix}
\nabla_\mu \mL \\ 
\begin{bmatrix} \vech (A_i)  \\  \vc(G_{gi})  \\  \end{bmatrix}_{i=1:n} \\ 
\vech(G_g)
\end{bmatrix} = \begin{bmatrix}
\Sigma \nabla_\mu \mL \\
\begin{bmatrix}
\vech (T_i \dH_i) \\
\vc( T_{gi}\dH_i + T_g T_g^\top G_{gi}) \\
\end{bmatrix}_{i=1:n} \\
\vech(T_g \dH_g)
\end{bmatrix},
\]
where $[a_i]_{i=1:n} = (a_1^\top  , \dots, a_n^\top  )^\top $. To compute the natural gradient, note that 
\begin{equation*}
H = T_d^\top  \bar{G} = \begin{bmatrix}
H_1 &  \dots & 0 & 0 \\
\vdots & \ddots & \vdots & \vdots \\
0 & \dots  & H_n & 0 \\
T_g^\top   G_{g1}  & \dots & T_g^\top G_{gn}  & H_g
\end{bmatrix}, \;\;
\dH = \begin{bmatrix}
\dH_1 &  \dots & 0 & 0 \\
\vdots & \ddots & \vdots & \vdots \\
0 & \dots  & \dH_n & 0 \\
T_g^\top   G_{g1}  & \dots & T_g^\top G_{gn}  & \dH_g
\end{bmatrix}
\end{equation*}
and
\[
T\dH =  \begin{bmatrix}
T_1 \dH_1 &  \dots & 0 & 0 \\
\vdots & \ddots & \vdots & \vdots \\
0 & \dots  & T_n \dH_n & 0 \\
T_{g1}\dH_1 + T_g T_g^\top  G_{g1}  & \dots & T_{gn}\dH_n +  T_g T_g^\top  G_{gn}  & T_g \dH_g
\end{bmatrix}.
\]
Thus the natural gradient update for $T$ is $T \leftarrow T + \rho_t T\dH$.

\subsection{Stochastic natural gradients}
Let $\theta_a = (\theta_1^\top , \dots, \theta_n^\top )^\top $,  $\mu_a = (\mu_1^\top , \dots, \mu_n^\top )^\top $, $v_a=(v_1^\top , \dots, v_n^\top )^\top $  and
\[
T = \begin{bmatrix} T_a & 0 \\ T_{ga} & T_g \end{bmatrix}, \quad 
T^{-1} = \begin{bmatrix} T_a^{-1} & 0 \\ - T_g^{-1} T_{ga}  T_a^{-1} & T_g^{-1} \end{bmatrix}, 
\]
where $T_a = \blockdiag(T_1, \dots, T_n)$ and $T_{ga} = [T_{g1} \dots T_{gn}]$. 
Note that 
\[
\begin{aligned}
v &= \begin{bmatrix} T_a^{-1} & 0 \\ - T_g^{-1} T_{ga}  T_a^{-1} & T_g^{-1} \end{bmatrix} \begin{bmatrix} \nabla_{\theta_a} h(\theta)  \\ \nabla_{\theta_g} h(\theta)  \end{bmatrix}
= \begin{bmatrix}
T_a^{-1} \nabla_{\theta_a} h(\theta) \\
T_g^{-1} \{\nabla_{\theta_g} h(\theta) - T_{ga}  T_a^{-1} \nabla_{\theta_a} h(\theta)\} \end{bmatrix}
= \begin{bmatrix} v_a \\ v_g \end{bmatrix}, \\
u &=T_d^{-\top}  T^\top  (\theta -\mu)  = \begin{bmatrix}
\begin{bmatrix} (\theta_i - \mu_i) + T_i^{-\top}  T_{gi} (\theta_g - \mu_g) \end{bmatrix}_{i=1:n} \\
\theta_g - \mu_g \end{bmatrix} = 
\begin{bmatrix} [u_i]_{i=1:n} \\ u_g \end{bmatrix}.
\end{aligned}
\]

First, we extract entries in $\G_2 = - (\theta - \mu) v^\top $ corresponding to nonzero entries in $T$. We have
\[
\begin{aligned}
\nabla_{\vech(T_i)} \mL &=- \E_q \vech \{ (\theta_i - \mu_i)  v_i^\top \} 
=- \E_q \vech \{ (u_i - T_i^{-\top}  T_{gi} u_g)  v_i^\top \} , \\
\nabla_{\vech(T_g)} \mL & = -\E_q \vech \{ (\theta_g - \mu_g) v_g^\top  \} 
= -\E_q \vech \{ u_g v_g^\top  \}, \\
\nabla_{\vc(T_{gi})} \mL &= -\E_q \vc\{  (\theta_g - \mu_g) v_i^\top  \} 
= -\E_q \vc\{  u_g v_i^\top  \}.
\end{aligned}
\]

Next, we extract entries in $\F_2 = -\Sigma \nabla_\theta^2 h(\theta) T^{-\top} $ corresponding to nonzero entries in $T$. If $\Sigma_g = T_g^{-\top}  T_g^{-1}$, $\Sigma_i = T_i^{-\top}  T_i$ and $\Sigma_a = T_a^{-\top}  T_a^{-1}$, then  
\[
\begin{aligned}
\Sigma = T^{-\top}  T^{-1} 
&=  \begin{bmatrix} T_a^{-\top}  &  - T_a^{-\top}   T_{ga}^\top   T_g^{-\top}  \\ 0 & T_g^{-\top}  \end{bmatrix} \begin{bmatrix} T_a^{-1} & 0 \\ - T_g^{-1} T_{ga}  T_a^{-1} & T_g^{-1} \end{bmatrix} \\
&= \begin{bmatrix} \Sigma_a  + T_a^{-\top}   T_{ga}^\top   \Sigma_g T_{ga}  T_a^{-1} &  - T_a^{-\top}   T_{ga}^\top   \Sigma_g  \\[1mm]  - \Sigma_g T_{ga} T_a^{-1} & \Sigma_g  \end{bmatrix} 
\end{aligned} 
\]
and 
\[
\begin{aligned}
\nabla_\theta^2 h(\theta) T^{-\top} & = \begin{bmatrix} \nabla_{\theta_a}^2 h(\theta)  &   \nabla_{\theta_a, \theta_g}^2 h(\theta)  \\  \nabla_{\theta_g, \theta_a}^2 h(\theta) &  \nabla_{\theta_g}^2 h(\theta)  \end{bmatrix}
\begin{bmatrix} T_a^{-\top}  &  - T_a^{-\top}   T_{ga}^\top   T_g^{-\top}  \\ 0 & T_g^{-\top}  \end{bmatrix} \\
&= \begin{bmatrix} \nabla_{\theta_a}^2 h(\theta) T_a^{-\top}    &   \{\nabla_{\theta_a, \theta_g}^2 h(\theta) - \nabla_{\theta_a}^2 h(\theta) T_a^{-\top}  T_{ga}^\top  \} T_g^{-\top}       \\  \nabla_{\theta_g, \theta_a}^2 h(\theta) T_a^{-\top}    &  \{\nabla_{ \theta_g}^2 h(\theta) - \nabla_{\theta_g, \theta_a}^2 h(\theta) T_a^{-\top}  T_{ga}^\top  \} T_g^{-\top}     \end{bmatrix}.
\end{aligned}
\]
Let 
\[
\begin{aligned}
U_{gi} &= \Sigma_g \{\nabla_{\theta_g, \theta_i}^2 h(\theta) - T_{gi} T_i^{-1} \nabla_{\theta_i}^2 h(\theta)\}T_i^{-\top} \quad (i=1, \dots, n), \\
U_{ga} &= \Sigma_g \{ \nabla_{\theta_g, \theta_a}^2 h(\theta) - T_{ga}  T_a^{-1} \nabla_{\theta_a}^2 h(\theta)  \} T_a^{-\top} 
= \begin{bmatrix} U_{g1} & \dots & U_{gn} \end{bmatrix}, \\
U_{gg} &= \Sigma_g  \{\nabla_{\theta_g}^2 h(\theta) - T_{ga}  T_a^{-1} \nabla_{\theta_a, \theta_g}^2 h(\theta)\}  T_g^{-\top}. 
\end{aligned}
\]
Then
\[
\begin{aligned}
\F_{2, 11} &= T_a^{-\top}  T_{ga}^\top U_{ga} - \Sigma_a \nabla_{\theta_a}^2 h(\theta)T_a^{-\top},
\\
\F_{2,21} &= - U_{ga} , \\
\F_{2,22} &= U_{ga} T_{ga}^\top T_g^{-\top} - U_{gg}.
\end{aligned}
\]
Thus, we have 
\[
\begin{aligned}
\nabla_{\vech(T_i)} \mL &= \E_q \vech (T_i^{-\top} T_{gi} u_g  v_i^\top - u_i  v_i^\top) = \E_q \vech (T_i^{-\top} T_{gi}^\top  U_{gi} - \Sigma_i  \nabla_{\theta_i}^2 h(\theta)  T_i^{-\top} ), \\
\nabla_{\vc(T_{gi})} \mL &= -\E_q \vc (u_g v_i^\top )
= -\E_q \vc(U_{gi}), \\
\nabla_{\vech(T_g)} \mL &= -\E_q \vech(u_g v_g^\top ) = \E_q \vech (U_{ga} T_{ga}^\top T_g^{-\top} - U_{gg}) .
\end{aligned}
\]

Using the above results, we can obtain unbiased estimates of the natural gradient in terms of the first derivative $\nabla_\theta h(\theta)$ by setting
\[
G_i = - u_i v_i^\top, \;\;
G_{gi} = -u_gv_i^\top, \;\;    
G_g = -u_g v_g^\top,
\]
for $i=1, \dots, n$. Note that setting $A_i = T_i^{-\top} T_{gi} u_g  v_i^\top - u_i  v_i^\top =  - T_i^{-\top} T_{gi} G_{gi} - u_i  v_i^\top$ leads to $G_i = A_i + T_i^{-\top} T_{gi}^\top G_{gi} = -u_i v_i^\top$. Hence an unbiased estimate of $G$ is 
\[ 
\begin{bmatrix}
-u_1 v_1^\top &  0 & \cdot & 0 \\
\vdots & \ddots & \vdots & \vdots \\
0 & \dots  & -u_n v_n^\top & 0 \\
-u_g v_1^\top  & \dots & -u_g v_n^\top & -u_g v_g^\top
\end{bmatrix},
\]
and an unbiased estimate of $\bar{G}$ can be obtained by setting elements in $-u v^\top$ which correspond to zero entries in $T$ to zero.

On the other hand, unbiased estimates in terms of the second derivative $\nabla_\theta^2 h(\theta)$ can be obtained by setting
\[
G_i = - \Sigma_i  \nabla_{\theta_i}^2 h(\theta)  T_i^{-\top}, \;\;
G_{gi} = -U_{gi}, \;\;
G_g = U_{ga} T_{ga}^\top T_g^{-\top} - U_{gg}, 
\]
for $i=1, \dots, n$. Note that setting $A_i = T_i^{-\top} T_{gi}^\top  U_{gi} - \Sigma_i  \nabla_{\theta_i}^2 h(\theta)  T_i^{-\top}= - T_i^{-\top} T_{gi}^\top G_{gi} - \Sigma_i  \nabla_{\theta_i}^2 h(\theta)  T_i^{-\top}$ leads to $G_i = A_i + T_i^{-\top} T_{gi}^\top G_{gi} = - \Sigma_i  \nabla_{\theta_i}^2 h(\theta)  T_i^{-\top}$. In addition, 
\[
\begin{aligned}
& T_d^{-\top} T^{-1} \nabla_\theta^2 h(\theta) T^{-\top}  \\
& = \begin{bmatrix}
\Sigma_a & 0 \\ -\Sigma_g T_{ga} T_a^{-1} & \Sigma_g
\end{bmatrix}
\begin{bmatrix} \nabla_{\theta_a}^2 h(\theta) T_a^{-\top}  & \{\nabla_{\theta_a, \theta_g}^2 h(\theta) - \nabla_{\theta_a}^2 h(\theta) T_a^{-\top}  T_{ga}^\top  \} T_g^{-\top} \\  
\nabla_{\theta_g, \theta_a}^2 h(\theta) T_a^{-\top} & \{\nabla_{ \theta_g}^2 h(\theta) - \nabla_{\theta_g, \theta_a}^2 h(\theta) T_a^{-\top}  T_{ga}^\top  \} T_g^{-\top}  
\end{bmatrix} \\
& = \begin{bmatrix}
\Sigma_a \nabla_{\theta_a}^2 h(\theta) T_a^{-\top} & \cdot \\
U_{ga}  & U_{ga} T_{ga}^\top T_g^{-\top} - U_{gg}
\end{bmatrix} \\
& = \begin{bmatrix}
\Sigma_1 \nabla_{\theta_1}^2 h(\theta) T_1^{-\top} & \dots & \cdot  & \cdot \\
\vdots & \ddots & \vdots & \vdots \\
\cdot & \dots & \Sigma_n \nabla_{\theta_n}^2 h(\theta) T_n^{-\top} & \cdot\\
U_{g1} & \dots & U_{gn} & U_{gg} - U_{ga} T_{ga}^\top T_g^{-\top} 
\end{bmatrix} \\
\end{aligned} 
\]
Hence an unbiased estimate of $\bar{G}$ can be obtained by setting elements in $-T_d^{-\top} T^{-1} \nabla_\theta^2 h(\theta) T^{-\top}$ which correspond to zero entries in $T$ to zero.

\section{Logistic regression}
Let $y=(y_1, \dots, y_n)^\top  $, $X= (x_1^\top  , \dots, x_n^\top  )$ and $w = (w_1, \dots, w_n)^\top  $, where $w_i = \exp(x_i^\top   \theta)/\{1+\exp(x_i^\top   \theta)\}$. Let $W$ be a diagonal matrix with the $i$th element given by $w_i(1-w_i)$ for $i=1, \dots, d$. Then
\[
\begin{gathered}
\log p(y, \theta) = y^\top   X \theta - \sum_{i=1}^n \log \{1 + \exp(x_i^\top   \theta)\} - \frac{d}{2} \log (2\pi \sigma_0^2) - \frac{\theta^\top   \theta}{2 \sigma_0^2}, \\
\nabla_\theta \log p(y, \theta)  = X^\top   (y- w) - \theta/\sigma_0^2, 
\quad
\nabla_\theta^2  \log p(y, \theta) = - X^\top  W X - I_d/\sigma_0^2,
\end{gathered}
\]

\section{Deep GLMs}
Consider $y_i$ following the Bernoulli or normal distribution. We first discuss treatment of the normal distribution as it has a dispersion parameter and is more complicated. The Bernoulli distribution can be treated similarly by removing terms associated to the dispersion parameter and replacing the likelihood function appropriately.

\subsection{Response from normal distribution}
Suppose $y_i \sim \N(\eta_i, v)$ for $i=1, \dots, n$ and $g(\cdot)$ is the identity link function. Let the prior for $v$ be an inverse gamma density with shape and scale parameters $a_0$ and $b_0$ respectively. We have
\[
\begin{aligned}
& \log p(y, \Theta|a_0, b_0, \gamma) \\
&= \sum_{i=1}^n p(y_i|\theta, v) + \log p(v|a_0, b_0)  + \log p(w|\gamma_w) + \sum_{j=1}^p \{\log p(w_{x_j}|\tau_j) + \log p(\tau_j |\gamma_j) \} \\
&= -\frac{n}{2}\log(2\pi) - \frac{n}{2} \log v - \frac{1}{2v}\sum_{i=1}^n (y_i - \eta_i)^2 + a_0 \log b_0 - \log \Gamma(a_0) - (a_0+1) \log v - \frac{b_0}{v} 
\end{aligned}
\]  
\[
\begin{aligned}
& \quad - \frac{d_w}{2}\log(2\pi) + \frac{d_w}{2} \log \gamma_w - \frac{\gamma_w}{2} w^\top w - \frac{mp}{2}\log(2\pi) - \frac{p(m+1)}{2} \log(2) \\
& \quad - p\log \Gamma(\tfrac{m+1}{2}) + \sum_{j=1}^p \left\{ (m+1)\log \gamma_j - \frac{w_{x_j}^\top w_{x_j}}{2\tau_j}  - \frac{1}{2}\log \tau_j - \frac{\gamma_j^2 \tau_j}{2} \right\}.
\end{aligned}
\]
The optimal density for $q(v)$ is  
\[
\begin{aligned}
q(v) &\propto \exp \E_{-v} \{\log p(y, \Theta|a_0, b_0, \gamma)\} \\
&\propto \exp \E_{-v} \left\{- \frac{n}{2} \log v - \frac{1}{2v}\sum_{i=1}^n (y_i - \eta_i)^2 - (a_0+1) \log v - \frac{b_0}{v} \right\} \\
&\propto \exp \E_{-v} \left\{ - \left(a_0+ \frac{n}{2} + 1 \right) \log v - \frac{b_0 + \sum_{i=1}^n (y_i - \eta_i)^2/2}{v} \right\}.
\end{aligned}
\]
Hence $q(v)$ is an inverse-gamma density with shape and scale parameters $a = a_0+ \frac{n}{2}$ and $b=b_0 + \sum_{i=1}^n \E_{q_\lambda}(y_i - \eta_i)^2/2$. Note that the update for $a$ need only be applied once at the beginning. In addition, $\E_q(\log v) = \log(b) - \psi(a)$ and $\E_q(1/v) = a/b$ where $\psi(\cdot)$ is the digamma function. The optimal density of $q(\tau_j)$ for $j=1, \dots, p$ is 
\[
\begin{aligned}
q(\tau_j) &\propto \exp \E_{-\tau_j} \{\log p(y, \Theta|a_0, b_0, \gamma)\} \\
&\propto \exp \left\{ - \frac{\E_{q_\lambda} (w_{x_j}^\top  w_{x_j})}{2\tau_j} - \frac{1}{2} \log \tau_j  - \frac{\gamma_j^2 \tau_j}{2} \right\} \\
&\propto \exp \left\{ - \frac{\mu_{x_j}^\top \mu_{x_j} + \sigma_{x_j}^\top  \sigma_{x_j}}{2\tau_j} - \frac{1}{2} \log \tau_j  - \frac{\gamma_j^2 \tau_j}{2} \right\},
\end{aligned}
\]
where $\mu_{x_j}$ and $\sigma^2_{x_j}$ are the variational means and variances of the weights that connect $x_j$ to hidden units in the first layer. The form of this density implies that the variational posterior of $1/\tau_j$ is the inverse Gaussian density with parameters, 
\[
\alpha_{\tau_j} = \frac{\gamma_j}{\sqrt{\mu_{x_j}^\top \mu_{x_j} + \sigma_{x_j}^\top  \sigma_{x_j}}}, \qquad 
\beta_{\tau_j} = \gamma_j^2.
\]
Note that $\E_q(1/\tau_j) = \alpha_{\tau_j}$ and $\E_q(\tau_j) = 1/\alpha_{\tau_j} + 1/\beta_{\tau_j}$. 

The evidence lower bound is 
\[
\begin{aligned}
\mL &= \E_q\{ \log p(y, \Theta|a_0, b_0, \gamma) \} - \E_q \{\log q(\theta)\} - \E_q \{\log q(v)\} - \sum_{j=1}^p \E_q \{\log q(\tau_j)\} \\
&= \frac{d + p(1-m) - n - d_w}{2}\log(2\pi) + \frac{p + d}{2} - p\log \Gamma(\tfrac{m+1}{2})- \frac{p(m+1)}{2} \log(2) + a_0 \log b_0 \\
& \quad - \log \Gamma(a_0) + \frac{1}{2} \log |\Sigma|  + \left(a - a_0 - \frac{n}{2} \right) \{\log(b) - \psi(a)\} - \frac{a}{2b}\sum_{i=1}^n \E_{q_\lambda}(y_i - \eta_i)^2  \\
& \quad + \frac{d_w}{2} \log \gamma_w - \frac{\gamma_w (\mu_w^\top \mu_w + \sigma_w^\top  \sigma_w )}{2} - a \log b + \log \Gamma(a) + a\left( 1 - \frac{b_0}{b}\right)\\
\end{aligned}
\]
\[
\begin{aligned}
& \quad  + \sum_{j=1}^p \left\{ (m+1)\log \gamma_j - \frac{1}{2} \log \beta_{\tau_j} - \frac{\alpha_{\tau_j} (\mu_{x_j}^\top \mu_{x_j} + \sigma_{x_j}^\top  \sigma_{x_j})}{2} - \frac{\gamma_j^2 }{2} \left( \frac{1}{\alpha_{\tau_j}} + \frac{1}{\beta_{\tau_j}} \right) \right\}.
\end{aligned}
\]
where $\mu_w$ and $\sigma^2_w$ are the variational means and variances corresponding to the weights $w$, which are not in the first hidden layer. Note that after performing the update $a= a_0 + n/2$, the term $\left(a - a_0 - \frac{n}{2} \right) \{\log(b) - \psi(a)\} = 0$ and can be omitted from $\mL$.

To determine appropriate values for the shrinkage parameters $\gamma$, we use the empirical Bayes approach proposed by \cite{Tran2020}, which can also be interpreted as selecting  $\gamma$ to be the values that maximize the evidence lower bound. We have 
\[
\begin{aligned}
\frac{\partial \mL}{\partial \gamma_w}
&= \frac{d_w}{2 \gamma_w} - \frac{\mu_w^\top \mu_w + \sigma_w^\top  \sigma_w}{2} = 0 \implies \gamma_w = \frac{d_w}{\mu_w^\top \mu_w + \sigma_w^\top  \sigma_w}, \\
\frac{\partial \mL}{\partial \gamma_j}
&= \frac{m+1}{\gamma_j} - \gamma_j \left( \frac{1}{\alpha_{\tau_j}} + \frac{1}{\beta_{\tau_j}} \right) = 0 \implies \gamma_j = \sqrt\frac{m+1}{  1/\alpha_{\tau_j} + 1/\beta_{\tau_j}}.
\end{aligned}
\]

Note that all the terms in the lower bound can be calculated analytically except $\sum_{i=1}^n \E_{q_\lambda}(y_i - \eta_i)^2$ because $\eta_i$ is the output of the deep neural network, which is a nonlinear function of $\theta$, the weights and biases in the neural network. At each iteration, a minibatch $B$ of observations is processed and we compute an unbiased estimate, $\frac{n}{|B|}\sum_{i\in B}(y_i - \widehat{m}_i)^2$, where $\hat{m}_i$ is the output of the neural network based on $\theta$ generated randomly from the existing variational approximation $q_\lambda (\theta)$. An unbiased estimate of the gradient of $\sum_{i=1}^n \E_{q_\lambda}(y_i - \eta_i)^2$ is computed using automatic differentiation.

\subsection{Response from Bernoulli distribution}
Let $\eta_i$ be the output of the neural network. If $y_i \sim \text{Bernoulli}(p_i)$, then the lower bound is 
\[
\begin{aligned}
\mL &= \frac{d + p(1-m) - d_w}{2}\log(2\pi) + \frac{p + d}{2} - p\log \Gamma(\tfrac{m+1}{2})- \frac{p(m+1)}{2} \log(2)  \\
& \quad + \frac{d_w}{2} \log \gamma_w - \frac{\gamma_w (\mu_w^\top \mu_w + \sigma_w^\top  \sigma_w )}{2}  + \frac{1}{2} \log |\Sigma| + \sum_{i=1}^n \E_{q_\lambda}\{ y_i \eta_i - \log(1 + \e^{\eta_i}) \}\\
& \quad  + \sum_{j=1}^p \left\{ (m+1)\log \gamma_j - \frac{1}{2} \log \beta_{\tau_j} - \frac{\alpha_{\tau_j} (\mu_{x_j}^\top \mu_{x_j} + \sigma_{x_j}^\top  \sigma_{x_j})}{2} - \frac{\gamma_j^2 }{2} \left( \frac{1}{\alpha_{\tau_j}} + \frac{1}{\beta_{\tau_j}} \right) \right\}.
\end{aligned}
\] 

\end{document}